\documentclass[aps,prb,twocolumn,showpacs]{revtex4-1}
\pdfoutput=1
\usepackage{amsmath,amssymb}
\usepackage[mathscr]{eucal}
\usepackage{graphicx}

\usepackage[dvipsnames]{color}
\definecolor{MyRed}{named}{BrickRed}
\definecolor{MyBlue}{named}{RoyalBlue}
\definecolor{MyGreen}{named}{ForestGreen}
\usepackage[unicode,
	colorlinks=true,
	linktocpage=true,
	linkcolor=red,
	citecolor=blue,
	urlcolor=blue,
	pagecolor=green
	]{hyperref}

\begin{document}

\title{Two components of critical current in YBa$_2$Cu$_3$O$_{7-\delta}$ films}
\author{A.~V. Kuznetsov, I.~I. Sanninkov and A.~A. Ivanov}
\affiliation{National Research Nuclear University ``MEPhI'' (Moscow Engineering Physics Institute), 
Kashirskoe sh. 31, 115409 Moscow, Russia}

\begin{abstract}
	Combined action of weak and strong pinning centers on the vortex lattice complicates magnetic
	behavior of a superconductor since temperature and magnetic field differently affect weak and
	strong pinning. In this paper we show that contributions of weak and strong pinning into
	magnetization of the layered superconductor YBa$_2$Cu$_3$O$_{7-\delta}$ can be separated and
	analyzed individually. We performed a careful analysis of temperature behavior of the relaxed
	superconducting current $J$ in YBa$_2$Cu$_3$O$_{7-\delta}$  films which revealed two components of
	the current  $J = J_1 +J_2$. A simple method of separation of the components and their  temperature
	dependence in low magnetic fields are discussed. We found that $J_1$ is produced by weak collective
	pinning on the oxygen vacancies in CuO$_2$ planes while $J_2$ is caused by strong pinning on the
	Y$_2$O$_3$ precipitates. $J_1$ component weakly changes with field and quasi-exponentially decays
	with temperature, disappearing at $T \simeq 30$--40~K. Rapid relaxation of $J_1$ causes formation
	of the normalized relaxation rate peak at $T \simeq 20$~K.  $J_2$ component is suppressed by field
	as $J_2\propto B^{-0.54}$ and decays with temperature following to the power law $J_2\propto(1 -
	T/T_\mathrm{dp} )^\alpha$ where $T_\mathrm{dp}$ is the depinning temperature. Detailed comparison
	of the experimental data with pinning theories is presented.
\end{abstract}
\pacs{74.25.Wx, 74.25.Sv, 74.72.-h, 74.78.-w}
\maketitle

\section{Introduction}\label{sec:Introduction}

Pinning of vortices on defects in type-II superconductors leads to formation of a critical state and
appearance of the critical current $J_c$.\cite{Campbell-AP-1972, Larkin-JLTP-1979, Blatter-RMP-1994,
Brandt-RPP-1995} Thermal fluctuations reduce the pinning strength and activate jumps of vortices
between pinning centers (defects).\cite{Blatter-RMP-1994, Feigelman-PRB-1990, Nelson-PRB-1993,
Brandt-RPP-1995}  High temperature  superconductors (HTSC) have a small activation energy and a 
high probability of thermal activation of vortices motion which leads to a giant magnetic flux creep and
decay of the superconducting current over time.\cite{Blatter-RMP-1994, Brandt-RPP-1995,
Yeshurun-RMP-1996} As a result the measured current $J$ becomes lower than
$J_c$.\cite{Yeshurun-RMP-1996}

For HTSC materials the maximal currents are achieved in a highly textured
YBa$_2$Cu$_3$O$_{7-\delta}$ films. 2G-tapes with a metal base and the superconducting
YBa$_2$Cu$_3$O$_{7-\delta}$ layer were developed for high-current
applications.\cite{Malozemoff-SST-2008, Senatore-SST-2016}  Great work was done on studying and
optimization of the defects landscape in 2G-tapes to obtain high currents in external magnetic
fields and at high temperatures, see Ref.~\onlinecite{Foltyn-NM-2007} for review. As a result,
several manufactures produce now long-length tapes with $J$ of several MA/cm$^2$ at liquid nitrogen
temperature.\cite{Senatore-SST-2016} Nevertheless the task of improving performance of the tapes by
combinations of artificial and natural pinning centers\cite{Maiorov-NM-2009} is still actual. 
Investigations of pinning on natural defects in standard YBa$_2$Cu$_3$O$_{7-\delta}$ films play a
major role in that work.

There is a large variety of defects in YBa$_2$Cu$_3$O$_{7-\delta}$ films such as vacancies,
substituting or extra atoms, dislocations, non-superconducting inclusions and so
on.\cite{Foltyn-NM-2007} The latter two act as \textit{strong} pinning centers. The
dislocations\cite{Dam-N-1999, Huijbregtse-PRB-2000, Pan-PRB-2006} are induced close to the
substrate-film interface and develop up to the film surface.\cite{Huijbregtse-PRB-2000}  The
nano-sized Y$_2$O$_3$ precipitations\cite{Catana-APL-1992, Selinder-PC-1992, Kastner-PC-1995,
Verbist-PC-1996, vanderBeek-PRB-2002, Ijaduola-PRB-2006, Maiorov-NM-2009, Miura-PRB-2011}
are spontaneously formed during deposition of YBa$_2$Cu$_3$O$_{7-\delta}$ films.  It was shown that
increase of the inclusion density rises the superconducting current\cite{Selinder-PC-1992} while
increase of the dislocation density reduces suppression of the current by magnetic field
$H$.\cite{Dam-N-1999, Huijbregtse-PRB-2000}

Point defects, mainly oxygen vacancies in superconducting CuO$_2$ planes, act as \textit{weak}
pinning centers. Weak pinning affects magnetic behavior of YBa$_2$Cu$_3$O$_{7-\delta}$ films at
temperatures below 30--40~K. For example,  the exponential dependence  $J\propto \exp(-T/T_0)$ was
observed in the range $ T< 60$~K\cite{Moraitakis-SST-1999, Peurla-SST-2005, Senatore-SST-2016,
Gutierres-APL-2007, Puig-SST-2008, Polat-PRB-2011}  while at high temperatures the current decay
follows a power law $J\propto [1-(T/T_c)^n]^\alpha$.\cite{Pashitskii-LTP-2001, Fedotov-LTP-2002,
Djupmyr-PRB-2005, Ijaduola-PRB-2006, Albrecht-JP:CM-2007, Miura-PRB-2011}. Here  $\alpha =1.2$--2,
$n=1$ (Refs.~\onlinecite{Fedotov-LTP-2002, Pashitskii-LTP-2001, Djupmyr-PRB-2005,
Albrecht-JP:CM-2007})  or 2 (Refs.~\onlinecite{Ijaduola-PRB-2006, Miura-PRB-2011}) and
$T_0\simeq17$--32~K.\cite{Moraitakis-SST-1999, Peurla-SST-2005, Senatore-SST-2016,
Gutierres-APL-2007, Puig-SST-2008, Polat-PRB-2011}

The magnetic flux creep also changes at low temperatures. A peak of the normalized relaxation rate
of the current $S=|d\ln J/d\ln t|$ was observed in YBa$_2$Cu$_3$O$_{7-\delta}$ films at $ T\simeq
20$~K\cite{Maiorov-NM-2009, Jung-SST-1999, Yan-PRB-2000} and the quantum creep was found  below
1~K.\cite{Yeshurun-RMP-1996, Fruchter-PRB-1991, vanDalen-PRB-1996, Hoekstra-PRB-1999,
Landau-PC-2000:2} The quantum creep and  a crossover to two-dimensional superconducting behavior
observed at $ T<80$~K\cite{Farrell-PRL-1990} revealed an importance of layered structure for
superconductivity in YBa$_2$Cu$_3$O$_{7-\delta}$.

The analysis of the critical state in HTSC is complicated by presence of weak and strong pinning and
the layered structure of HTSC materials. If pinning is weak, the elastic forces of the vortex
lattice dominates over the pinning forces.\cite{Larkin-JLTP-1979, Blatter-RMP-1994, Brandt-RPP-1995,
Blatter-PRL-2004} In this case the concerted action of many weak pins on the elastic vortex lattice
is described by the collective pinning theory (CP theory).\cite{Blatter-RMP-1994} The collective
pinning depends only slightly on parameters of individual pinning centers therefore CP theory is
easy to generalize. If pinning is strong, defects acts individually and introduce plastic
deformations in the vortex system.\cite{Larkin-JLTP-1979, Blatter-RMP-1994, Blatter-PRL-2004} In
this case pinning depends on parameters of the defects so many various models were developed for
different kinds of defects.\cite{Blatter-RMP-1994, Feigelman-PRB-1990, Ovchinnikov-PRB-1991,
Nelson-PRB-1993, Gurevich-PRB-1998, Pashitskii-LTP-2001, Pan-PRB-2006}

Strong pinning in a layered superconductor, which contains point defects in the superconducting
planes and three-dimensional defects in the bulk, was analyzed by Ovchinnikov and
Ivlev\cite{Ovchinnikov-PRB-1991} (OI theory). They found that the critical current $J_c$ of such
superconductor consists of two components produced by in-plane and in-volume pinning. Further
developing the OI theory, van~der Beek \textit{et al.} considered in-volume  pinning and calculated
the dependence of $J_c$ on film thickness $d$ and temperature.\cite{vanderBeek-PRB-2002} The
dependence $J(d)$ for thin YBa$_2$Cu$_3$O$_{7-\delta}$ films was successfully
described\cite{vanderBeek-PRB-2002, Ijaduola-PRB-2006} in the frame of extended OI
theory\cite{vanderBeek-PRB-2002} by pinning on Y$_2$O$_3$ inclusions. At the same time the extended
theory agree with experimental data on $J(T)$ and $H^*(T)$ only for
$T>30$~K\cite{vanderBeek-PRB-2002, Ijaduola-PRB-2006} since the in-plane pinning was completely
ignored by van~der Beek \textit{et al.} Here $H^*$ is the crossover field above which $J$ becomes
field dependent.

There exists the model describing strong pinning on edge dislocations at low-angle boundaries of
crystallites in YBa$_2$Cu$_3$O$_{7-\delta}$ films (EDP model).\cite{Pan-PRB-2006, Fedotov-LTP-2002,
Pan-IEEETAS-2003}  For some samples this model  approximates well the field dependence of the
current in a wide range of fields.\cite{Pan-PRB-2006, Pan-IEEETAS-2003, Kosse-SST-2008}  At the same
time the EDP model is not universal for all YBa$_2$Cu$_3$O$_{7-\delta}$ films, some conditions are
necessary for its correct application.\cite{Kosse-SST-2008} Restriction of the EDP model may be
caused by neglecting of weak pinning which may influence $J(H)$ behavior at low temperatures. 
$J(T)$ behavior that follows from the EDP model haven't been tested yet.

To clarify the role of weak and strong pinning we performed a careful analysis of temperature
behavior of the relaxed superconducting current in different magnetic fields  in
YBa$_2$Cu$_3$O$_{7-\delta}$ films. The analysis allowed us to separate and describe the current
components produced by weak and strong pinning.

The paper is organized as follows. Samples and details of $J(T)$ measurements are discussed in
Sec.~\ref{sec:Experimental}. Experimental results are presented in Sec.~\ref{sec:Results}. At first
we show that $J(T)$ behavior  observed in our experiments is common for YBa$_2$Cu$_3$O$_{7-\delta}$
films. Then analyzing experimental data we separate currents produced by in-plane and in-volume
pinning. The separated current components are analyzed in Sec.~\ref{seq:Discussion} We  discuss a
relationship between low-temperature peak of the relaxation rate and component of the current
produced by weak pinning. Then we show that this component is caused by single-vortex collective
pinning in Cu-O$_2$ planes and try to describe it in the frame of CP theory. At the end we consider
the component produced by strong pinning and show that is well described by OI theory extended for
strong pinning on Y$_2$O$_3$ inclusions. Our conclusions are presented in Sec.~\ref{seq:Summary}.
The critical current following from OI theory for magnetic field applied along a normal to the
superconducting planes is calculated in Appendix.
 
\section{Experimental details}\label{sec:Experimental}

Thin epitaxial films of YBa$_2$Cu$_3$O$_{7-\delta}$ were prepared by pulsed laser deposition
technique using KrF excimer laser. Disk-shaped single crystal plates of SrTiO$_3$ (100) were
used as substrates. The deposition took place at substrate temperature about 750$^\circ$~C, the
oxidizer pressure (N$_2$O or O$_2$) varied from 400 to 800~mtorr in different experiments. The
velocity filter was used to select the fine part of the ablation plume (atoms  and clusters of small
size) and obtain better quality of the film surface.\cite{Ivanov-PC-1991, Pechen-APL-1995} 

The film structure was analyzed by XRD at D8 Discover diffractometer (Bruker) using Cu-$K_\alpha$
radiation. The study confirmed that films were epitaxial and $c$-oriented. No additional phase was
detected. The peaks (002), (005) and (007)  were used to determine the $c$ lattice parameter and
estimate  the values of coherent scattering regions and  microdeformation. The  $c$ lattice
parameter was in the range 11.70--11.73~\AA, the rocking curve widths $\omega$ of the (005) Bragg
peak for best samples was less than 0.2 degree. The oxygen content varied depending on oxidation
condition and brought about the variation of $c$-parameter.  As follows from the values of the
structure parameters  presented in Table~\ref{tab:samples}, the films had high-quality crystalline
structure with small microdeformation and disorientation.

The critical temperatures of the superconducting transition $T_c=90$--91~K were obtained in
resistivity measurements performed on witness-samples made in the same deposition process. The
samples demonstrated sharp transitions with width of about 1~K. SQUID-magnetometry and study of 
the magnetic susceptibility in an alternating magnetic field  were used to measure temperature of the
magnetic transition $T_c^M$. Obtained $T_c$ and $T_c^M$ values are presented in the
Table~\ref{tab:samples}.

\begin{table*}
	\caption{Parameters of the YBa$_2$Cu$_3$O$_{7-\delta}$ thin film samples.\footnote{ 
		The  (002), (005) and  (007) Bragg peaks were used to obtain the lattice parameter $c$, size of
		the coherent scattering regions (CDB)  and the microdeformation $\varepsilon_\text{micro}$.\\
		The full width on half maximum  (FWHM) and the rocking curve width $\omega$ were
		   measured for the (005) Bragg peak.\\ 
		   $T_\mathrm{dp}$ and $J$ was measured in field $H=910$~Oe for samples Y1--Y3 and 
		   1530~Oe for sample Y4.\\
		   $J$ was taken at $T=4.21$~K. $J_1$ and $J_2$ are presented for zero temperature.\\ 
		   $D_\mathit{iz}$, $n_i$ and $n_i^*$ were calculated for $D_i$ and $n_i d_\mathit{iz}^{9/4} = 
		   n_i(D_\mathit{iz}/\xi_0)^{9/4}$ obtained via fit of experimental curves.\\
		   $\langle L\rangle$ was calculated for $B=910$~G.}}
		\label{tab:samples}
	\begin{ruledtabular}
		\begin{tabular}{ccc|ccccc|cccccc|ccccc|cc}
\# & $D$ & $d$ & $c$    & FWHM & $\omega$ & CDB
   &$\varepsilon_\text{micro}$&$T_c$&$T_c^M$&$T_\mathrm{dp}$ &$J$ & $J_1$ & $J_2$  
   & $D_i$ & $n_id_\mathit{iz}^{9/4}$ & $D_\mathit{iz}$ & $n_i$ & $n_i^*$ & $r_d$ & $\langle L\rangle$ \\   
& mm  & nm  & \AA    & \multicolumn{2}{c}{degree}  & nm  & \%  & K   & K   & K    &
\multicolumn{3}{c|}{$10^6$ A/cm$^2$}	& nm & $10^{16}\text{cm}^{-3}$ & 
nm & \multicolumn{2}{c|}{$10^{15}\text{cm}^{-3}$} & \AA & nm \\ \hline
Y1& 2.1& 550& 11.707& 0.096& 0.112& 969& 0.11& 90 & 88   & 84& 13.2& 8.0& 5.9
&$ 5 \pm 1$  & $2.4 \pm 0.4$ & 3.9 & 4 & 2 & 4.6 & 284\\
Y2& 1.8& 300& 11.697& 0.129& 0.198& 206& 0.10& 90 & 89   & 88& 10.4& 3.8& 7.2
& $ 14 \pm 2$ & $0.7 \pm 0.2$ & 1.8 & 7 & 0.4 & 4.3 & 37\\
Y3& 1.8& 280& 11.713& 0.299& 0.601& 96  & 0.23&90.5&89.5& 82& 9.1  & 1.2& 8.3
& $ 14 \pm 2$ & $1.1 \pm 0.3$ & 1.8 & 11 & 0.7 & 5.3 & 46\\
Y4& 2.0& 380& 11.725& 0.277& 0.729& 150& 0.26& 91 & 89   & 75& 12.8& 8.3& 5.2
& $ 1.8 \pm 1$ & $80 \pm 40$ & 14 & 7 & 790 & 4.6 & ---
		\end{tabular}
	\end{ruledtabular}
\end{table*}

Measurements of a persistent current induced in YBa$_2$Cu$_3$O$_{7-\delta}$ film under change of
magnetic field were performed using home-built SQUID magnetometer.\cite{Trofimov92:1,
Kuznetsov-PRB-1995} During the measurements a sample was placed inside a copper tube  isolated 
from
the LHe bath by a vacuum jacket. Temperature of the sample varied in the range from 4.21 to 300~K
via heating of the tube  filled with exchange-helium gas. Magnetic field was produced by a NbTi tube
enclosed in NbTi solenoid. To apply a field the solenoid was supplied by current and the tube was
warmed above $T_c$ by a short heat pulse. After freezing of the field in the tube the current was
withdrawn out the solenoid to minimize noises. Magnetic field up to 2100~Oe can be frozen in the
tube of 0.3~mm wall thickness. High fields were applied step by step to prevent overheating of the
superconducting films by the current induced under the abrupt change of the external magnetic
field.\cite{Landau-PRB-2001} At each step the field increment twice exceeded the characteristic
field for flux penetration into the film\cite{Mikheenko-PC-1993, Clem-PRB-1994, Brandt-RPP-1995} to
make sure that the induced current is high enough to create the critical state throughout the
sample.  When measurements were performed in zero applied field, a high field was applied at first
and the sample was maintained several minutes in this field. Then the field was decreased step by
step and at last step the solenoid was warmed together with the tube to remove a magnetic flux
frozen in its wire. Due to strong demagnetization effect a self demagnetizing field is produced by
current flowing in a superconducting film when field is applied perpendicular to the film
plane.\cite{Babaei_Brojeny-SST-2005, Bernstein-JAP-2012, Kuznetsov-IEEETAS-2016} In the critical
state this self field exists even after complete removal of external field. The measurements were
performed for applied fields of 910 and  1530~Oe and in self-field after removing field of 2090~Oe.
These fields were enough to form the homogeneous critical state in samples  at  all temperatures.

\begin{figure}
\centering
\includegraphics[scale=.8]{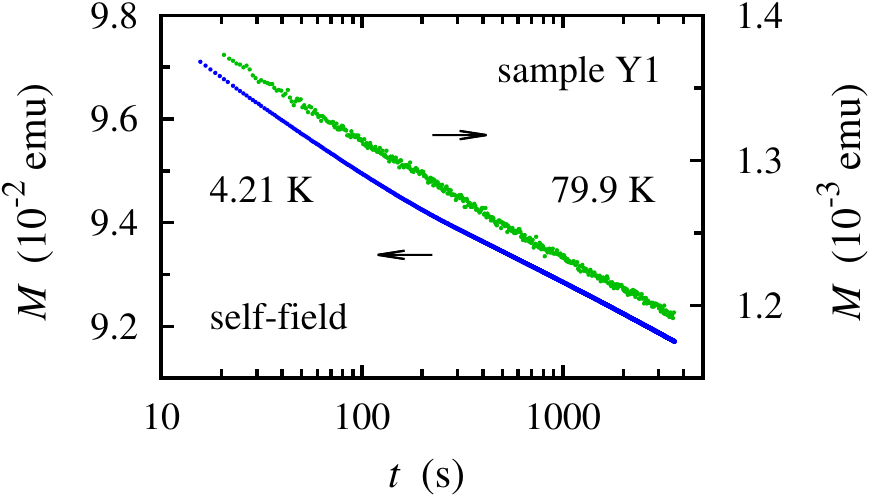}
	\caption{(Color online) Relaxation curves of remanent moment measured for sample Y1 at low and 
	high temperatures. The moment decays by about 5\% and 14\% in time-window of the relaxation 
	measurements.}
	\label{fig:M(t)}
\end{figure}

A method of SQUID magnetometry with motionless sample\cite{Beasley-PR-1969, Kuznetsov-PRB-1995,
Niderost-PRB-1996, Krylov-PRB-1998, Thompson-SST-2005} was used in our experiments. The 
measurements
were performed as follows. The film locked in one of pick-up coils of a superconducting flux
transformer was warmed above $T_c$ and cooled in zero field to a desired temperature (ZFC
procedure). After that a magnetic field was applied perpendicular to the film plane and a signal
caused by change of film magnetic moment with time $\delta M(t)$ due to relaxation was measured for
one hour. Then magnetometer indications were reset and the sample was warmed above $T_c$ in 
order to
record its residual moment $M_*$. Combining $M_*$ with data on  $\delta M(t)$ we precisely obtained
the time dependence of the magnetic moment $M(t)$.\cite{Sannikov-BLPI-2014, 
Kuznetsov-IEEETAS-2016}
Examples of $M(t)$ curves obtained at low and high temperatures are shown in Figure~\ref{fig:M(t)}.
As seen, the noise of the curves is considerably lower than the change of the moment due to
relaxation.

Preliminary results of the relaxation experiments were published elsewhere.\cite{Sannikov-BLPI-2014,
Sannikov-PP-2015-1, Kuznetsov-IEEETAS-2016} In the present work we analyze the current obtained
from the $M_*(T)$ dependences so let us consider this issue in detail.

\begin{figure}
\centering
\includegraphics[scale=.8]{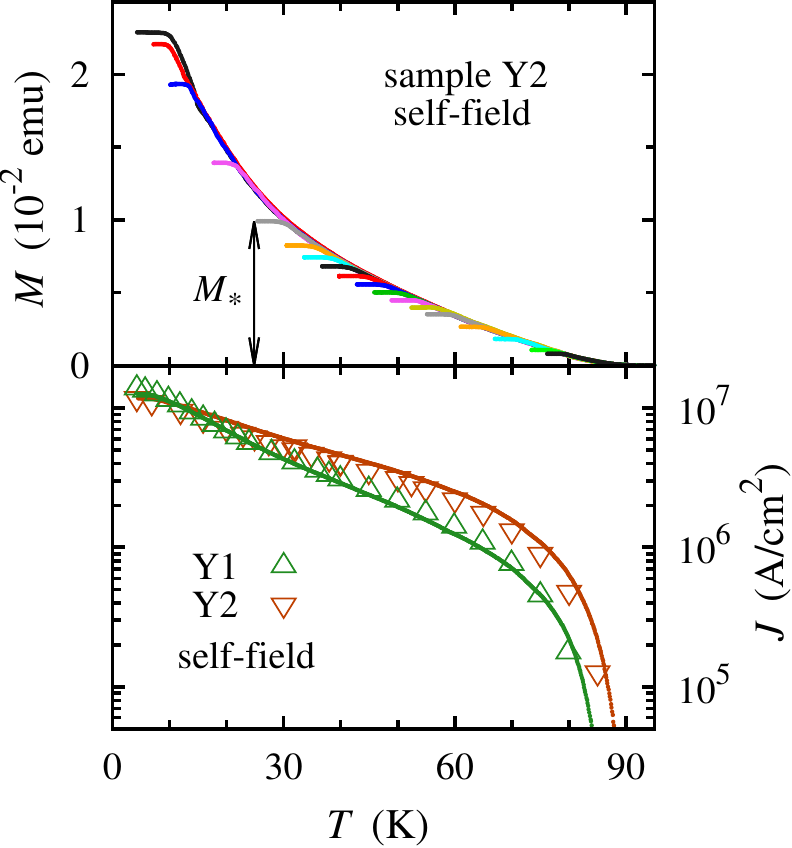}
	\caption{(Color online) \underline{\emph{Top:}} Temperature dependences of remanent moment 
	measured for sample Y2 	after relaxation for 1 hour. Amplitude of the residual moment $M_*$
	obtained for $T=24.8$~K is shown by arrow. 
	\underline{\emph{Bottom:}} $J(T)$ dependences obtained from the relaxed moments $M_*$
	(triangles) and measured under warming of films during temperature sweep (continuous curves).}
	\label{fig:M(T)}
\end{figure}

The sample was heated at the rate of 5~K/min up to $T=95$~K during warming and a signal produced 
by
the magnetometer background and the film moment was recorded in steps of 0.1~K. The background
signal measured without sample was subtracted from the total one to separate the signal produced by
the film only. Inaccuracy of obtaining the film magnetic moment $M$ due to the subtraction did not
exceed 0.5\% at $T\simeq80$~K and was considerably smaller at low temperatures. $M(T)$ dependences
obtained in such a manner are shown in top Fig.~\ref{fig:M(T)}. $M(T)$ curves begin with a plateau
caused by preceding relaxation of the moment. $M$ values at the plateaus are equal to the residual
moments $M_*(T)$.

In bottom Fig.~\ref{fig:M(T)}  we also presented temperature dependences of the current density
calculated as\cite{Mikheenko-PC-1993, Clem-PRB-1994, Brandt-RPP-1995} $J=24Mc/(\pi D^3d)$ where $c$
is the light velocity, $D$ and $d$ are the film's diameter and thickness. Two types of $J(T)$ curves
are shown for comparison. The first type, obtained from the $M_*(T)$ values, corresponds to a
long-time relaxed current. The second one, recorded under film warming immediately after magnetic
field removal, presents a short-time relaxed current. As seen in the bottom Fig.~\ref{fig:M(T)},
shapes of the short- and long-time relaxed curves slightly differ each other. At some temperatures
the long-time relaxed current for sample Y1 is greater than the short-time relaxed one while it
obviously should be smaller. This  artifact is caused by fast temperature sweep during $M(T)$
recording.\cite{hysteresis} Since temperature measurement error can affect the $J(T)$ dependences
recorded under film warming they are presented below mainly as illustrations. At the same time the
$J(T)$ curves obtained from the residual moments correspond to equilibrium temperatures at which the
sample was kept more than hour. These temperatures were stabilized and measured with accuracy 
better
than 0.05~K

\section{Results}\label{sec:Results}

We start from comparison of measured $J(T)$ dependences with published data to clarify 
which features of $J(T)$ behavior are common for  YBa$_2$Cu$_3$O$_{7-\delta}$ films. However the
current $J$ itself should be elucidated first. The critical current $J_c$ determined by pinning
theories cannot be measured directly because of huge Joule heat dissipated in
YBa$_2$Cu$_3$O$_{7-\delta}$ films.\cite{EcJc} 
Therefore either a current  $J_T$ measured in transport experiments or a current
induced by applied magnetic field are used to characterize the superconducting current. $J_T$ is
maintained during measurement by a current source so it does not relax. On the contrary the
persistent current $J$ is affected by creep, therefore it is lower than $J_T$.\cite{E-J} 
Moreover, dependences of $J$ and $J_T$ on $T$ and $H$ can differ especially at
high temperatures and fields. Therefore only data on the persistent current measured in
self-field\cite{ Pashitskii-LTP-2001, Fedotov-LTP-2002, Djupmyr-PRB-2005, Albrecht-JP:CM-2007,
Ijaduola-PRB-2006, Miura-PRB-2011, Lairson-PRB-1990, Moraitakis-SST-1999, Peurla-SST-2005,
Senatore-SST-2016, Polat-PRB-2011, Gutierres-APL-2007, Puig-SST-2008} were chosen for
verification.\cite{Josephson}

\begin{figure*}
\centering
\includegraphics[scale=.8]{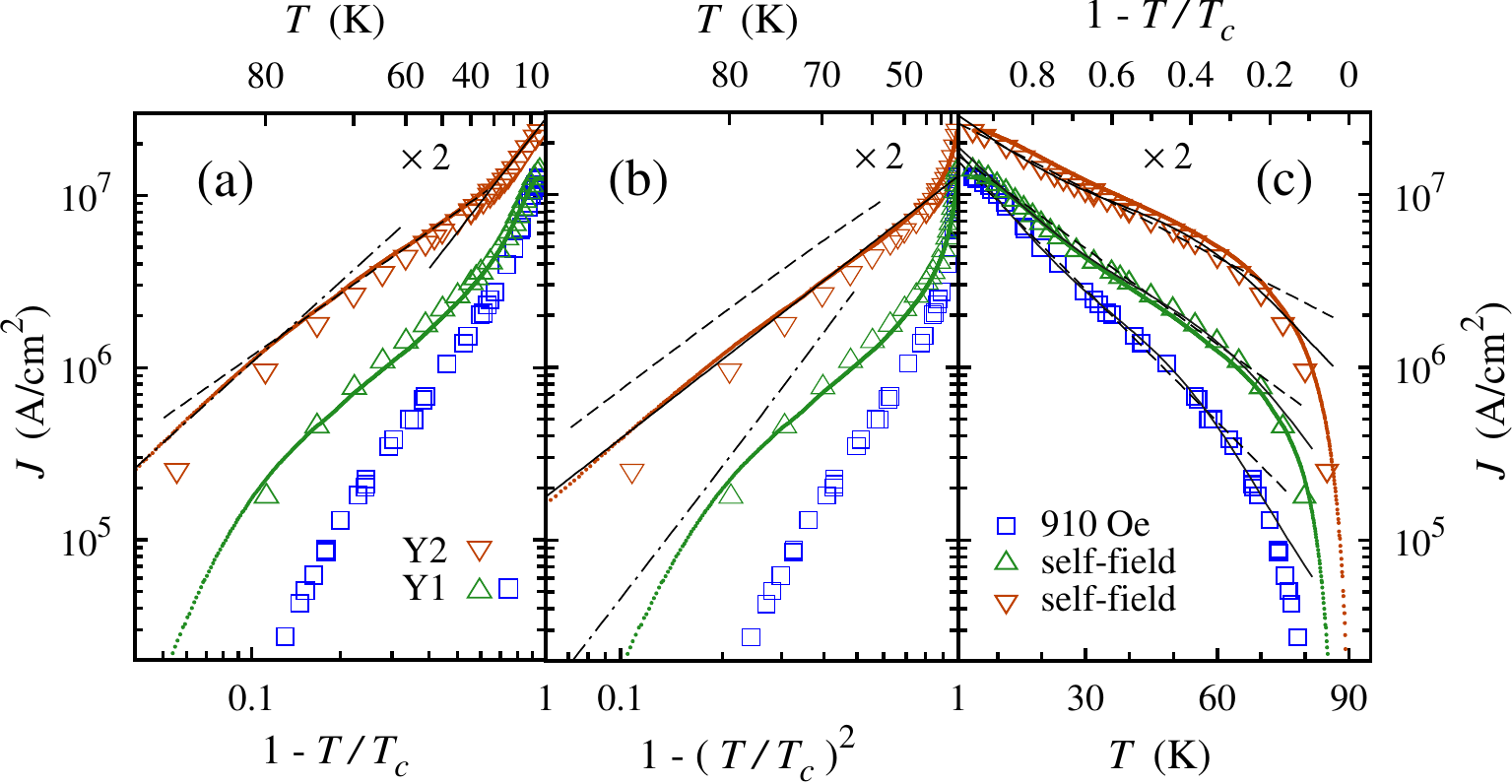}
	\caption{(Color online) Temperature dependences of the current density of
	YBa$_2$Cu$_3$O$_{7-\delta}$ films in different scales. The curves for sample Y2 are shifted 
	(multiplied by factor 2) to avoid a crossing with ones for sample Y1. \textit{\underline{Symbols:}} 
	$J(T)$ obtained after relaxation for 1 hour in
	self-field (triangles) and in field of 910~Oe (squares). Curves with small dots were measured in
	self-field under warming the samples at sweep rate of 5~K/min.
	\textit{\underline{Panel a:}} The lines are approximations $J \propto (1-T/T_c)^\alpha$ with
	$\alpha=2.2$ (solid), 1.2 (dashed) and 1.55 (dash-dotted).
	\textit{\underline{Panel b:}} Solid line is an approximation  $J \propto
	\tau_-^\alpha$ with $\alpha = 1.52$. Dashed and dash-dotted lines are calculated  for strong
	pinning on large $J_c \propto \tau_-^{3/2} \tau_+^{1/2}$ and small $J_c \propto
	\tau_-^{5/2}\tau_+^{-1/2}$ defects.\protect\cite{Klaassen-PRB-2001}
	\textit{\underline{Panel c:}} Dashed lines are fits by $J \propto e^{-T/T_0}$ 
	with $T_0=33$, 23.5~K for Y2, Y1 in self-field and  $T_0=17$~K for $H=910$~Oe. 
	Solid lines are fits by the dependence~\eqref{eq:Jc-exponential}: $T_w=18$, 24~K for Y1, Y2 and
	$T_s=52$~K for both samples in self-field; $T_w=14$~K and $T_s=43$~K for $H=910$~Oe.
	}\label{fig:J(T)_published}
\end{figure*}
	
Representative $J(T)$ curves measured for our samples are shown in Fig.~\ref{fig:J(T)_published}
in different scales to display $J$ behavior in different temperature ranges. To illustrate a field
influence, the curves obtained in self and external field of 910~Oe are shown for sample Y1.

A power law $J_c\propto (1-T/T_c)^\alpha$ is expected for pinning of vortices on boundaries between
crystallites in the films.\cite{Gurevich-PRB-1998, Pashitskii-LTP-2001} Three temperature ranges
with different $\alpha $ values were found for YBa$_2$Cu$_3$O$_{7-\delta}$
films.\cite{Fedotov-LTP-2002, Pashitskii-LTP-2001, Djupmyr-PRB-2005, Albrecht-JP:CM-2007} The powers
$\alpha \simeq1.2$--2, 0.9--1.2 and 1.4--2.5 were obtained respectively at
high\cite{Pashitskii-LTP-2001, Fedotov-LTP-2002} $T\gtrsim77$~K, elevated\cite{Djupmyr-PRB-2005,
Albrecht-JP:CM-2007}  $36\text{ K} \lesssim T\lesssim72$~K and lower\cite{Djupmyr-PRB-2005,
Albrecht-JP:CM-2007} $12\text{ K} \lesssim T\lesssim35$~K  temperatures. As shown in
Fig.~\ref{fig:J(T)_published}(a), our results well agree with the published data. Fitting curves for
sample Y2 demonstrate good approximation by the power law with $ \alpha=1.55$, 1.2 and 2.2 in the
above mentioned ranges. Relaxation slightly affects $J(T)$ at  low and elevated temperatures and 
increases the power at high $T$. External field smoothes $J(T)$ and rises the powers in all
ranges. Summing up we conclude that our results are consistent with published data.

Since pinning parameters depend on the penetration depth $\lambda (T) = \lambda_0/\sqrt{1-(T/T_c)^4}
= \lambda_0/\sqrt{\tau_+\tau_-}$ and the coherence length $\xi (T) =
\xi_0\sqrt{(1+(T/T_c)^2)/(1-(T/T_c)^2)} = \xi_0\sqrt{\tau_+/\tau_-}$ of the
superconductor\cite{Blatter-RMP-1994} one can assume that their temperature change determine $J(T)$
behavior. Here we denoted $\tau_+=1+(T/T_c)^2$, $\tau_-=1-(T/T_c)^2$ and $\lambda_0 = \lambda(0)$,
$\xi_0 = \xi(0)$. In the frame of CP model\cite{Blatter-RMP-1994} Griessen \textit{et al.} obtained
that $J_c \propto \tau_-^{7/6}\tau_+^{5/6}$ for $\delta T_c$ pinning and $J_c \propto \tau_-^{5/2}
\tau_+^{-1/2}$ for $\delta \ell$ pinning.\cite{Griessen-PRL-1994, dTc_dl} Similar expressions were
calculated by Klaassen \textit{et al.} for strong pinning on inclusions of large, $J_c \propto
\tau_-^{3/2} \tau_+^{1/2}$, and small, $J_c \propto \tau_-^{5/2} \tau_+^{-1/2}$,
size.\cite{Klaassen-PRB-2001}

The dependence $J\propto \tau_-^\alpha$ with $\alpha=1.2$--1.4 in self-field at $T\gtrsim50$~K was
observed experimentally in YBa$_2$Cu$_3$O$_{7-\delta}$ films.\cite{Ijaduola-PRB-2006} In field of
100~Oe  for $T\gtrsim40$~K the power $\alpha=1.53$ was  found while at high temperatures
$T\gtrsim83$~K a more rapid decay of $J$ was observed.\cite{Miura-PRB-2011} Our results presented in
Fig.~\ref{fig:J(T)_published}(b) are consistent with the published data. For example, for sample Y2
in self-field for  $T\gtrsim40$~K we obtained $\alpha=1.52$. The range in which a rapid decay of $J$
is observed shifts to lower temperatures in external field. Thus we conclude again that our results
well agree with published data.

The exponential decay $J\propto\exp(-T/T_0)$ was observed  at $T< 50$--60~K in HTSC single
crystals,\cite{Schneemeyer-PRB-1987, Senoussi-PRB-1988, Christen-N-1993, Martinez-PRB-1996}
YBa$_2$Cu$_3$O$_{7-\delta}$ films\cite{Lairson-PRB-1990, Moraitakis-SST-1999, Peurla-SST-2005} and
2G-tapes.\cite{Senatore-SST-2016} Such behavior was attributed to oxygen vacancies acting as weak
pinning centers.  The scaling temperature depends on field, for example $T_0=25$--32~K in
self-field\cite{Peurla-SST-2005, Senatore-SST-2016} and  $T_0=17$--25~K in
$H=1$--200~kOe.\cite{Lairson-PRB-1990, Peurla-SST-2005, Senatore-SST-2016} Both
increase\cite{Senatore-SST-2016} and decrease\cite{Peurla-SST-2005, Senatore-SST-2016} of $T_0$ was
observed in lower fields. As shown in Fig.~\ref{fig:J(T)_published}(c), our results are in good
agreement again with published temperatures. In the range $ T < 60$~K we obtained $ T_0 = 33 $,
23.5~K for self-field and 17~K for $H=910$~Oe.

Studying magnetization of solidified YBa$_2$Cu$_3$O$_{7-\delta}$-Y$_2$BaCuO$_5$ composites
Mart\'{\i}nez \textit{et~al.}\cite{Martinez-PRB-1996}  in the range $40\text{ K} \leqslant
T\leqslant 80 $~K found the dependence $J\propto\exp[-3(T/T^*)^2]$  caused by strong pinning on
nonsuperconducting Y$_2$BaCuO$_5$ precipitates. Here $T^*$ is a characteristic temperature. Authors
also concluded that at low temperatures both weak and strong pinning centers were
effective.\cite{Martinez-PRB-1996} Following this conclusion Plain
\textit{et~al.}\cite{Plain-PRB-2002} proposed the approximation
\begin{equation}\label{eq:Jc_exponential}
	J=J_w \exp(-T/T_w) + J_s \exp\left[-3(T/T_s)^2\right]
\end{equation}
\noindent  for the current in YBa$_2$Cu$_3$O$_{7-\delta}$ films. Here $w$ and $s$ mark the current
components produced by weak and strong pinning. This expression extends the range of the 
exponential
approximation for $J(T)$ to $T\lesssim75$~K. The temperatures $T_w=8$--13~K, $T_s=78$--93~K were
found for YBa$_2$Cu$_3$O$_{7-\delta}$ films.\cite{Polat-PRB-2011, Gutierres-APL-2007, Puig-SST-2008}

The dependence $J_c\propto\exp[-3(T/T^*)^2]$ was calculated in theory of strong pinning on columnar
pins (\textit{line} correlated disorder).\cite{Nelson-PRB-1993} We found that the current $J_c
\propto (T/T^*)^2 \exp\left[-(T/T^*)^3\right]$, calculated for compact pins (\textit{point}
correlated disorder),\cite{Feigelman-PRB-1990} gives a better approximation for standard
YBa$_2$Cu$_3$O$_{7-\delta}$ films. As shown in Fig.~\ref{fig:J(T)_published}(c), in the range $T
\lesssim75$~K the $J(T)$ curves are well fitted by the dependence
\begin{equation}\label{eq:Jc-exponential}
	J_c=J_w \exp(-T/T_w) + J_s \left(\frac{T}{T_s}\right)^2 \exp\left[-(T/T_s)^3\right].
\end{equation}
\noindent We obtained $T_w=18$ and 24~K and $T_s=52$~K in self-field. Approximation of the curves by
Eq.~\eqref{eq:Jc_exponential} gave lower $T_w=8$--10~K and higher $T_s=85$--93~K  values which 
excellently agree again with published data.

The above analysis confirms validity of all proposed earlier approximations for $J(T)$ for our
samples in restricted temperature ranges. The common features of this behavior are a slow quasi
exponential decay at low temperatures and a more rapid power-law decay at high ones. We assumed 
that at least two components are needed to describe $J(T)$ in the whole temperature range. Thermal
fluctuations must also be taken into account at high temperatures since they reduce the effective
pinning strength and lead to depinning of vortices at some temperature $T_\mathrm{dp}$ which is less
than $T_c$.\cite{Blatter-RMP-1994,Deak-PRB-1994} Above the  depinning temperature $T_\mathrm{dp}$
the critical state is destroyed, the persistent current disappears and its relaxation rate becomes
zero. We supposed that in vicinity of   $T_\mathrm{dp}$ the current depends on difference
$(T_\mathrm{dp}-T)$ or its powers and found the depinning temperatures for our samples to check this
point.

\begin{figure}
	\centering \includegraphics[scale=.8]{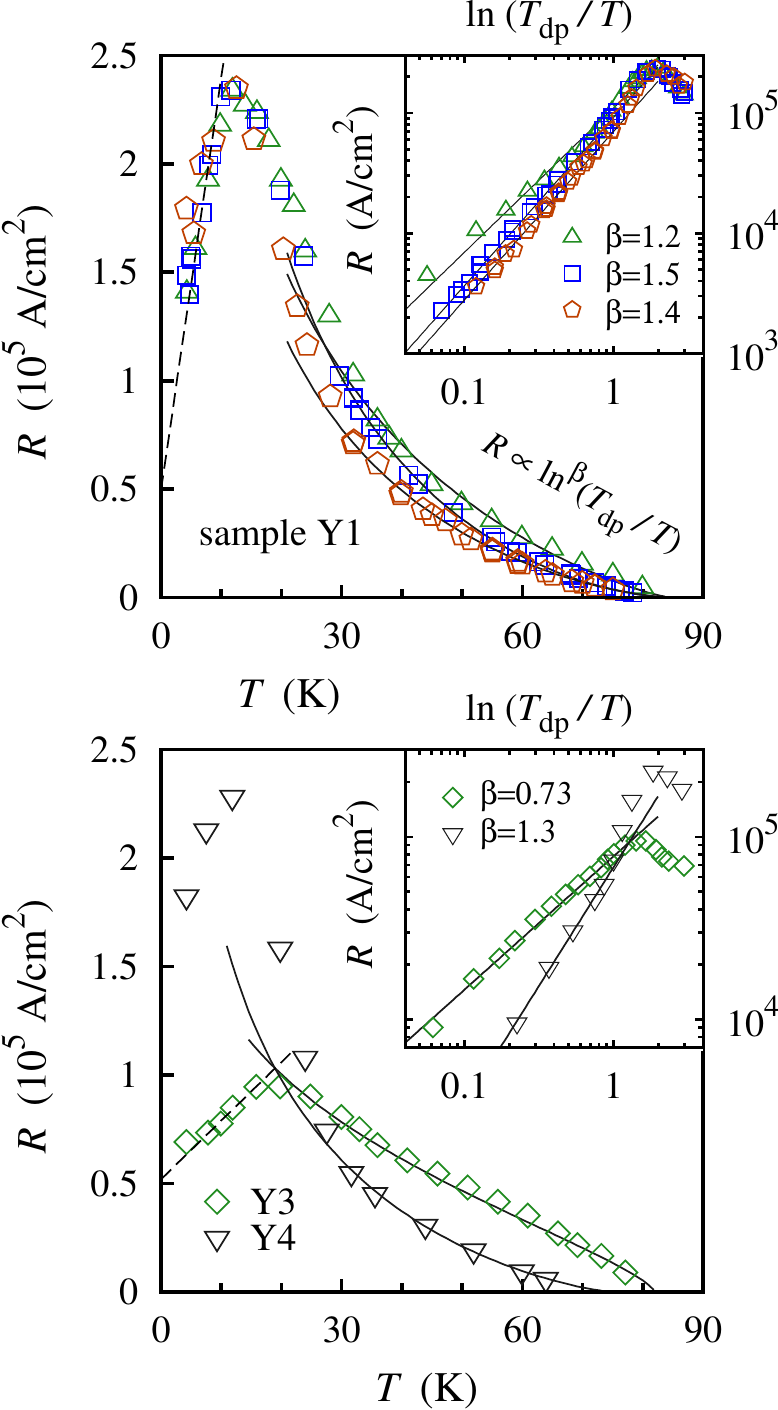} 
	\caption{(Color online) $|dJ/d\ln t|\textit{ vs } T$ for
	YBa$_2$Cu$_3$O$_{7-\delta}$ films obtained after relaxation for 1 hour in fields of 1530~Oe
	(pentagons and triangles down), 910~Oe (squares and diamonds) and in self-field
	(triangles). The dashed lines extrapolate $R$ to zero temperature. The continuous lines are fits
	$R=\mathscr{R}\ln^\beta(T_\mathrm{dp}/T)$. \underline{\emph{Inset:}} $R$	\emph{vs}
	$\ln(T_\mathrm{dp}/T)$ in logarithmic scales.  See text for details.}
	\label{fig:R(T)}
\end{figure}

The relaxation rate $R\equiv|dJ/d\ln t|$ for our films is presented in Fig.~\ref{fig:R(T)}. $R(T)$
curves demonstrate a well-known maximum at low temperatures\cite{Yeshurun-RMP-1996} behind which
they smoothly decrease down to zero. We found that above 30~K the rate is well fitted by the
dependence $R =\mathscr{R}\ln^\beta(T_\mathrm{dp}/T)$. To obtain the fitting parameters
$\mathscr{R}$, $T_\mathrm{dp}$ and $\beta$ we plotted $R$ \emph{vs} $\ln(T_\mathrm{dp}/T)$ in
logarithmic scales and varied $T_\mathrm{dp}$ to straighten the curves as shown in insets of
Fig.~\ref{fig:R(T)}. Then $\mathscr{R}$ and $\beta$ were got as shifts and inclination factors of
the fitting lines. Obtained values of $T_\mathrm{dp}$ and $\beta$ slightly depend on magnetic field.
For sample Y1 we found $T_\mathrm{dp}=84.5 \pm0.5$~K, $\beta=1.2 \pm0.1$ in self-field and
$T_\mathrm{dp}=84 \pm0.5$~K, $\beta=1.4 \pm0.1$ for $H=910$ and 1530~Oe. For sample Y2 the same
$T_\mathrm{dp}=88\pm0.5$~K and $\beta=1.0\pm0.05$ were found for all fields. The obtained depinning
temperatures are presented in Table~\ref{tab:samples}. While the critical temperatures are of the
same order for all samples, their $T_\mathrm{dp}$ strongly differ. For example, $T_\mathrm{dp}
\simeq T_c^M$ for sample Y2 but $T_c^M-T_\mathrm{dp}\simeq14$~K for Y4. To be sure in
$T_\mathrm{dp}$ evaluation we checked their maximal and minimal values by direct measurements of 
the magnetization thermal hysteresis.\cite{Deak-PRB-1994}

\begin{figure}\centering
	\includegraphics[scale=.8]{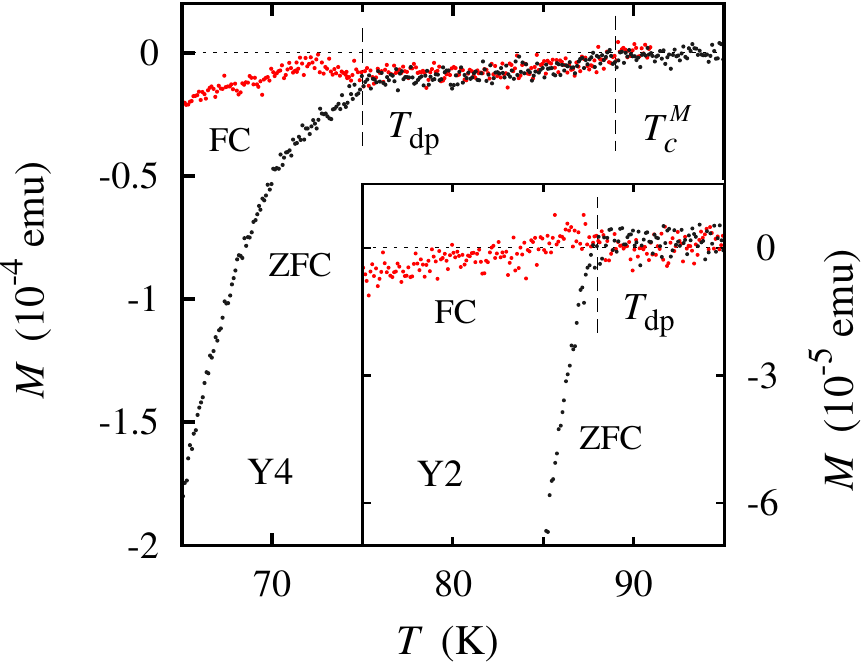}
	\caption{(Color online) $M(T)$  measured in field of 910~Oe under warming (ZFC) and cooling 
	(FC) of samples Y4 and Y2 (inset). Dashed lines mark 	depinning and critical temperatures. To 
	avoid a faulty hysteresis\cite{hysteresis} the low sweep rate of 3~K/min was used in 
	measurements.}
	\label{fig:ZFC-FC}
\end{figure}

$M(T)$ curves measured in field of 910~Oe for samples Y4 and Y2 are presented in
Fig.~\ref{fig:ZFC-FC}. The data were obtained after ZFC procedure under warming and subsequent
cooling of samples in field of 910~Oe. A thermal hysteresis caused by pinning of vortices  is observed
below the depinning temperature in Fig.~\ref{fig:ZFC-FC}. A reversible magnetization is distinctly
seen  above $T_\mathrm{dp}$ up to the critical temperature $T_c^M$ for sample Y4 while for Y2 it is
indiscernible because of small difference between $T_\mathrm{dp}$ and $T_c^M$. The measured
depinning temperatures coincide with $T_\mathrm{dp}$ obtained by fit of $R(T)$ dependences.

Taking obtained $T_\mathrm{dp}$ we found that at high temperatures the current density follows the
power law
\begin{equation}\label{eq:J2(T)}
	J_2= J_2(0)(1-T/T_\mathrm{dp})^\alpha,
\end{equation}
therefore we plotted $J$ \emph{vs}  $1-T/T_\mathrm{dp}$ in logarithmic scales and additionally fit
$T_\mathrm{dp}$ as it was done for the relaxation rate. $T_\mathrm{dp}$ obtained by $R(T)$ and
$J(T)$ fits mostly coincided or differed in the error range of 0.5~K.

\begin{figure}\centering
	\includegraphics[scale=.8]{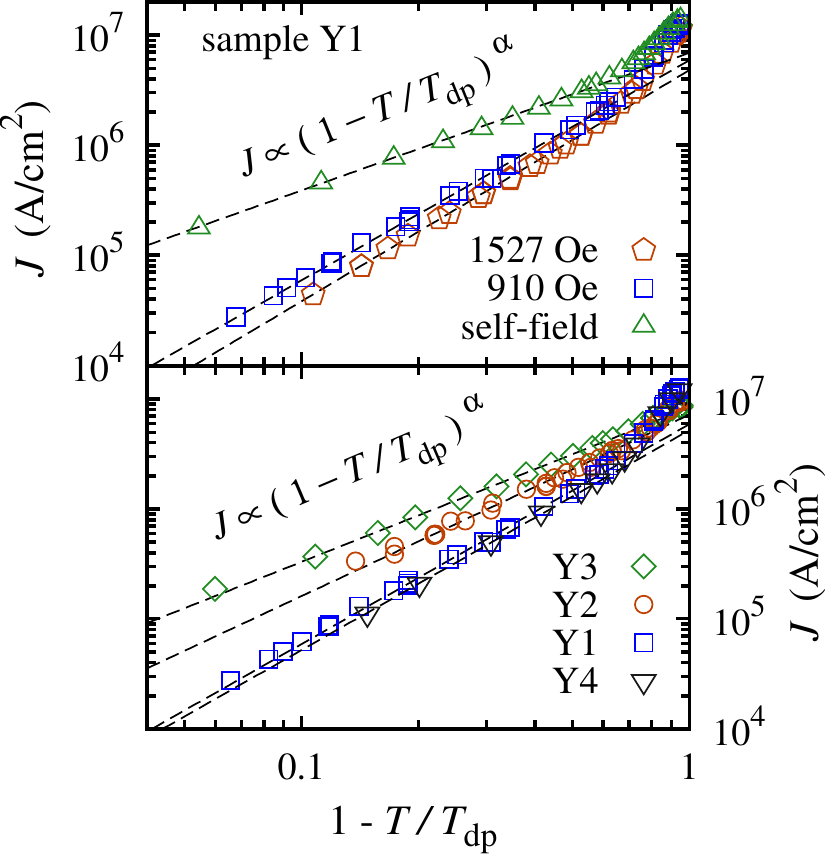}
	\caption{(Color online) $J \textit{ vs } 1-T/T_\mathrm{dp}$ for YBa$_2$Cu$_3$O$_{7-\delta}$ films.
	Dashed lines are fits by Eq.~\protect\eqref{eq:J2(T)}. The curves in bottom panel were measured in 
	field of 910~Oe for samples Y1--Y3 and  1530~Oe for Y4. See Fig.~\protect\ref{fig:J(T)_Y1},
	\protect\ref{fig:J(T)} for  $\alpha$ and $T_\mathrm{dp}$ values.}
	\label{fig:J(T)_Td}
\end{figure}

\begin{figure}\centering
	\includegraphics[scale=.8]{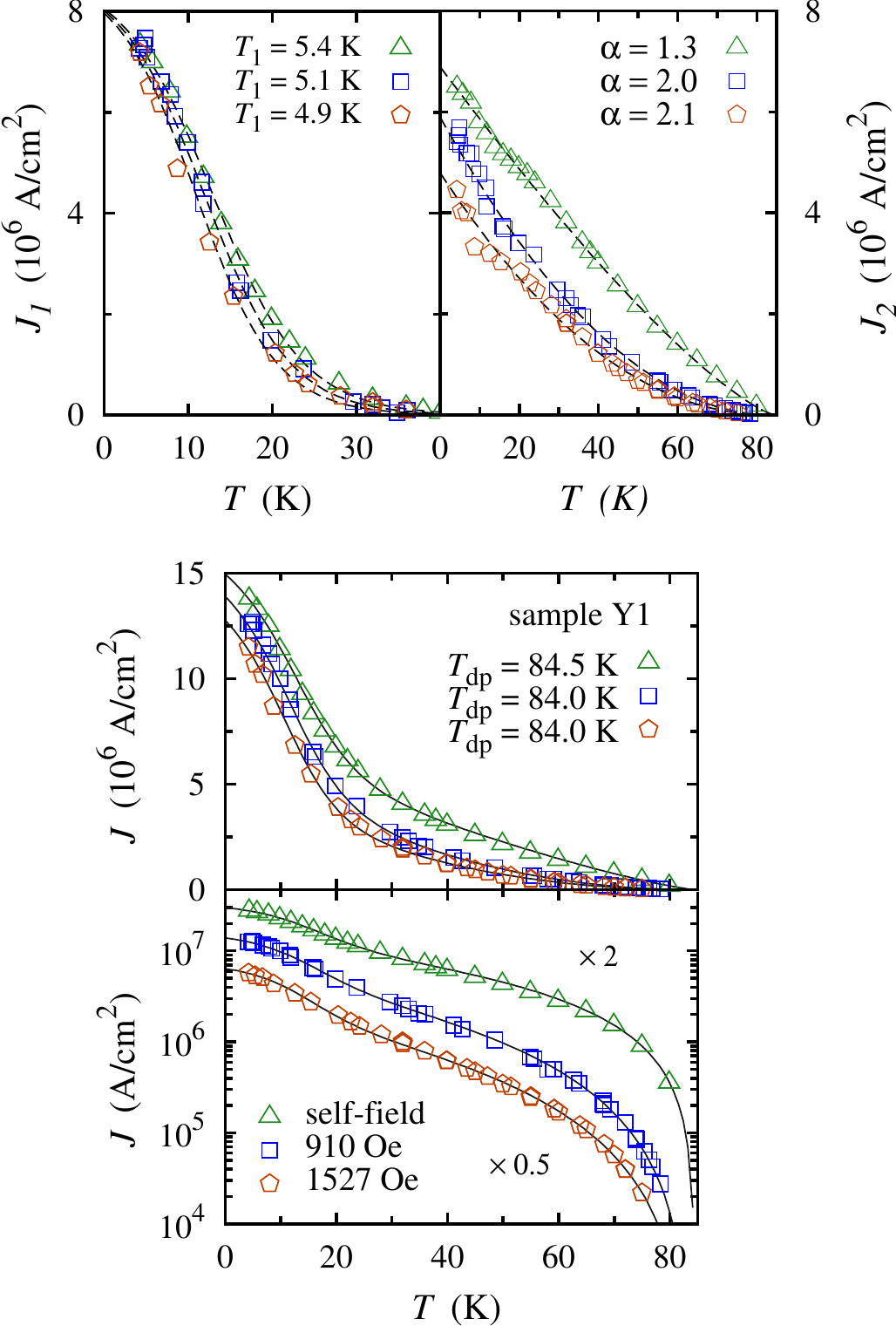}
	\caption{(Color online) Temperature dependences of the current density of the
	YBa$_2$Cu$_3$O$_{7-\delta}$ film (sample Y1) in
	different magnetic fields. \underline{\emph{Top:}} current components $J_1=J-J_2$ (left) 
	and $J_2=J-J_1$ (right). Dashed lines are fits by
	dependences~\protect\eqref{eq:J2(T)} and~\protect\eqref{eq:J1(T)}. 
	\underline{\emph{Bottom:}} Curves $J(T)$ are shown in standard and semilogarithmic scales in order
	to bring out both low and high temperature behavior of  $J$. The curves are shifted (multiplied by
	shown factors)  to avoid a crowding. Continuous lines are sums of fitting curves presented in top
	panels. }
	\label{fig:J(T)_Y1}
\end{figure}

$J$ \emph{vs}  $1-T/T_\mathrm{dp}$ dependences are shown in Fig.~\ref{fig:J(T)_Td} in logarithmic
scales. The curves demonstrate a pronounced linear part at high temperatures. Top panel of
Fig.~\ref{fig:J(T)_Td} shows that the curve obtained in self-field differs from ones measured
in external fields which are quite similar. The curves obtained in fields of 910 and 1530~Oe  for
sample Y1 are approximated  by the same $T_\mathrm{dp}=84$~K, difference of $\alpha\simeq2\pm0.1$ is
within the error and only $J_2(0)$ values differ by 19\%  (see Fig.~\ref{fig:J(T)_Y1}). In the
self-field the current demonstrates a weaker temperature dependence with $\alpha =1.3 \pm0.1$ and
slightly higher $T_\mathrm{dp}=84.5$~K.

\begin{figure}\centering
	\includegraphics[scale=.8]{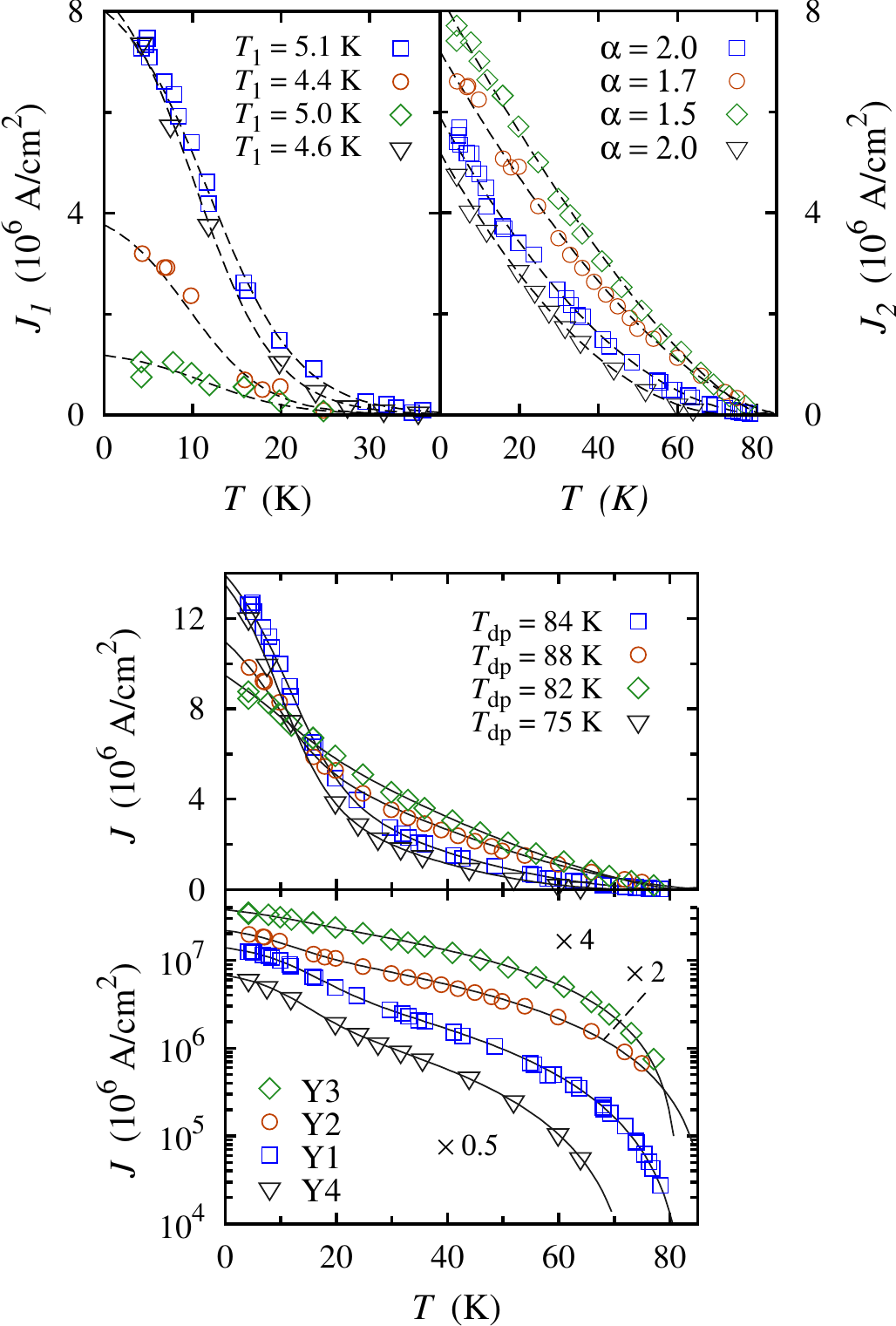}
	\caption{(Color online) Temperature dependences of the current density of
	YBa$_2$Cu$_3$O$_{7-\delta}$ films in magnetic field of 910~Oe for samples Y1-Y3 and 1530~Oe for 
	Y4.
	\underline{\emph{Top:}} current components $J_1=J-J_2$ (left) and $J_2=J-J_1$ (right). Dashed lines
	are fits by dependences~\protect\eqref{eq:J2(T)}
	and~\protect\eqref{eq:J1(T)}.
	\underline{\emph{Bottom:}} Curves $J(T)$ are shown in standard and semilogarithmic scales in order
	to bring out both low and high temperature behavior of  $J$. The curves are shifted (multiplied by
	shown factors)  to avoid a crossing. Continuous lines are sums of fitting curves presented in top
	panels.}
	\label{fig:J(T)}
\end{figure}

Analyzing $J(T)$ obtained for different samples we found that $T_\mathrm{dp}$ and $\alpha$ values do
not correlate with each other, at the same time the higher current densities $J_2(0)$ correspond to the
lower powers $\alpha$ (see Fig.~\ref{fig:J(T)}). It can be seen in bottom panel of
Fig.~\ref{fig:J(T)_Td} where the curves measured in external field are presented for all samples.
For example, for samples Y1 and Y4 the fitting lines demonstrate the same inclination $ \alpha =2.0
\pm0.1$, but $T_\mathrm{dp}$ differ by 9~K (see Table~\ref{tab:samples} and Fig.~\ref{fig:J(T)}).
	
Below 30~K the measured current deviates from the power law~\eqref{eq:J2(T)}.
Following to Ovchinnikov and Ivlev\cite{Ovchinnikov-PRB-1991} we supposed that the current consists
of two components and subtracted the dependence~\eqref{eq:J2(T)} from the experimental data to 
analyze a low-temperature behavior. Results of subtraction are shown in left top panels of
Figures~\ref{fig:J(T)_Y1} and~\ref{fig:J(T)}. As seen, the current $J_1=J-J_2$ strongly changes in
the range $T\lesssim 30 $~K. At low temperatures $J_1(T)$ dependence moderates and above 20~K the
current gradually falls down to zero at $T\sim40$~K. We found that low-temperature component of the
current can be approximated by an empiric dependence

\begin{equation}\label{eq:J1(T)}
	J_1=\frac{J_1^*}{1+\exp(T/T_1)/2T_1},
\end{equation}

\noindent where  the parameter $T_1$ is in Kelvins in the exponent power and dimensionless in its
divisor. As seen in Figures~\ref{fig:J(T)_Y1} and~\ref{fig:J(T)} the exponential
law~\eqref{eq:J1(T)} well fits $J_1(T)$ dependences. In low field the parameter
$J_1^*$ is field-independent and $T_1$ slightly decreases with $H$ (see left top panel in
Fig.~\ref{fig:J(T)_Y1}).

The current component $J_2=J-J_1$ is also plotted for comparison in right top panels of
Figures~\ref{fig:J(T)_Y1} and~\ref{fig:J(T)}.  The ratio of the components $J_1$ and $J_2$ is sample
dependent and changes with temperature. For example,  $J_2>J_1$ at all temperatures for samples Y2
and Y3  while $J_1$ becomes more than $J_2$ for Y1 and Y4 at low temperatures. The ratio determines
temperature behavior of total current. Though $J$ is higher in samples Y2 and Y3 at elevated
temperatures, at $T\lesssim15$~K  it becomes higher in Y1 and Y4 because of rapid increase of large
$J_1$ component.

$J(T)$ curves and sum of $J_1(T)$ and $J_2(T)$ fits are presented in bottom panels of
Figures~\ref{fig:J(T)_Y1} and~\ref{fig:J(T)}. As seen, the current change by about three orders of
magnitude is well approximated. Thus analysis of the separated current components allows us
empirically describe $J(T)$ at all temperatures.

\section{Discussion}\label{seq:Discussion}

\begin{figure}\centering
	\includegraphics[scale=.8]{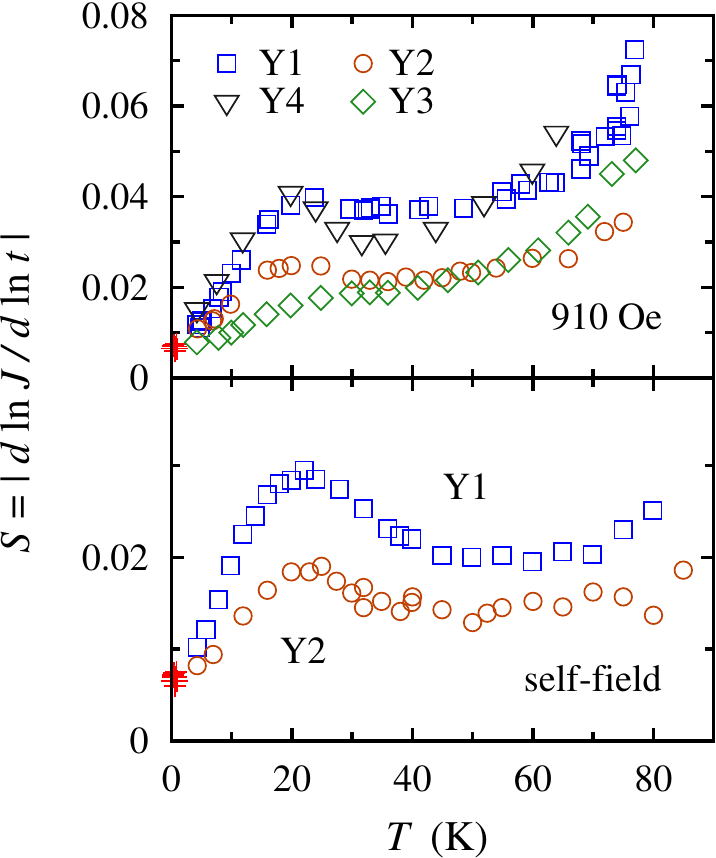} 
	\caption{(Color online)  $|d\ln J/d\ln t|\textit{ vs } T$  for YBa$_2$Cu$_3$O$_{7-\delta}$ films
	obtained after relaxation during 1 hour in field of 910~Oe  (1530~Oe for sample Y4). Crosses are
	data by Fruchter \emph{et al.}~\protect\cite{Fruchter-PRB-1991} obtained in the range $0.09\text{
	K}\leq T \leq0.9$~K at $H=2$~kOe.}
	\label{fig:S(T)}
\end{figure}

Let us consider a relation between the components and relaxation of the current. Comparison of the
relaxation rates  $R(T)$ with $J_1$ components in Figs.~\ref{fig:R(T)} and~\ref{fig:J(T)}
demonstrates a correlation: the larger $J_1$ the larger $R(T)$ maxima. $J_1$ rapidly decays with
both time and temperature therefore it is evidently produced by weak pinning on point defects
having a small pinning energy.

The normalized relaxation rate $S\equiv R/J$ \emph{vs} $T$ is plotted in Fig.~\ref{fig:S(T)}. At low
temperatures $S$ rises due to $R$ increase and $J$ decrease. When temperature rises $R$ passes
though maximum and then decreases. Since $J$ also decreases, the well known $S(T)$
plateau\cite{Yeshurun-RMP-1996} is observed. At elevated temperatures $J$ decreases more
rapidly than $R$ so $S$ rises again.

Before the plateau a maximum of $S(T)$ is often observed for YBa$_2$Cu$_3$O$_{7-\delta}$
films\cite{Yan-PRB-2000, Maiorov-NM-2009} in the temperature range where the $J_1$ component exists.
As seen in Figs.~\ref{fig:J(T)} and~\ref{fig:S(T)}, the  $S(T)$ peak is pronounced for sample Y4
having large $J_1$ and small $J_2$ but it is absent for sample Y3 having an inverse ratio of the
components. Therefore we suppose that the peak is caused by fast relaxation of the $J_1$ component.
Because of field suppression of both $R$ and $J_2$ (see Figs.~\ref{fig:R(T)} and~\ref{fig:J(T)_Y1}),
value of $S \simeq R/J_2$ at the plateau depends on $H$ and proves to be smallest in self-field. At
the same time, $H$ slightly affects $J_1$ and amplitude of $R(T)$ maximum therefore field influence 
on $S(T)=R/(J_1+J_2)$ is reduced in the temperature range of the peak location. As a result, the
peak is more pronounced in self-field as illustrated in bottom panel of Fig.~\ref{fig:S(T)}.

In Ref.~\onlinecite{Maiorov-NM-2009} the peak was  attributed to a synergetic combination of two
types of pinning centers present in the films, namely artificial columnar BaZrO$_3$ inclusions aligned
along $c$ axis and the Y$_2$O$_3$ nanoparticles horizontally aligned in $ab$ plane. There are no
artificial inclusions in our films. As discussed below, pinning in our samples is apparently
produced by the Y$_2$O$_3$ participates and oxygen vacancies. Therefore we suppose that the $S(T)$
peak is caused by combination of these pinning centers inherent to YBa$_2$Cu$_3$O$_{7-\delta}$
films.

As seen in Figures~\ref{fig:R(T)} and~\ref{fig:S(T)}, the relaxation rates $R$ and $S$ extrapolated
to zero temperature don't vanish.  The extrapolated $S$ well agrees with the value obtained for
YBa$_2$Cu$_3$O$_{7-\delta}$ single crystal at $T<1$~K.\cite{Fruchter-PRB-1991} The nonzero rate is
caused by the quantum tunneling of vortices which occurs in layered
superconductors.\cite{Blatter-RMP-1994, Yeshurun-RMP-1996} In YBa$_2$Cu$_3$O$_{7-\delta}$ films the
quantum creep affects vortices dynamics at temperatures up to $5-10$~K.\cite{vanDalen-PRB-1996,
Hoekstra-PRB-1999, Landau-PC-2000:2}  Precise torque measurements of YBa$_2$Cu$_3$O$_{7-\delta}$
single crystal, which revealed a crossover to two-dimensional superconducting behavior at
$T<80$~K,\cite{Farrell-PRL-1990} as well as the quantum creep  testify importance of layered
structure for superconductivity in YBa$_2$Cu$_3$O$_{7-\delta}$.

Let's consider now the theoretical basis for two-component current in  YBa$_2$Cu$_3$O$_{7-\delta}$
films. In Appendix we reduced the general solution of OI theory\cite{Ovchinnikov-PRB-1991} for the
case of magnetic field applied normally to the superconducting planes and calculated the critical
current density

\begin{gather}
		J_c=J_{c1}+J_{c2}, \nonumber\\
		J_{c1}=J_p\left[ 1-\exp\left\{ - 
			a_1 \frac{(J_p/J_0)^{5/4}n_p s^{5/4}\xi^{3/4}}{[\varepsilon^2 b\ln(1/b)]^{5/8}}
		\right\}\right], \label{eq:Jc1}\\
		J_{c2} = a_2 J_0 \frac{(F_v/\varepsilon_0)^{9/4}n_v\xi^3}{[\varepsilon^2 b\ln(1/b)]^{5/8}},
		\label{eq:Jc2}\\
	   	J_0=  J_0(0)\tau_-^{3/2}\tau_+^{1/2},\qquad b=\xi^2B/\Phi_0, \nonumber\\
	   	J_0(0)=\frac{c\Phi_0}{12\sqrt3\pi^2\xi_0\lambda_0^2}\simeq300\text{ MA/cm}^2.  \nonumber
\end{gather}

\noindent The current component $J_{c1}$ is produced by pinning in the superconducting layers. Here
$J_{p} \equiv cF_p/\Phi_0 s$ is the characteristic in-plane current density, $F_p$ and $n_p$ are
values of maximal pinning force and concentration of point pinning centers, $s$ is a distance
between the planes, $\Phi_0$ is the magnetic flux quantum, and $a_1=0.5203$ is the numerical factor.
The  component $J_{c2}$ is caused by an anisotropic pinning in the superconductor volume. Here $F_v$
and $n_v$ are values of maximal pinning force and concentration of pinning centers, $\varepsilon_0 =
(\Phi_0/4\pi\lambda)^2$ determines the self-energy of the vortex-lines, and $a_2=0.9273$ is the
numerical factor. Both components depend on the depairing current density
$J_0$~\eqref{eq:J_depairing}, dimensionless reduced magnetic field $b$ and  the anisotropy parameter
$\varepsilon$. Using $\lambda_0 =1400$~\AA,\cite{Schilling-PC-1990, Zimmermann-PRB-1995} $\xi_0
=17.2$~\AA\cite{Hao91:1} we estimated $J_0$ at $T=0$ for YBa$_2$Cu$_3$O$_{7-\delta}$.

As mentioned above, the field dependence of the $J_1$ component of the measured current is weak. For
example, at $T=0$ for sample Y1 we obtained the same value $J_1(0)=8.0$~MA/cm$^2$ with accuracy of
0.6\% for all fields. Let us estimate  $J_{c1}$ from its field dependence using the ratio
$J_{c1}(B_1) / J_{c1}(B_2) \lesssim 1.01$ of the order of $J_1(0)$ uncertainty for fields
$B_1=910$~G and $B_2=1530$~G. From~\eqref{eq:Jc1} we obtained $J_{c1}(B_1) / J_{c1}(B_2) =
\{1-\exp[-xf(b_1)]\} / \{1-\exp[-xf(b_2)]\}$ where  $f(b)=[b\ln(1/b)]^{-5/8}$ and
$x=a_1n_p\xi^2(sJ_p/\varepsilon\xi J_0)^{5/4}$. Taking the above ratio we calculated $x\gtrsim
0.0851$. For oxygen deficiency $\delta\gtrsim0.03$ in our films the concentration of randomly
distributed vacancies  in CuO$_2$ planes is estimated as $n_p =(4/7)\delta/ab \simeq1.15
\cdot10^{-3}\text{ \AA}^{-2}$ where $a \simeq3.82$~\AA\ and $b \simeq3.89$~\AA\ are the orthorhombic
lattice cell parameters.\cite{Jorgensen-PRB-1990} Two distances separate pairs of CuO$_2$ planes in
YBa$_2$Cu$_3$O$_{7-\delta}$: the intra-pair distance $s_p=3.37$~\AA\ and the inter-pair one of
8.32~\AA.\cite{Jorgensen-PRB-1990} Using $s=8.32$~\AA{} and $\varepsilon = 1/6.5$
(Ref.~\onlinecite{Bosma-PRB-2011})  we estimated lower limits for both the ratio $J_p/J_0 \gtrsim
0.177$ and the current $J_{c1}\gtrsim 53$~MA at $T=0$.

As follows from the estimation, a weak field dependence of $J_{c1}$ is realized for large currents
which are much more than $J_1(0)$.  $J_{c1}$ is really more than $J_1$
due to the quantum creep, but at $T\to0$ the relaxation rate is small (see Fig.~\ref{fig:S(T)}) and a
difference between $J_{c1}(0)$ and $J_1(0)$ must also be small. Thus the field dependence of $J_{c1}$
following from expression~\eqref{eq:Jc1} contradicts to that of $J_1$. The failure is caused by high
concentration of defects in the CuO$_2$ planes. Since a number of defects per square of vortex core
exceeds unity, $N_p \simeq \pi\xi^2n_p \simeq1.07$, the core contains a defect  at any site. In such
conditions only fluctuations of defect density pin vortices and the pinning becomes
\textit{collective}\cite{Blatter-RMP-1994,Blatter-PRL-2004} while  expression~\eqref{eq:Jc1} is
obtained for \textit{strong} pinning.

In the case of the collective pinning the current is independent of field in the single-vortex
pinning regime which is realized if $s<L_c^c<\varepsilon a_0$. In magnetic field directed along the
$c$ axis a length of the collective pinning segment  for YBa$_2$Cu$_3$O$_{7-\delta}$ is estimated as
$L_c^c\simeq10\varepsilon\xi_0 \simeq26.5$~\AA.\cite{Blatter-RMP-1994} Both an inter-vortex distance
$a_0 \simeq (2\Phi_0/\sqrt{3}B)^{1/2}\gtrsim1100$~\AA{} and its product $\varepsilon
a\gtrsim170$~\AA{} exceeded $L_c^c$  in our experiments so conditions for field independence of the
current were fulfilled. Therefore we suppose that in-plane pinning is produced by the collective
action of oxygen vacancies. Let us compare $J_1$ with $J_c$ obtained in CP theory for a layered
superconductor.
 
  \begin{figure}\centering
  	\includegraphics[scale=.8]{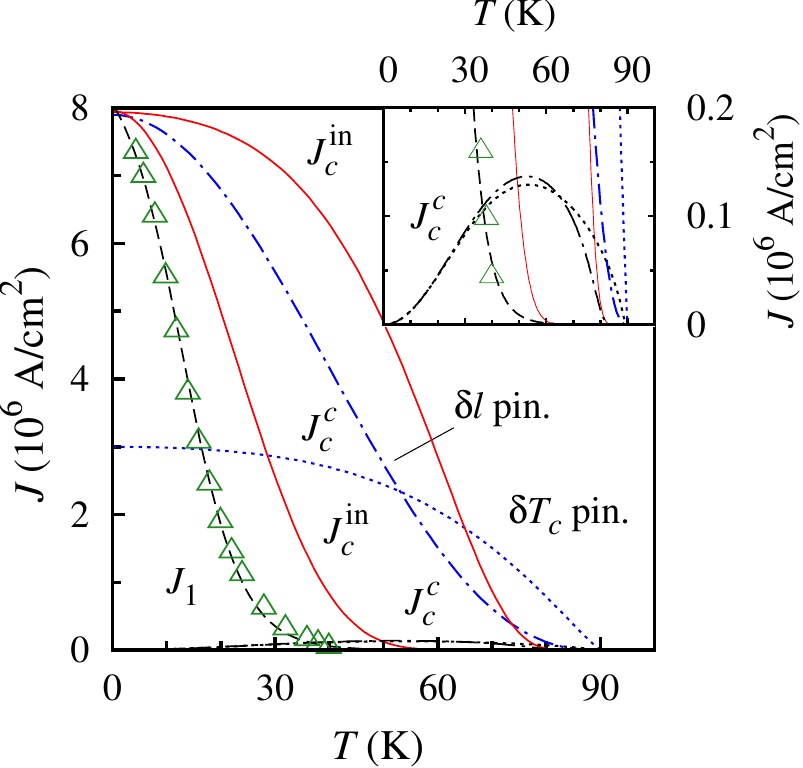} 
  	\caption{(Color online)  $J_1(T)$ obtained in self-field for sample Y1 (triangles) and its
  	approximations by Eq.~\eqref{eq:J1(T)} --- dashed line, $J_c^\text{in}$
  	Eq.~\eqref{eq:Jc_intrinsic}  --- continuous lines (left for $\beta=0.62$ and right for
  	$\beta=0.1$), $J_c^c$ Eqs.~\eqref{eq:Jc_CP-LT} and~\eqref{eq:Jc_CP-HT} --- dotted lines for
  	$\delta T_c$ pinning and dash-dotted lines for $\delta\ell$ pinning. The currents are magnified
  	in inset to illustrate $J_c^c(T)$ following from Eqs.~\eqref{eq:Jc_CP-HT}.} \label{fig:J1(T)}
  \end{figure}
 
 In field applied along normal to superconducting planes the critical current coincides for
 layered and anisotropic superconductors. For single vortex collective pinning  it is expressed
 as\cite{Blatter-RMP-1994}
\begin{subequations}\label{eq:Jc_CP-LT}
\begin{eqnarray}
	 J_c^c&=&J_0\left[\frac{\delta_m}{\varepsilon}\right]^{2/3}\mspace{-22mu}
			=J_0(0)\left[\frac{\delta_m(0)}{\varepsilon}\right]^{2/3}\mspace{-16mu}
			\tau_+^{-1/2}\tau_-^{5/2} \quad\delta\ell\text{ pin. }\quad\\
	 J_c^c&=&J_0\left[\frac{\delta_\alpha}{\varepsilon}\right]^{2/3}\mspace{-22mu}
		 =J_0(0)\left[\frac{\delta_\alpha(0)}{\varepsilon}\right]^{2/3}\mspace{-16mu}
		 \tau_+^{5/6}\tau_-^{7/6} \quad\text{ $\delta T_c$ pin. }
\end{eqnarray} 
\end{subequations}
\noindent The dimensionless pinning parameters for oxygen vacancies in YBa$_2$Cu$_3$O$_{7-\delta}$
are estimated as $\delta_m(0)/\varepsilon \simeq (0.2-1) 10^{-2}$ for $\delta\ell$ pinning and
$\delta_\alpha(0)/\varepsilon \simeq 10^{-3}$ for $\delta T_c$ pinning,\cite{Blatter-RMP-1994} and
the corresponding currents are $J_c^c(0) \simeq (5-14)$~MA/cm$^2$ and $J_c^c(0) \simeq 3$~MA/cm$^2$.
The dependence $J_1(T)$ obtained in self-field for sample Y1 as well as fitting curves for
Eqs.~\eqref{eq:Jc_CP-LT} are presented in Fig.~\ref{fig:J1(T)}. As seen $J_1(T)$ disagree with 
$J_c^c(T)$ curves, moreover for $\delta T_c$ pinning the current $J_c^c(0)$ is about two times lower
than $J_1(0)$.

Eqs.~\eqref{eq:Jc_CP-LT} does not take into account thermal fluctuations suppressing the critical
current at high temperatures\cite{Blatter-RMP-1994}
 \begin{equation}
  	 J_c^c = \frac{c(k_\mathrm{B}T)^2}{\Phi_0\varepsilon_0\xi^3}\exp\left[
 		-\frac{3w}{2\delta_{\alpha,m}}\left(\frac{k_\mathrm{B}T}{\varepsilon_0\xi}\right)^3\right].
   \end{equation}
\noindent Here $w$ is a factor of the order of unity. Selecting temperature dependences of
quantities\cite{Griessen-PRL-1994} we write
\begin{align}\label{eq:Jc_CP-HT}
  	J_c^c &= J_c^c(0)\frac{\tau^2\tau_-^{1/2}}{\tau_+^{5/2}}
		  	\exp\left[-\frac{3w}{2}\left(\frac{T}{T^*}\right)^3\frac{1}{f_c^c(T)}\right],\\
 		 	  f_c^c(T) &=
 			\begin{cases}
 				\tau_+^3\tau_-^3 &\text{for } \delta\ell\text{ pinning},\\
 				\tau_+^5\tau_- &\text{for }  \delta T_c\text{ pinning},\\
 			\end{cases} 	\nonumber \\
  		 	 J_c^c(0) &=\frac{3\sqrt3}{4}J_0(0) \left[\frac{k_\mathrm{B}T_c}{\varepsilon_0(0)\xi_0}\right]^2
  		 	 \simeq 1.07 \text{ MA/cm}^2, \nonumber\\
  		 	 T^* &= \frac{\varepsilon_0(0)\xi_0\delta_{\alpha,m}^{1/3}(0)}{k_\mathrm{B}}
  		 	\simeq
 			\begin{cases}(198-116)\text{ K} & \delta\ell\text{ pin.},\\
 			92\text{ K}  & \delta T_c \text{ pin}.\\
 			\end{cases}	 \nonumber
\end{align}
\noindent Due to fluctuations the current is strongly suppressed at temperatures above the depinning
temperature which is calculated from the equation $T_\mathrm{dp}^3 =
{T^*}^3{f_c^c}(T_\mathrm{dp})$.\cite{Blatter-RMP-1994} The temperatures $T_\mathrm{dp}\simeq89$~K
and 71--79~K  calculated for $\delta T_c$ and  $\delta\ell$ pinning are considerably higher than
temperatures at which $J_1$ disappears. In Fig.~\ref{fig:J1(T)}  dependences~\eqref{eq:Jc_CP-HT} are
shown. As seen, a magnitude of $J_1(T)$ a lot more than maximal values of  $J_c^c(T)$ and the curves
lie in different temperature ranges. Thus we conclude that neither Eqs.~\eqref{eq:Jc_CP-LT}
nor~\eqref{eq:Jc_CP-HT} describe $J_1$ component of the measured current.

In magnetic field parallel to a superconducting layers the intrinsic pinning takes place in a layered
superconductor.\cite{Blatter-RMP-1994, Barone-JS-1990, Ivlev-JLTP-1990}  Kinks of
vortices\cite{Feinberg-MPLB-1990, Feinberg-PRl-1990} also lead to  the intrinsic pinning. Though
our experiments were performed in a transverse field, a demagnetizing effect, which was strong
because of low fields and large demagnetizing factor of films, produced a tangential component of
field\cite{Brandt-PRB-1998-1} directed along superconducting layers in the films. Therefore the
critical current produced by the intrinsic pinning should be also considered. Its temperature
dependence has the form\cite{Blatter-RMP-1994, Ivlev-JLTP-1990}
\begin{equation}
	J_c^\mathrm{in} = J_0\left(\frac{8\varepsilon\xi}{s}\right)^2\left(1-\frac{B}{B_{c2}}\right)
	\exp\left[-8\left(\frac{\varepsilon\xi}{s}\right)^2\right].
\end{equation}
Neglecting the field dependence, since $B\ll B_{c2}$ in our experiments, we write it as
\begin{gather}\label{eq:Jc_intrinsic}
	J_c^\mathrm{in} =64\beta J_0(0)\tau_+^{3/2}\tau_-^{1/2}\exp[-8\beta\tau_+/\tau_-],\\
	\beta = (\varepsilon\xi_0/s)^2 = 0.62-0.1.\nonumber
\end{gather}
\noindent Here we estimated $\beta$ for the above mentioned distances between CuO$_2$ planes.
$J_c^\mathrm{in}(T)$ curves calculated for $\beta=0.62$ and 0.1 and normalized by factors
0.094 and  0.0092 respectively are presented in Fig.~\ref{fig:J1(T)}. While quasi-exponential shape of
$J_c^\mathrm{in}(T)$ is more appropriate to $J_1(T)$, the intrinsic current decreases more slowly
and disappears at higher temperatures. In addition at low temperatures $J_c^\mathrm{in}$ is one or
two orders more than $J_1$.

Summing up we conclude that $J_1$ component of the current is caused by the collective pinning of
vortices on oxygen vacancies in the single vortex pinning regime. However we failed to find an
appropriate approximation for $J_1(T)$ in the frame of CP theory. Apparently, because of smallness
of pinning energy, $J_1$ rapidly relaxes and its temperature dependence is strongly affected by
creep.
 
Turning to volume pinning we begin with a remark about pinning on the dislocations. In EDP
model\cite{Pan-PRB-2006} the critical current depends on an average size of the crystallites, i.~e.
CDB size in the diffraction experiments, as well as on a misorientation angle $\omega$ between
them.\cite{Kosse-SST-2008} As seen  in Table~\ref{tab:samples}, in our samples the CDB size changes 
by
one order and the angle $\omega$ varies almost seven times, but these parameters do not correlate
with $J_2$. Despite absence of the correlation, we compare below $J_2(T)$ with $J_c(T)$ calculated
for pinning on both non-superconducting inclusions and the edge dislocations for analysis to be
comprehensive.

In low fields the critical current of YBa$_2$Cu$_3$O$_{7-\delta}$ films is independent of
field.\cite{Dam-N-1999, Klaassen-PRB-2001, vanderBeek-PRB-2002, Peurla-SST-2005, Foltyn-NM-2007,
Miura-PRB-2011} 

\begin{figure}\centering
\includegraphics[scale=.8]{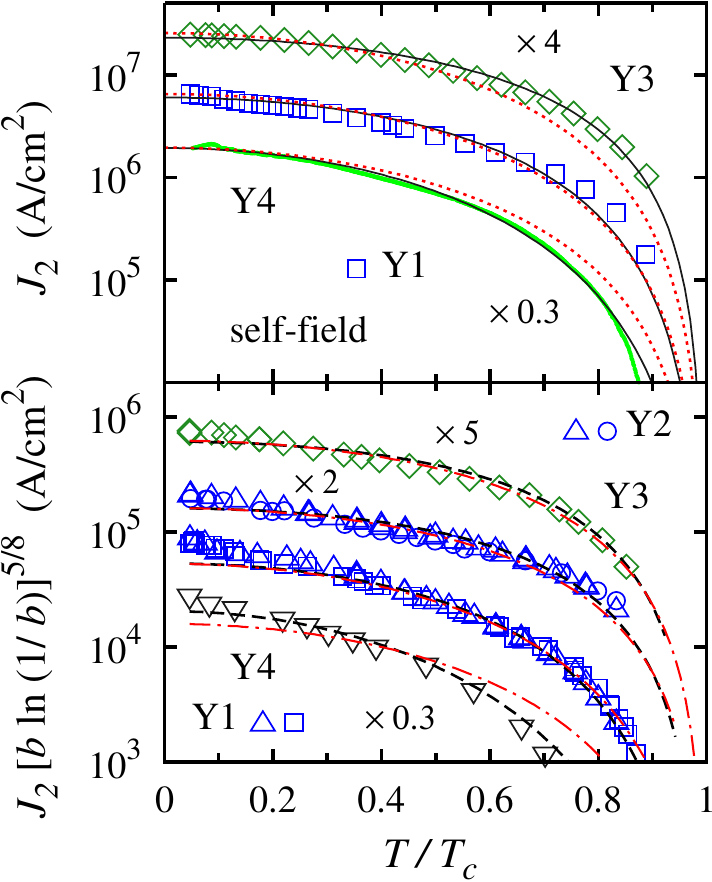}
	\caption{(Color online) \underline{\emph{Top:}} $J_2(T)$ for YBa$_2$Cu$_3$O$_{7-\delta}$ films
	obtained in self-field. Experimental curve for sample Y4 was recorded under warming of film
	right after magnetic field removing. Dotted and continuous lines are fits by
	dependences~\eqref{eq:Jd(T0)} and~\eqref{eq:Ji(T0)}. \underline{\emph{Bottom:}} Scaled $J_2(T)$
	dependences obtained in fields 1530~Oe (triangles) and 910~Oe (other symbols).
	Dashed 	and dash-dotted lines are fits by dependences~~\eqref{eq:Ji(TB)} and~\eqref{eq:Jd(TB)}. 
	The curves 	are shifted (multiplied by shown factors) to avoid a crossing. See text for details.}
	\label{fig:J2(T)}
\end{figure}

The field-independent current produced by pinning on the edge dislocations is calculated
as\cite{Pan-PRB-2006}
 \begin{gather}\label{eq:Jd(T0)}
       	J_{cd}(T,0)\simeq  \frac{3\sqrt3}{16\sqrt2}\frac{c\varepsilon_0}{\Phi_0}
       	\frac{r_d^2}{\xi^3}
	       	=J_{cd}^0\left(\frac{r_d}{\xi_0}\right)^2\frac{\tau_-^{5/2}}{\tau_+^{1/2}},\\
	       	J_{cd}^0 =  \frac{27}{64\sqrt2}J_0(0)=89.5 \text{ MA/cm}^2,\nonumber
\end{gather}
\noindent where $r_d$ is the radius of dislocation normal core. Dependences $J_2(T)$ obtained in
self-field and their fits by Eqs.~\eqref{eq:Jd(T0)} are presented in Fig.~\ref{fig:J2(T)}.  From the
fits we obtained $r_d\simeq 4.6$, 5.0, 5.3 and 4.6~\AA{} respectively for samples Y1--Y4.
Eq.~\eqref{eq:Jd(T0)} provides the same temperature dependence for all samples scaled by $r_d$
values while $J_2(T)$ curves differ for different samples. The dependence  $J_{cd}(T,0)$
satisfactory approximates  $J_2(T)$ only for sample Y1 while for other ones a discrepancy of the
fitting curves and the experimental data is clearly seen at $T/T_c \gtrsim 0.5$.

The field-independent current caused by pinning on inclusions\cite{vanderBeek-PRB-2002}
 \begin{equation*}
   	J_{ci}(T,0)	\simeq \frac{3J_0}{4}\sqrt\frac{3n_i}{\pi \varepsilon^2}
   		\left(\frac{F_i\xi}{\varepsilon_0}\right)^{3/2}
 \end{equation*}
 depends on inclusion density $n_i$ and the pinning force $F_i$ approximated
 as\cite{Blatter-RMP-1994, vanderBeek-PRB-2002}
 \begin{equation}\label{eq:Fp}
 \begin{gathered}
 		\frac{F_i\xi}{\varepsilon_0} \simeq  \frac{D_\mathit{iz}}{4}\mathcal{F}(T,d_i),\\
 		\mathcal{F}(T,d_i)=\ln\left(1+\frac{D_i^2}{2\xi^2}\right)
 			=\ln\left(1+\frac{d_i^2\tau_-}{2\tau_+}\right). 
 \end{gathered}
 \end{equation}
 \noindent Here $D_i$ is an average extent of  an inclusion, $D_\mathit{iz}$ is its extent along the field
 direction and $d_i=D_i/\xi_0$. Selecting temperature dependences of quantities we write the current 
 in the form 
 \begin{gather}\label{eq:Ji(T0)}
    	J_{ci}(T,0)\simeq J_{ci}^0 
     	[\mathcal{F}(T,d_i)\tau_-]^{3/2}\tau_+^{1/2},\\
    	J_{ci}^0 \simeq \frac{3\sqrt{3} }{32\sqrt\pi}\frac{J_0(0)}{\varepsilon}\sqrt{n_iD _\mathit{iz}^3}
     	\simeq\sqrt{n_iD _\mathit{iz}^3} \cdot 179\text{ MA/cm}^2.\nonumber
 \end{gather}

$J_2(T)$ curves were fitted by the dependence~\eqref{eq:Ji(T0)} via parameters $J_{ci}^0$ and $D_i$.
The currents $J_{ci}^0=2.7$, 1.1 0.88 and 25~MA/cm$^2$ were respectively obtained for samples
Y1--Y4. $D_i$ values are presented in Table~\ref{tab:samples}. The fitting curves shown in
Fig.~\ref{fig:J2(T)} well agree with $J_2(T)$ for samples Y1 and Y3 though for Y1 the measured
current decays more slowly at $T/T_c \gtrsim 0.75$. For sample Y4 the measured and fitting curves
coincide up to $T\sim T_\mathrm{dp}$.

\begin{figure}\centering
\includegraphics[scale=.8]{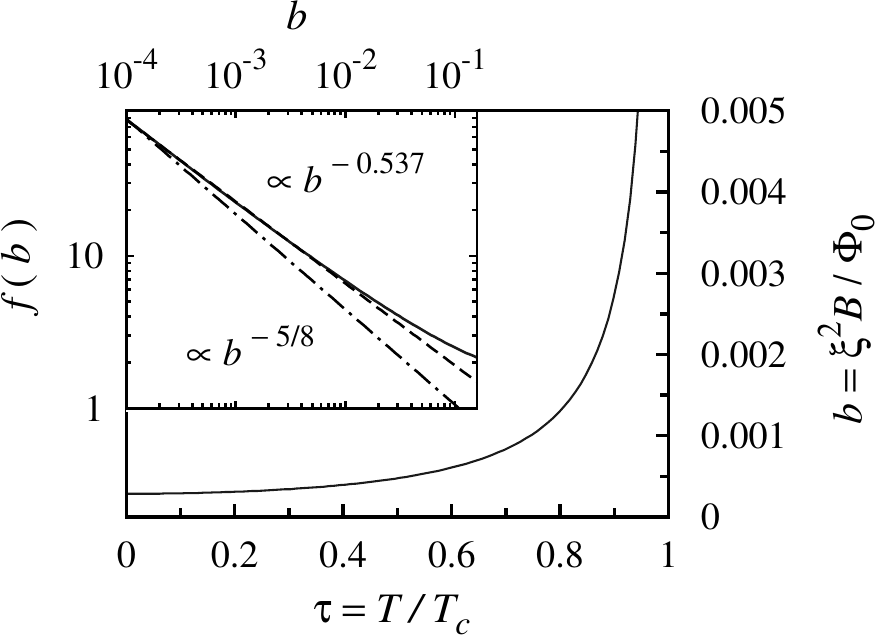}
	\caption{Temperature dependence of the parameter $b=B\xi^2/\Phi_0 =
	B\xi_0^2\tau_+/\Phi_0\tau_-$ calculated for $B=2$~kG and $\xi_0=17.2$~\AA. 
	\underline{\emph{Inset:}} function $f = [b\ln(1/b)]^{-5/8}$ (continuous line) and its
	approximations $f = 0.5552 \cdot b^{-0.537}$ (dashed), $f = 0.25 \cdot b^{-5/8}$
	(dash-dot).}
	\label{fig:J(B)_theory}
\end{figure}

Thus pinning on inclusions well describes  $J_2(T)$ of YBa$_2$Cu$_3$O$_{7-\delta}$ films in
self-field. Interaction of vortices suppresses the critical current when a vortex density
$\tilde{n}\simeq B/\Phi_0$ increases. Let us proceed with analysis of field dependence of $J_{c2}$
which is determined by the function $f = [b\ln(1/b)]^{-5/8}$, see Eq.~\eqref{eq:Jc2}. Because of
large $B_{c2}$ in YBa$_2$Cu$_3$O$_{7-\delta}$ the parameter $b=B/2\pi B_{c2}$ is small. In
Figure~\ref{fig:J(B)_theory} we plotted $b(T)$ for $B=2$~kG exceeding maximum field in our
experiments. As seen, $b$ is less than 0.005 for $T/T_c\lesssim0.95$. In the inset of
Fig.~\ref{fig:J(B)_theory} the function $f$ is plotted in the range up to $b=1/2\pi$ corresponding
to $B=B_{c2}$. For small $b$ it follows a power law and in the range $10^{-4}\le b\le 0.005$ is
approximated as $f = 0.5552 \cdot b^{-0.537}$ with the accuracy of $\pm1.2$\%.  The dependence
$J\propto B^{-\alpha}$ with $\alpha\simeq0.4$--0.8 was often observed for
YBa$_2$Cu$_3$O$_{7-\delta}$ films\cite{Dam-N-1999, Klaassen-PRB-2001, vanderBeek-PRB-2002,
Peurla-SST-2005, Foltyn-NM-2007, Miura-PRB-2011, Kuznetsov-IEEETAS-2016} and
2G-tapes.\cite{Senatore-SST-2016}

As seen in Fig.~~\ref{fig:J(B)_theory}, at $b\gtrsim0.01$ the function $f(b)$ moderates and should
be approximated by $f\propto b^{-\alpha}$ with a lower $\alpha$. However strengthening $J(B)$
dependence in high fields was reliably established for YBa$_2$Cu$_3$O$_{7-\delta}$ films in numerous
experiments.\cite{Dam-N-1999, Klaassen-PRB-2001, vanderBeek-PRB-2002, Fedotov-LTP-2002,
Peurla-SST-2005, Ijaduola-PRB-2006, Maiorov-NM-2009,Miura-PRB-2011} Expression~\eqref{eq:Jc2} is
valid only if the lateral displacement of the vortex lines $u_0$ is small in comparison with the
inter-vortex distance $a_0$. In high fields when $a_0\propto B^{-1/2} $ becomes larger than $u_0$ a
more strong suppression of the current $J_{c2}\propto B^{-1}$ is expected.\cite{vanderBeek-PRB-2002}

According to Eq.~\eqref{eq:Jc2} the curves $J_{c2}[b\ln(1/b)]^{5/8}$ should be independent of
field. Indeed, as seen in bottom Fig.~\ref{fig:J2(T)}, the data obtained for samples Y1 and Y2 in
different fields are joined into common curves under such scaling. Therefore we compare  
$J_c(T,B)[b\ln(1/b)]^{5/8}$ dependences with the scaled data collected for different fields.

In the EDP model the field-dependent critical current\cite{Pan-PRB-2006}
\begin{gather}\label{eq:Jd(TB)}
     	J_{cd}(T,B)= J_{cd}(T,0)\frac{\tilde{n}_p}{\tilde{n}},\\
      \frac{\tilde{n}_p(T,B)}{\tilde{n}(B)} = 1- 
      \frac{[\Gamma(\nu,\eta)-\eta\Gamma(\nu-1,\eta)]^2}{\Gamma^2(\nu)},\nonumber\\
      	\nu=\left[\frac{\langle L\rangle}{\sigma}\right]^2,\quad \eta(T,B)
	       	=\frac{r_d}{\langle L\rangle}	
	       	\frac{2\nu}{\xi_0}\sqrt{\frac{\Phi_0}{B}\frac{\tau_-}{\tau_+}}, \label{eq:rd_L}
 \end{gather}
\noindent  is determined by a relative number of pinned vortices  $\tilde{n}_p/\tilde{n}$ expressed
via complete and incomplete Euler's gamma functions $\Gamma(x)$ and $\Gamma(x,y)$.\cite{EDP} Here
$\sigma$ is the dispersion of the crystallite size distribution function around the mean value
$\langle L\rangle$.

The scaled currents were fitted by  $J_{cd}(T,B)[b\ln(1/b)]^{5/8}$ via  $J_{cd}(0,0)$, $\nu$ and
$\eta$ using $B=910$~G as parameter. The fitting  curves, shown in bottom Fig.~\ref{fig:J2(T)},
agree with experimental data  for samples Y1--Y3 though a systematic deviation to a lower current is
observed at low temperatures. At the same time the fit badly approximates data for sample Y4. From
the fit we obtained $\eta \simeq 2$ for all samples, $\nu \simeq 7$ for Y2, Y3 and  $\nu \simeq 1$
for Y1, Y4. Then from~\eqref{eq:rd_L} for $B=910$~G the ratio $r_d/\langle L\rangle$ was estimated
as $1.1\cdot10^{-2}$ for samples Y2, Y3 and $1.6\cdot10^{-3}$ for Y1,Y4. From $J_{cd}(0,0)$ values
we calculated $r_d$ and then obtained $\langle L\rangle$ presented  in Table~\ref{tab:samples}. As
seen, the fit gives a correct order for $r_d$ and $\langle L\rangle$ values, however lengths $\langle
L\rangle$ does not correlate with average sizes of crystallites CDB obtained in the diffraction
experiments. Neither $r_d$ nor $\langle L\rangle$ correlate with the measured current $J$ or its
components.

For pinning on inclusions the field-dependent current  calculated from Eqs.~\eqref{eq:Jc2}
and~\eqref{eq:Fp} takes the form\cite{vdBeek}

\begin{subequations}\label{eq:Ji(TB)}
\begin{gather}
		J_{ci}(T,B) \simeq \frac{J_{ci}^B}{[ b\ln(1/b)]^{5/8}}\mathrm{F}(T,d_i), \label{eq:Ji(B)} \\
		\mathrm{F}(T,d_i)=\ln^{9/4}\left(1+\frac{d_i^2\tau_-}{2\tau_+}\right)\tau_-^{9/8}\tau_+^{7/8},\\
		J_{ci}^B =  \frac{3^{3/4}J_0(0)n_i 
		D_\mathit{iz}^{9/4}\xi_0^{3/4}}{16\cdot2^{3/4}\pi^{5/8}\varepsilon^{5/4}}
		\simeq n_i D_\mathit{iz}^{9/4}\xi_0^{3/4}\cdot129 \text{ MA/cm}^2,\nonumber
\end{gather}
\end{subequations}

The scaled currents were fitted by the dependence $J_{ci}^B\mathrm{F}(T,d_i)$ via $J_{ci}^B$ and
$D_i$. Since size of inclusions is independent of field, we used the same $D_i$ to fit $J_2$ by
both~\eqref{eq:Ji(T0)} and~\eqref{eq:Ji(TB)}. An effective density of inclusions
$n_id_\mathit{iz}^{9/4} = n_i(D_\mathit{iz}/\xi_0)^{9/4}$ was calculated from $J_{ci}^B$ values. 
The fitting  curves  $J_{ci}^B\mathrm{F}(T,d_i)$, presented in bottom Fig.~\ref{fig:J2(T)}, agree
with experimental data  for all samples.

Obtained $D_i$ and $n_id_\mathit{iz}^{9/4}$ values are presented 
in Table~\ref{tab:samples}. The average extent of inclusions $D_i$ varying in the range $2-14$~nm
well agrees with size of Y$_2$O$_3$ precipitates in YBa$_2$Cu$_3$O$_{7-\delta}$
films.\cite{Catana-APL-1992, Selinder-PC-1992, Kastner-PC-1995, Verbist-PC-1996} $D_i$ and
$n_id_\mathit{iz}^{9/4}$ values correlate with $J_2$ component of the current. The larger inclusion
extent the more $J_2$. For inclusions with the same $D_i$ the current rises with increase of the
effective inclusion density $n_id_\mathit{iz}^{9/4}$.

As follows form Eqs.~\eqref{eq:Ji(T0)} and~\eqref{eq:Ji(TB)}, the extent of inclusion along the
field direction can be obtained from the ratio ${(J_{ci}^0})^2/J_{ci}^B = (D_\mathit{iz}/\xi_0)^{3/4}
\cdot 248$~A/cm$^2$. Then $n_i$ is simply calculated from $J_{ci}^B$ or $J_{ci}^0$. Values of
$D_\mathit{iz}$ and $n_i$ found in such a procedure are presented in Table~\ref{tab:samples}.

Parameters of pinning centers obtained for our films well agree with  $D_i=15$~nm and $n_i = (1-3)
\cdot 10^{15}$~cm$^{-3}$ found in magnetic experiments in Ref.~\onlinecite{vanderBeek-PRB-2002}. At
the same time a lower density of inclusions was found in direct measurements by means of the
electron microscopy.\cite{Catana-APL-1992, Selinder-PC-1992, Kastner-PC-1995, Verbist-PC-1996}

Among microstructure defects in YBa$_2$Cu$_3$O$_{7-\delta}$ films\cite{Foltyn-NM-2007,
Kastner-PC-1995, Verbist-PC-1996} the precipitates\cite{Catana-APL-1992, Selinder-PC-1992,
Kastner-PC-1995, Verbist-PC-1996} [001]-Y$_2$0$_3$ and [110]-Y$_2$0$_3$ have dimensions close to 
our estimations of $D_i$. The [110]-Y$_2$0$_3$ precipitates are small cubes or rectangles with sides
ranging from 3 to 5 nm.\cite{Verbist-PC-1996} The [001]-Y$_2$0$_3$ precipitates have extension of 10
to 20~nm in the $ab$-plane and about 6 to 8~nm along the $c$ axis.\cite{Catana-APL-1992,
Selinder-PC-1992, Kastner-PC-1995, Verbist-PC-1996}  There are no data on density and shape of
inclusions with size smaller than 2~nm since such small inclusions are hard to recognize even in
high-resolution electron microscopy (HREM) micrographs.\cite{Kastner-PC-1995, Verbist-PC-1996} Our
results for sample Y4 demonstrate presence of such inclusions which we classified as the
[110]-Y$_2$0$_3$ precipitates. Taking $D_\mathit{iz}=D_i$ for samples Y1, Y4 and
$D_\mathit{iz}=6$~nm for Y2, Y4, from the effective density of inclusions obtained above we
estimated densities $n_i^*$ presented in Table~\ref{tab:samples}. For samples Y1--Y3 estimated
$n_i^*$ values are only twice less than that observed by direct HREM method in laser-deposited
YBa$_2$Cu$_3$O$_{7-\delta}$ films.\cite{Verbist-PC-1996} Since a density of Y$_2$0$_3$ precipitates
depends on both method and conditions of the deposition process\cite{Kastner-PC-1995,
Verbist-PC-1996, Foltyn-NM-2007, Maiorov-NM-2009} such agreement seems quite satisfactory. Note
also that the measured relaxed persistent current is less than $J_c$ so a lower limit for the
inclusion densities was estimated in our experiment.

Summing up we conclude that the $J_2$ component of the measured current is well described by
pinning on the Y$_2$0$_3$ inclusions. The pinning is strong and its efficiency rises with increase of
inclusions size.

\section{Summary}\label{seq:Summary} 

In this paper we confirmed experimentally that the critical current of laser-deposited
YBa$_2$Cu$_3$O$_{7-\delta}$ films consists of two components caused by in-plane pinning of vortices
by oxygen vacancies in superconducting CuO$_2$ planes and by anisotropic pinning on the Y$_2$0$_3$
precipitates in the superconductor volume.\cite{Ovchinnikov-PRB-1991} We proposed a simple method 
to separate the current components and found their temperature dependences~\eqref{eq:J2(T)}
and~\eqref{eq:J1(T)}. Analysis of the current components led us to the following conclusions.

The component produced by the in-plane pinning is described as single-vortex collective pinning
however we failed to find an appropriate theoretical dependence to approximate its temperature
behavior. This component slightly depends on field and rapidly relaxes.  The in-plane pinning is
substantial only at low temperatures $T\lesssim30$~K but in this temperature range its contribution
into the critical current and vortices dynamics should not be neglected.

The component produced by the volume pinning  is well described in the frame of OI
theory\cite{Ovchinnikov-PRB-1991} further developed by van der Beek \emph{et
al}.\cite{vanderBeek-PRB-2002} We confirmed that in laser-deposited YBa$_2$Cu$_3$O$_{7-\delta}$
films the strong anisotropic volume pinning is produced by the nano-size Y$_2$O$_3$ precipitates.
Varying inclusion sizes in different films causes difference in the depinning temperatures and
parameters of $J(T)$ dependence. Rather low magnetic field of about $ 1 $~kOe applied normally
to the film plane affects this current component.

Different ratio of the current components and variation of size of the Y$_2$O$_3$ inclusions lead to
a wide variety of $J(T)$ dependences in standard YBa$_2$Cu$_3$O$_{7-\delta}$ films. Addition of
artificial defects further complicates $J(T,\mathbf{B})$ behavior. Nevertheless films produced by
different techniques demonstrate some common features discussed in beginning Sec.~\ref{sec:Results}.

While in-plane and volume defects act simultaneously and provide additive components of the current,
combining several types of volume defects is not simply additive. Therefore engineering the pinning
landscape in 2G/3G-tapes is a very complex problem.\cite{Foltyn-NM-2007, Maiorov-NM-2009} We hope
that separation and correct analysis of the additive components demonstrating strongly different
temperature, field and angle behavior help in solving this actual problem.

\appendix*
\section{}\label{sec:Appendix}

An  inhomogeneous layered superconductor with an axial anisotropy and the mass anisotropy ratio
$\varepsilon^2 = m/M = \lambda/\lambda_c$ was considered by Ovchinnikov and
Ivlev.\cite{Ovchinnikov-PRB-1991} Here $m$ and $\lambda$ are the effective mass of carriers and
the penetration depth in isotropic superconducting planes and $M,\;\lambda_c$ are the corresponding
parameters along normal to the planes. General case of magnetic field $B$ applied along the
direction forming an angle $\theta$ with the planes was analyzed  and  the critical current density
was calculated. We simplify the results obtained by Ovchinnikov and Ivlev for the case
$\theta=\pi/2$, when the field is directed normally to the planes, and rewrite the values in the
notations of Ref.~\onlinecite{Blatter-RMP-1994} commonly used at present.

As shown by Ovchinnikov and Ivlev, the critical current consists of two parts 

\begin{equation}
	J_c = J_{c1}  + J_{c2} 
\end{equation}
 
 \noindent caused by in-plane pinning on point defects in the superconducting planes and  by
 anisotropic pinning of vortices by ``macro-defects'' in the superconductor volume.

The anisotropic component of the current is written as follows \cite{Ovchinnikov-PRB-1991}

\begin{equation}\label{eq:Jc_OI}
	J_{c2}  =  \frac{cn_vF_v^2}{\Phi_0\alpha^2\sqrt{\epsilon_{xx}C_{xx}}} \left[
	\frac{128 F_v\xi^3}{27\sqrt{\epsilon_{yy}C_{yy}}}\right]^{1/4},
\end{equation}
\noindent where $F_v$ and $n_v$ are maximum values of pinning force and concentration of pinning
centers in volume of the superconductor, $\xi$ is the in-plane coherent length, $\Phi_0$ is the
magnetic flux quantum. Taking into account the units $\hbar=c=1$ used by Ovchinnikov and Ivlev
\cite{Ovchinnikov-PRB-1991} we multiplied the right hand side by the light velocity  $c$. The
function $\alpha^2 = \sin^2\theta +  \varepsilon^2\cos^2\theta$ is equal to unity in our case. The
quantities

\begin{equation}\label{eq:Cxx}
C_{yy} = \alpha^2C_{xx} = C_{xx} =  \frac{\Phi_0 B}{8\pi\lambda^2}=\frac{2\pi  b\varepsilon_0}{\xi^2}
\end{equation}

\noindent we express via the energy $\varepsilon_0=(\Phi_0/4\pi\lambda)^2$ which determines the
self-energy of the vortex-lines\cite{Blatter-RMP-1994} and the parameter $b=\xi^2B/\Phi_0$.  The
quantities $\epsilon_{xx}$ and  $\epsilon_{yy}$ are written as

\begin{equation}\label{eq:Exx}
\epsilon =  \frac{\Phi_0^2}{8\pi^2\lambda^2}
		\ln\left[ \frac{\sqrt{\alpha\Phi_0/B}}{d\cos\theta +\alpha\xi}\right]\tilde{\epsilon}
	= \varepsilon_0\ln(1/b)\tilde{\epsilon}.
\end{equation}

\noindent From bulky but simple expressions for  $\tilde{\epsilon}_{xx}(\theta,\varepsilon)$
and  $\tilde{\epsilon}_{yy}(\theta,\varepsilon)$ in Ref.~\onlinecite{Ovchinnikov-PRB-1991} we
calculated that in our case $\tilde{\epsilon}_{xx} =\tilde{\epsilon}_{yy} =\tilde{\epsilon} =
\varepsilon^2/2$.

Substituting all obtained values in \eqref{eq:Jc_OI}, and using expression for the
depairing current density\cite{Blatter-RMP-1994}
\begin{equation}\label{eq:J_depairing}
	J_0 =  \frac{4}{3\sqrt3}\frac{c\varepsilon_0}{\Phi_0\xi},
\end{equation}
\noindent after simple algebraic transformations we get the anisotropic component of the critical
current density in the form

\begin{equation}\label{eq:Jc_anisotropic}
	\begin{gathered}
	J_{c2}  = a_2 J_0	\left(\frac{F_v}{\varepsilon_0}\right)^{9/4} 
	\frac{n_v\xi^3}{[\varepsilon^2 b\ln(1/b)]^{5/8}},\\
				a_2 =  \frac{3^{3/4}}{2^{1/4}\pi^{5/8}} =0.9373.
	\end{gathered}
\end{equation}

The current component caused by the layered structure of superconductor is written
as\cite{Ovchinnikov-PRB-1991}
\begin{widetext}
	\begin{equation}\label{eq:Jc-layered}
			 J_{c1}  =  \frac{cF_p}{\Phi_0 s}\left[
				1-\exp\left\{-\frac{n_pF_p}{\sin\theta\sqrt{\epsilon_{xx}C_{xx}}}\left(
			\frac{128 F_p\xi^3}{27\sqrt{\epsilon_{yy}C_{yy}}}
			\right)^{1/4}\right\}\right],
	\end{equation}

\noindent where $F_p$ and $n_p$ are maximum values of pinning force and concentration of pinning
centers in the superconducting planes, $s$ is a distance between the planes. Substituting all values
obtained above, taking into account that $\sin\theta=1$ in our case and denoting $J_{p}=cF_p/(\Phi_0
s)$,  after simple algebraic transformations we rewrite \eqref{eq:Jc-layered} as

	\begin{equation}\label{eq:Jc_layered}
			 J_{c1}  =  J_{p}\left[ 1-\exp\left\{ - a_1 \left(\frac{J_{p}}{J_0}\right)^{5/4}
			\frac{n_ps^{5/4}\xi^{3/4}}{[\varepsilon^2 b\ln(1/b)]^{5/8}}
			\right\}\right], \qquad a_1 =\frac{16\cdot2^{1/4}}{9\cdot(3\pi)^{5/8}}=0.5203.
	\end{equation}
\end{widetext}

\begin{acknowledgments}
	 The authors would like to thank O.~A. Krymskaya for the structure measurements, O.~A Churkin for
	 technical assistance and A.~P. Menushenkov for fruitful discussions. The work was supported by
	 Russian Science Foundation grant 14-22-00098.
\end{acknowledgments}


\begin{thebibliography}{89}%
\makeatletter
\providecommand \@ifxundefined [1]{%
 \@ifx{#1\undefined}
}%
\providecommand \@ifnum [1]{%
 \ifnum #1\expandafter \@firstoftwo
 \else \expandafter \@secondoftwo
 \fi
}%
\providecommand \@ifx [1]{%
 \ifx #1\expandafter \@firstoftwo
 \else \expandafter \@secondoftwo
 \fi
}%
\providecommand \natexlab [1]{#1}%
\providecommand \enquote  [1]{``#1''}%
\providecommand \bibnamefont  [1]{#1}%
\providecommand \bibfnamefont [1]{#1}%
\providecommand \citenamefont [1]{#1}%
\providecommand \href@noop [0]{\@secondoftwo}%
\providecommand \href [0]{\begingroup \@sanitize@url \@href}%
\providecommand \@href[1]{\@@startlink{#1}\@@href}%
\providecommand \@@href[1]{\endgroup#1\@@endlink}%
\providecommand \@sanitize@url [0]{\catcode `\\12\catcode `\$12\catcode
  `\&12\catcode `\#12\catcode `\^12\catcode `\_12\catcode `\%12\relax}%
\providecommand \@@startlink[1]{}%
\providecommand \@@endlink[0]{}%
\providecommand \url  [0]{\begingroup\@sanitize@url \@url }%
\providecommand \@url [1]{\endgroup\@href {#1}{\urlprefix }}%
\providecommand \urlprefix  [0]{URL }%
\providecommand \Eprint [0]{\href }%
\providecommand \doibase [0]{http://dx.doi.org/}%
\providecommand \selectlanguage [0]{\@gobble}%
\providecommand \bibinfo  [0]{\@secondoftwo}%
\providecommand \bibfield  [0]{\@secondoftwo}%
\providecommand \translation [1]{[#1]}%
\providecommand \BibitemOpen [0]{}%
\providecommand \bibitemStop [0]{}%
\providecommand \bibitemNoStop [0]{.\EOS\space}%
\providecommand \EOS [0]{\spacefactor3000\relax}%
\providecommand \BibitemShut  [1]{\csname bibitem#1\endcsname}%
\let\auto@bib@innerbib\@empty
\bibitem [{\citenamefont {Campbell}\ and\ \citenamefont
  {Evetts}(1972)}]{Campbell-AP-1972}%
  \BibitemOpen
  \bibfield  {author} {\bibinfo {author} {\bibfnamefont {A.~M.}\ \bibnamefont
  {Campbell}}\ and\ \bibinfo {author} {\bibfnamefont {J.~E.}\ \bibnamefont
  {Evetts}},\ }\href {\doibase 10.1080/00018737200101288} {\bibfield  {journal}
  {\bibinfo  {journal} {Adv. Phys.}\ }\textbf {\bibinfo {volume} {21}},\
  \bibinfo {pages} {199} (\bibinfo {year} {1972})}\BibitemShut {NoStop}%
\bibitem [{\citenamefont {Larkin}\ and\ \citenamefont
  {Ovchinnikov}(1979)}]{Larkin-JLTP-1979}%
  \BibitemOpen
  \bibfield  {author} {\bibinfo {author} {\bibfnamefont {A.~I.}\ \bibnamefont
  {Larkin}}\ and\ \bibinfo {author} {\bibfnamefont {Y.~N.}\ \bibnamefont
  {Ovchinnikov}},\ }\href {\doibase 10.1007/BF00117160} {\bibfield  {journal}
  {\bibinfo  {journal} {J.~Low Temp. Phys.}\ }\textbf {\bibinfo {volume}
  {34}},\ \bibinfo {pages} {409} (\bibinfo {year} {1979})}\BibitemShut
  {NoStop}%
\bibitem [{\citenamefont {Blatter}\ \emph {et~al.}(1994)\citenamefont
  {Blatter}, \citenamefont {Feigel'man}, \citenamefont {Geshkenbein},
  \citenamefont {Larkin},\ and\ \citenamefont {Vinokur}}]{Blatter-RMP-1994}%
  \BibitemOpen
  \bibfield  {author} {\bibinfo {author} {\bibfnamefont {G.}~\bibnamefont
  {Blatter}}, \bibinfo {author} {\bibfnamefont {M.~V.}\ \bibnamefont
  {Feigel'man}}, \bibinfo {author} {\bibfnamefont {V.~B.}\ \bibnamefont
  {Geshkenbein}}, \bibinfo {author} {\bibfnamefont {A.~I.}\ \bibnamefont
  {Larkin}}, \ and\ \bibinfo {author} {\bibfnamefont {V.~M.}\ \bibnamefont
  {Vinokur}},\ }\href {\doibase 10.1103/RevModPhys.66.1125} {\bibfield
  {journal} {\bibinfo  {journal} {Rev. Mod. Phys.}\ }\textbf {\bibinfo {volume}
  {66}},\ \bibinfo {pages} {1125} (\bibinfo {year} {1994})}\BibitemShut
  {NoStop}%
\bibitem [{\citenamefont {Brandt}(1995)}]{Brandt-RPP-1995}%
  \BibitemOpen
  \bibfield  {author} {\bibinfo {author} {\bibfnamefont {E.~H.}\ \bibnamefont
  {Brandt}},\ }\href {http://iopscience.iop.org/0034-4885/58/11/003/}
  {\bibfield  {journal} {\bibinfo  {journal} {Rep. Prog. Phys.}\ }\textbf
  {\bibinfo {volume} {58}},\ \bibinfo {pages} {1465} (\bibinfo {year}
  {1995})}\BibitemShut {NoStop}%
\bibitem [{\citenamefont {Feigel'man}\ and\ \citenamefont
  {Vinokur}(1990)}]{Feigelman-PRB-1990}%
  \BibitemOpen
  \bibfield  {author} {\bibinfo {author} {\bibfnamefont {M.~V.}\ \bibnamefont
  {Feigel'man}}\ and\ \bibinfo {author} {\bibfnamefont {V.~M.}\ \bibnamefont
  {Vinokur}},\ }\href {\doibase 10.1103/PhysRevB.41.8986} {\bibfield  {journal}
  {\bibinfo  {journal} {Phys. Rev.~B}\ }\textbf {\bibinfo {volume} {41}},\
  \bibinfo {pages} {8986} (\bibinfo {year} {1990})}\BibitemShut {NoStop}%
\bibitem [{\citenamefont {Nelson}\ and\ \citenamefont
  {Vinokur}(1993)}]{Nelson-PRB-1993}%
  \BibitemOpen
  \bibfield  {author} {\bibinfo {author} {\bibfnamefont {D.~R.}\ \bibnamefont
  {Nelson}}\ and\ \bibinfo {author} {\bibfnamefont {V.~M.}\ \bibnamefont
  {Vinokur}},\ }\href {\doibase 10.1103/PhysRevB.48.13060} {\bibfield
  {journal} {\bibinfo  {journal} {Phys. Rev.~B}\ }\textbf {\bibinfo {volume}
  {48}},\ \bibinfo {pages} {13060} (\bibinfo {year} {1993})}\BibitemShut
  {NoStop}%
\bibitem [{\citenamefont {Yeshurun}\ \emph {et~al.}(1996)\citenamefont
  {Yeshurun}, \citenamefont {Malozemoff},\ and\ \citenamefont
  {Shaulov}}]{Yeshurun-RMP-1996}%
  \BibitemOpen
  \bibfield  {author} {\bibinfo {author} {\bibfnamefont {Y.}~\bibnamefont
  {Yeshurun}}, \bibinfo {author} {\bibfnamefont {A.~P.}\ \bibnamefont
  {Malozemoff}}, \ and\ \bibinfo {author} {\bibfnamefont {A.}~\bibnamefont
  {Shaulov}},\ }\href {\doibase 10.1103/RevModPhys.68.911} {\bibfield
  {journal} {\bibinfo  {journal} {Rev. Mod. Phys.}\ }\textbf {\bibinfo {volume}
  {68}},\ \bibinfo {pages} {911} (\bibinfo {year} {1996})}\BibitemShut
  {NoStop}%
\bibitem [{\citenamefont {Malozemoff}\ \emph {et~al.}(2008)\citenamefont
  {Malozemoff}, \citenamefont {Fleshler}, \citenamefont {Rupich}, \citenamefont
  {Thieme}, \citenamefont {Li}, \citenamefont {Zhang1}, \citenamefont {Otto},
  \citenamefont {Maguire}, \citenamefont {Folts}, \citenamefont {Yuan},
  \citenamefont {Kraemer}, \citenamefont {Schmidt}, \citenamefont {Wohlfart},\
  and\ \citenamefont {Neumueller}}]{Malozemoff-SST-2008}%
  \BibitemOpen
  \bibfield  {author} {\bibinfo {author} {\bibfnamefont {A.~P.}\ \bibnamefont
  {Malozemoff}}, \bibinfo {author} {\bibfnamefont {S.}~\bibnamefont
  {Fleshler}}, \bibinfo {author} {\bibfnamefont {M.}~\bibnamefont {Rupich}},
  \bibinfo {author} {\bibfnamefont {C.}~\bibnamefont {Thieme}}, \bibinfo
  {author} {\bibfnamefont {X.}~\bibnamefont {Li}}, \bibinfo {author}
  {\bibfnamefont {W.}~\bibnamefont {Zhang1}}, \bibinfo {author} {\bibfnamefont
  {A.}~\bibnamefont {Otto}}, \bibinfo {author} {\bibfnamefont {J.}~\bibnamefont
  {Maguire}}, \bibinfo {author} {\bibfnamefont {D.}~\bibnamefont {Folts}},
  \bibinfo {author} {\bibfnamefont {J.}~\bibnamefont {Yuan}}, \bibinfo {author}
  {\bibfnamefont {H.-P.}\ \bibnamefont {Kraemer}}, \bibinfo {author}
  {\bibfnamefont {W.}~\bibnamefont {Schmidt}}, \bibinfo {author} {\bibfnamefont
  {M.}~\bibnamefont {Wohlfart}}, \ and\ \bibinfo {author} {\bibfnamefont
  {H.-W.}\ \bibnamefont {Neumueller}},\ }\href {\doibase
  10.1088/0953-2048/21/3/034005} {\bibfield  {journal} {\bibinfo  {journal}
  {Supercond. Sci. Technol.}\ }\textbf {\bibinfo {volume} {21}},\ \bibinfo
  {pages} {034005} (\bibinfo {year} {2008})}\BibitemShut {NoStop}%
\bibitem [{\citenamefont {Senatore}\ \emph {et~al.}(2016)\citenamefont
  {Senatore}, \citenamefont {Barth}, \citenamefont {Bonura}, \citenamefont
  {Kulich},\ and\ \citenamefont {Mondonico}}]{Senatore-SST-2016}%
  \BibitemOpen
  \bibfield  {author} {\bibinfo {author} {\bibfnamefont {C.}~\bibnamefont
  {Senatore}}, \bibinfo {author} {\bibfnamefont {C.}~\bibnamefont {Barth}},
  \bibinfo {author} {\bibfnamefont {M.}~\bibnamefont {Bonura}}, \bibinfo
  {author} {\bibfnamefont {M.}~\bibnamefont {Kulich}}, \ and\ \bibinfo {author}
  {\bibfnamefont {G.}~\bibnamefont {Mondonico}},\ }\href {\doibase
  10.1088/0953-2048/29/1/014002} {\bibfield  {journal} {\bibinfo  {journal}
  {Supercond. Sci. Technol.}\ }\textbf {\bibinfo {volume} {29}},\ \bibinfo
  {pages} {014002} (\bibinfo {year} {2016})}\BibitemShut {NoStop}%
\bibitem [{\citenamefont {Foltyn}\ \emph {et~al.}(2007)\citenamefont {Foltyn},
  \citenamefont {Civale}, \citenamefont {MacManus-Driscoll}, \citenamefont
  {Jia}, \citenamefont {Maiorov}, \citenamefont {Wang},\ and\ \citenamefont
  {Maley}}]{Foltyn-NM-2007}%
  \BibitemOpen
  \bibfield  {author} {\bibinfo {author} {\bibfnamefont {S.~R.}\ \bibnamefont
  {Foltyn}}, \bibinfo {author} {\bibfnamefont {L.}~\bibnamefont {Civale}},
  \bibinfo {author} {\bibfnamefont {J.~L.}\ \bibnamefont {MacManus-Driscoll}},
  \bibinfo {author} {\bibfnamefont {Q.~X.}\ \bibnamefont {Jia}}, \bibinfo
  {author} {\bibfnamefont {B.}~\bibnamefont {Maiorov}}, \bibinfo {author}
  {\bibfnamefont {H.}~\bibnamefont {Wang}}, \ and\ \bibinfo {author}
  {\bibfnamefont {M.}~\bibnamefont {Maley}},\ }\href {\doibase
  10.1038/nmat1989} {\bibfield  {journal} {\bibinfo  {journal} {Nat. Mater.}\
  }\textbf {\bibinfo {volume} {6}},\ \bibinfo {pages} {631} (\bibinfo {year}
  {2007})}\BibitemShut {NoStop}%
\bibitem [{\citenamefont {Maiorov}\ \emph {et~al.}(2009)\citenamefont
  {Maiorov}, \citenamefont {Baily}, \citenamefont {Zhou}, \citenamefont
  {Ugurlu}, \citenamefont {Kennison}, \citenamefont {Dowden}, \citenamefont
  {Holesinger}, \citenamefont {Foltyn},\ and\ \citenamefont
  {Civale}}]{Maiorov-NM-2009}%
  \BibitemOpen
  \bibfield  {author} {\bibinfo {author} {\bibfnamefont {B.}~\bibnamefont
  {Maiorov}}, \bibinfo {author} {\bibfnamefont {S.~A.}\ \bibnamefont {Baily}},
  \bibinfo {author} {\bibfnamefont {H.}~\bibnamefont {Zhou}}, \bibinfo {author}
  {\bibfnamefont {O.}~\bibnamefont {Ugurlu}}, \bibinfo {author} {\bibfnamefont
  {J.~A.}\ \bibnamefont {Kennison}}, \bibinfo {author} {\bibfnamefont {P.~C.}\
  \bibnamefont {Dowden}}, \bibinfo {author} {\bibfnamefont {T.~G.}\
  \bibnamefont {Holesinger}}, \bibinfo {author} {\bibfnamefont {S.~R.}\
  \bibnamefont {Foltyn}}, \ and\ \bibinfo {author} {\bibfnamefont
  {L.}~\bibnamefont {Civale}},\ }\href {\doibase 10.1038/NMAT2408} {\bibfield
  {journal} {\bibinfo  {journal} {Nat. Mater.}\ }\textbf {\bibinfo {volume}
  {8}},\ \bibinfo {pages} {398} (\bibinfo {year} {2009})}\BibitemShut {NoStop}%
\bibitem [{\citenamefont {Dam}\ \emph {et~al.}(1999)\citenamefont {Dam},
  \citenamefont {Huijbregtse}, \citenamefont {Klaassen}, \citenamefont {van~der
  Geest}, \citenamefont {Doornbos}, \citenamefont {Rector}, \citenamefont
  {Testa}, \citenamefont {Freisem}, \citenamefont {Martinez}, \citenamefont
  {St\"auble-P\"umpin},\ and\ \citenamefont {Griessen}}]{Dam-N-1999}%
  \BibitemOpen
  \bibfield  {author} {\bibinfo {author} {\bibfnamefont {B.}~\bibnamefont
  {Dam}}, \bibinfo {author} {\bibfnamefont {J.~M.}\ \bibnamefont
  {Huijbregtse}}, \bibinfo {author} {\bibfnamefont {F.~C.}\ \bibnamefont
  {Klaassen}}, \bibinfo {author} {\bibfnamefont {R.~C.~F.}\ \bibnamefont
  {van~der Geest}}, \bibinfo {author} {\bibfnamefont {G.}~\bibnamefont
  {Doornbos}}, \bibinfo {author} {\bibfnamefont {J.~H.}\ \bibnamefont
  {Rector}}, \bibinfo {author} {\bibfnamefont {A.~M.}\ \bibnamefont {Testa}},
  \bibinfo {author} {\bibfnamefont {S.}~\bibnamefont {Freisem}}, \bibinfo
  {author} {\bibfnamefont {J.~C.}\ \bibnamefont {Martinez}}, \bibinfo {author}
  {\bibfnamefont {B.}~\bibnamefont {St\"auble-P\"umpin}}, \ and\ \bibinfo
  {author} {\bibfnamefont {R.}~\bibnamefont {Griessen}},\ }\href {\doibase
  10.1038/20880} {\bibfield  {journal} {\bibinfo  {journal} {Nature}\ }\textbf
  {\bibinfo {volume} {399}},\ \bibinfo {pages} {439} (\bibinfo {year}
  {1999})}\BibitemShut {NoStop}%
\bibitem [{\citenamefont {Huijbregtse}\ \emph {et~al.}(2000)\citenamefont
  {Huijbregtse}, \citenamefont {Dam}, \citenamefont {van~der Geest},
  \citenamefont {Klaassen}, \citenamefont {Elberse}, \citenamefont {Rector},\
  and\ \citenamefont {Griessen}}]{Huijbregtse-PRB-2000}%
  \BibitemOpen
  \bibfield  {author} {\bibinfo {author} {\bibfnamefont {J.~M.}\ \bibnamefont
  {Huijbregtse}}, \bibinfo {author} {\bibfnamefont {B.}~\bibnamefont {Dam}},
  \bibinfo {author} {\bibfnamefont {R.~C.~F.}\ \bibnamefont {van~der Geest}},
  \bibinfo {author} {\bibfnamefont {F.~C.}\ \bibnamefont {Klaassen}}, \bibinfo
  {author} {\bibfnamefont {R.}~\bibnamefont {Elberse}}, \bibinfo {author}
  {\bibfnamefont {J.~H.}\ \bibnamefont {Rector}}, \ and\ \bibinfo {author}
  {\bibfnamefont {R.}~\bibnamefont {Griessen}},\ }\href {\doibase
  10.1103/PhysRevB.62.1338} {\bibfield  {journal} {\bibinfo  {journal} {Phys.
  Rev.~B}\ }\textbf {\bibinfo {volume} {62}},\ \bibinfo {pages} {1338}
  (\bibinfo {year} {2000})}\BibitemShut {NoStop}%
\bibitem [{\citenamefont {Pan}\ \emph {et~al.}(2006)\citenamefont {Pan},
  \citenamefont {Cherpak}, \citenamefont {Komashko}, \citenamefont {Pozigun},
  \citenamefont {Tretiatchenko}, \citenamefont {Semenov}, \citenamefont
  {Pashitskii},\ and\ \citenamefont {Pan}}]{Pan-PRB-2006}%
  \BibitemOpen
  \bibfield  {author} {\bibinfo {author} {\bibfnamefont {V.}~\bibnamefont
  {Pan}}, \bibinfo {author} {\bibfnamefont {Y.}~\bibnamefont {Cherpak}},
  \bibinfo {author} {\bibfnamefont {V.}~\bibnamefont {Komashko}}, \bibinfo
  {author} {\bibfnamefont {S.}~\bibnamefont {Pozigun}}, \bibinfo {author}
  {\bibfnamefont {C.}~\bibnamefont {Tretiatchenko}}, \bibinfo {author}
  {\bibfnamefont {A.}~\bibnamefont {Semenov}}, \bibinfo {author} {\bibfnamefont
  {E.}~\bibnamefont {Pashitskii}}, \ and\ \bibinfo {author} {\bibfnamefont
  {A.~V.}\ \bibnamefont {Pan}},\ }\href {\doibase 10.1103/PhysRevB.73.054508}
  {\bibfield  {journal} {\bibinfo  {journal} {Phys. Rev.~B}\ }\textbf {\bibinfo
  {volume} {73}},\ \bibinfo {pages} {054508} (\bibinfo {year}
  {2006})}\BibitemShut {NoStop}%
\bibitem [{\citenamefont {Catana}\ \emph {et~al.}(1992)\citenamefont {Catana},
  \citenamefont {Broom}, \citenamefont {Bednorz}, \citenamefont {Mannhart},\
  and\ \citenamefont {Schlom}}]{Catana-APL-1992}%
  \BibitemOpen
  \bibfield  {author} {\bibinfo {author} {\bibfnamefont {A.}~\bibnamefont
  {Catana}}, \bibinfo {author} {\bibfnamefont {R.~F.}\ \bibnamefont {Broom}},
  \bibinfo {author} {\bibfnamefont {J.~G.}\ \bibnamefont {Bednorz}}, \bibinfo
  {author} {\bibfnamefont {J.}~\bibnamefont {Mannhart}}, \ and\ \bibinfo
  {author} {\bibfnamefont {D.~G.}\ \bibnamefont {Schlom}},\ }\href {\doibase
  10.1063/1.106507} {\bibfield  {journal} {\bibinfo  {journal} {Appl. Phys.
  Lett.}\ }\textbf {\bibinfo {volume} {60}},\ \bibinfo {pages} {1016} (\bibinfo
  {year} {1992})}\BibitemShut {NoStop}%
\bibitem [{\citenamefont {Selinder}\ \emph {et~al.}(1992)\citenamefont
  {Selinder}, \citenamefont {Helmersson}, \citenamefont {Z.~Hart},
  \citenamefont {Sj\"ostr\"om},\ and\ \citenamefont
  {Wallenberg}}]{Selinder-PC-1992}%
  \BibitemOpen
  \bibfield  {author} {\bibinfo {author} {\bibfnamefont {T.~I.}\ \bibnamefont
  {Selinder}}, \bibinfo {author} {\bibfnamefont {U.}~\bibnamefont
  {Helmersson}}, \bibinfo {author} {\bibfnamefont {J.-E.~S.}\ \bibnamefont
  {Z.~Hart}}, \bibinfo {author} {\bibfnamefont {H.}~\bibnamefont
  {Sj\"ostr\"om}}, \ and\ \bibinfo {author} {\bibfnamefont {L.~R.}\
  \bibnamefont {Wallenberg}},\ }\href {\doibase 10.1016/0921-4534(92)90297-P}
  {\bibfield  {journal} {\bibinfo  {journal} {Physica~C}\ }\textbf {\bibinfo
  {volume} {202}},\ \bibinfo {pages} {69} (\bibinfo {year} {1992})}\BibitemShut
  {NoStop}%
\bibitem [{\citenamefont {K\"astner}\ \emph {et~al.}(1995)\citenamefont
  {K\"astner}, \citenamefont {Hesse}, \citenamefont {Scholz}, \citenamefont
  {Koch}, \citenamefont {Ludwig}, \citenamefont {Lorenz},\ and\ \citenamefont
  {Kittel}}]{Kastner-PC-1995}%
  \BibitemOpen
  \bibfield  {author} {\bibinfo {author} {\bibfnamefont {G.}~\bibnamefont
  {K\"astner}}, \bibinfo {author} {\bibfnamefont {D.}~\bibnamefont {Hesse}},
  \bibinfo {author} {\bibfnamefont {R.}~\bibnamefont {Scholz}}, \bibinfo
  {author} {\bibfnamefont {H.}~\bibnamefont {Koch}}, \bibinfo {author}
  {\bibfnamefont {F.}~\bibnamefont {Ludwig}}, \bibinfo {author} {\bibfnamefont
  {M.}~\bibnamefont {Lorenz}}, \ and\ \bibinfo {author} {\bibfnamefont
  {H.}~\bibnamefont {Kittel}},\ }\href {\doibase 10.1016/0921-4534(95)00011-9}
  {\bibfield  {journal} {\bibinfo  {journal} {Physica~C}\ }\textbf {\bibinfo
  {volume} {243}},\ \bibinfo {pages} {281 } (\bibinfo {year}
  {1995})}\BibitemShut {NoStop}%
\bibitem [{\citenamefont {Verbist}\ \emph {et~al.}(1996)\citenamefont
  {Verbist}, \citenamefont {K\"uhle},\ and\ \citenamefont
  {Vasiliev}}]{Verbist-PC-1996}%
  \BibitemOpen
  \bibfield  {author} {\bibinfo {author} {\bibfnamefont {K.}~\bibnamefont
  {Verbist}}, \bibinfo {author} {\bibfnamefont {A.}~\bibnamefont {K\"uhle}}, \
  and\ \bibinfo {author} {\bibfnamefont {A.~L.}\ \bibnamefont {Vasiliev}},\
  }\href {\doibase 10.1016/0921-4534(96)00437-6} {\bibfield  {journal}
  {\bibinfo  {journal} {Physica~C}\ }\textbf {\bibinfo {volume} {269}},\
  \bibinfo {pages} {131 } (\bibinfo {year} {1996})}\BibitemShut {NoStop}%
\bibitem [{\citenamefont {van~der Beek}\ \emph {et~al.}(2002)\citenamefont
  {van~der Beek}, \citenamefont {Konczykowski}, \citenamefont {Abal'oshev},
  \citenamefont {Abal'osheva}, \citenamefont {Gierlowski}, \citenamefont
  {Lewandowski}, \citenamefont {Indenbom},\ and\ \citenamefont
  {Barbanera}}]{vanderBeek-PRB-2002}%
  \BibitemOpen
  \bibfield  {author} {\bibinfo {author} {\bibfnamefont {C.~J.}\ \bibnamefont
  {van~der Beek}}, \bibinfo {author} {\bibfnamefont {M.}~\bibnamefont
  {Konczykowski}}, \bibinfo {author} {\bibfnamefont {A.}~\bibnamefont
  {Abal'oshev}}, \bibinfo {author} {\bibfnamefont {I.}~\bibnamefont
  {Abal'osheva}}, \bibinfo {author} {\bibfnamefont {P.}~\bibnamefont
  {Gierlowski}}, \bibinfo {author} {\bibfnamefont {S.~J.}\ \bibnamefont
  {Lewandowski}}, \bibinfo {author} {\bibfnamefont {M.~V.}\ \bibnamefont
  {Indenbom}}, \ and\ \bibinfo {author} {\bibfnamefont {S.}~\bibnamefont
  {Barbanera}},\ }\href {http://link.aps.org/doi/10.1103/PhysRevB.66.024523}
  {\bibfield  {journal} {\bibinfo  {journal} {Phys. Rev.~B}\ }\textbf {\bibinfo
  {volume} {66}},\ \bibinfo {pages} {024523} (\bibinfo {year}
  {2002})}\BibitemShut {NoStop}%
\bibitem [{\citenamefont {Ijaduola}\ \emph {et~al.}(2006)\citenamefont
  {Ijaduola}, \citenamefont {Thompson}, \citenamefont {Feenstra}, \citenamefont
  {Christen}, \citenamefont {Gapud},\ and\ \citenamefont
  {Song}}]{Ijaduola-PRB-2006}%
  \BibitemOpen
  \bibfield  {author} {\bibinfo {author} {\bibfnamefont {A.~O.}\ \bibnamefont
  {Ijaduola}}, \bibinfo {author} {\bibfnamefont {J.~R.}\ \bibnamefont
  {Thompson}}, \bibinfo {author} {\bibfnamefont {R.}~\bibnamefont {Feenstra}},
  \bibinfo {author} {\bibfnamefont {D.~K.}\ \bibnamefont {Christen}}, \bibinfo
  {author} {\bibfnamefont {A.~A.}\ \bibnamefont {Gapud}}, \ and\ \bibinfo
  {author} {\bibfnamefont {X.}~\bibnamefont {Song}},\ }\href {\doibase
  10.1103/PhysRevB.73.134502} {\bibfield  {journal} {\bibinfo  {journal} {Phys.
  Rev.~B}\ }\textbf {\bibinfo {volume} {73}},\ \bibinfo {pages} {134502}
  (\bibinfo {year} {2006})}\BibitemShut {NoStop}%
\bibitem [{\citenamefont {Miura}\ \emph {et~al.}(2011)\citenamefont {Miura},
  \citenamefont {Maiorov}, \citenamefont {Baily}, \citenamefont {Haberkorn},
  \citenamefont {Willis}, \citenamefont {Marken}, \citenamefont {Izumi},
  \citenamefont {Shiohara},\ and\ \citenamefont {Civale}}]{Miura-PRB-2011}%
  \BibitemOpen
  \bibfield  {author} {\bibinfo {author} {\bibfnamefont {M.}~\bibnamefont
  {Miura}}, \bibinfo {author} {\bibfnamefont {B.}~\bibnamefont {Maiorov}},
  \bibinfo {author} {\bibfnamefont {S.~A.}\ \bibnamefont {Baily}}, \bibinfo
  {author} {\bibfnamefont {N.}~\bibnamefont {Haberkorn}}, \bibinfo {author}
  {\bibfnamefont {J.~O.}\ \bibnamefont {Willis}}, \bibinfo {author}
  {\bibfnamefont {K.}~\bibnamefont {Marken}}, \bibinfo {author} {\bibfnamefont
  {T.}~\bibnamefont {Izumi}}, \bibinfo {author} {\bibfnamefont
  {Y.}~\bibnamefont {Shiohara}}, \ and\ \bibinfo {author} {\bibfnamefont
  {L.}~\bibnamefont {Civale}},\ }\href {\doibase 10.1103/PhysRevB.83.184519}
  {\bibfield  {journal} {\bibinfo  {journal} {Phys. Rev.~B}\ }\textbf {\bibinfo
  {volume} {83}},\ \bibinfo {pages} {184519} (\bibinfo {year}
  {2011})}\BibitemShut {NoStop}%
\bibitem [{\citenamefont {Moraitakis}\ \emph {et~al.}(1999)\citenamefont
  {Moraitakis}, \citenamefont {Pissas}, \citenamefont {Kallias},\ and\
  \citenamefont {Niarchos}}]{Moraitakis-SST-1999}%
  \BibitemOpen
  \bibfield  {author} {\bibinfo {author} {\bibfnamefont {E.}~\bibnamefont
  {Moraitakis}}, \bibinfo {author} {\bibfnamefont {M.}~\bibnamefont {Pissas}},
  \bibinfo {author} {\bibfnamefont {G.}~\bibnamefont {Kallias}}, \ and\
  \bibinfo {author} {\bibfnamefont {D.}~\bibnamefont {Niarchos}},\ }\href
  {\doibase 10.1088/0953-2048/12/10/306} {\bibfield  {journal} {\bibinfo
  {journal} {Supercond. Sci. Technol.}\ }\textbf {\bibinfo {volume} {12}},\
  \bibinfo {pages} {682} (\bibinfo {year} {1999})}\BibitemShut {NoStop}%
\bibitem [{\citenamefont {Peurla}\ \emph {et~al.}(2005)\citenamefont {Peurla},
  \citenamefont {Huhtinen},\ and\ \citenamefont {Paturi}}]{Peurla-SST-2005}%
  \BibitemOpen
  \bibfield  {author} {\bibinfo {author} {\bibfnamefont {M.}~\bibnamefont
  {Peurla}}, \bibinfo {author} {\bibfnamefont {H.}~\bibnamefont {Huhtinen}}, \
  and\ \bibinfo {author} {\bibfnamefont {P.}~\bibnamefont {Paturi}},\ }\href
  {\doibase 10.1088/0953-2048/18/5/009} {\bibfield  {journal} {\bibinfo
  {journal} {Supercond. Sci. Technol.}\ }\textbf {\bibinfo {volume} {18}},\
  \bibinfo {pages} {628} (\bibinfo {year} {2005})}\BibitemShut {NoStop}%
\bibitem [{\citenamefont {Guti\'errez}\ \emph {et~al.}(2007)\citenamefont
  {Guti\'errez}, \citenamefont {Puig},\ and\ \citenamefont
  {Obradors}}]{Gutierres-APL-2007}%
  \BibitemOpen
  \bibfield  {author} {\bibinfo {author} {\bibfnamefont {J.}~\bibnamefont
  {Guti\'errez}}, \bibinfo {author} {\bibfnamefont {T.}~\bibnamefont {Puig}}, \
  and\ \bibinfo {author} {\bibfnamefont {X.}~\bibnamefont {Obradors}},\ }\href
  {\doibase 10.1063/1.2728757} {\bibfield  {journal} {\bibinfo  {journal}
  {Appl. Phys. Lett.}\ }\textbf {\bibinfo {volume} {90}},\ \bibinfo {pages}
  {162514} (\bibinfo {year} {2007})}\BibitemShut {NoStop}%
\bibitem [{\citenamefont {Puig}\ \emph {et~al.}(2008)\citenamefont {Puig},
  \citenamefont {Guti\'errez}, \citenamefont {Pomar}, \citenamefont
  {Llord\'es}, \citenamefont {G\'azquez}, \citenamefont {Ricart}, \citenamefont
  {Sandiumenge},\ and\ \citenamefont {Obradors}}]{Puig-SST-2008}%
  \BibitemOpen
  \bibfield  {author} {\bibinfo {author} {\bibfnamefont {T.}~\bibnamefont
  {Puig}}, \bibinfo {author} {\bibfnamefont {J.}~\bibnamefont {Guti\'errez}},
  \bibinfo {author} {\bibfnamefont {A.}~\bibnamefont {Pomar}}, \bibinfo
  {author} {\bibfnamefont {A.}~\bibnamefont {Llord\'es}}, \bibinfo {author}
  {\bibfnamefont {J.}~\bibnamefont {G\'azquez}}, \bibinfo {author}
  {\bibfnamefont {S.}~\bibnamefont {Ricart}}, \bibinfo {author} {\bibfnamefont
  {F.}~\bibnamefont {Sandiumenge}}, \ and\ \bibinfo {author} {\bibfnamefont
  {X.}~\bibnamefont {Obradors}},\ }\href
  {http://stacks.iop.org/0953-2048/21/i=3/a=034008} {\bibfield  {journal}
  {\bibinfo  {journal} {Supercond. Sci. Technol.}\ }\textbf {\bibinfo {volume}
  {21}},\ \bibinfo {pages} {034008} (\bibinfo {year} {2008})}\BibitemShut
  {NoStop}%
\bibitem [{\citenamefont {Polat}\ \emph {et~al.}(2011)\citenamefont {Polat},
  \citenamefont {Sinclair}, \citenamefont {Zuev}, \citenamefont {Thompson},
  \citenamefont {Christen}, \citenamefont {Cook}, \citenamefont {Kumar},
  \citenamefont {Chen},\ and\ \citenamefont {Selvamanickam}}]{Polat-PRB-2011}%
  \BibitemOpen
  \bibfield  {author} {\bibinfo {author} {\bibfnamefont {O.}~\bibnamefont
  {Polat}}, \bibinfo {author} {\bibfnamefont {J.~W.}\ \bibnamefont {Sinclair}},
  \bibinfo {author} {\bibfnamefont {Y.~L.}\ \bibnamefont {Zuev}}, \bibinfo
  {author} {\bibfnamefont {J.~R.}\ \bibnamefont {Thompson}}, \bibinfo {author}
  {\bibfnamefont {D.~K.}\ \bibnamefont {Christen}}, \bibinfo {author}
  {\bibfnamefont {S.~W.}\ \bibnamefont {Cook}}, \bibinfo {author}
  {\bibfnamefont {D.}~\bibnamefont {Kumar}}, \bibinfo {author} {\bibfnamefont
  {Y.}~\bibnamefont {Chen}}, \ and\ \bibinfo {author} {\bibfnamefont
  {V.}~\bibnamefont {Selvamanickam}},\ }\href {\doibase
  10.1103/PhysRevB.84.024519} {\bibfield  {journal} {\bibinfo  {journal} {Phys.
  Rev.~B}\ }\textbf {\bibinfo {volume} {84}},\ \bibinfo {pages} {024519}
  (\bibinfo {year} {2011})}\BibitemShut {NoStop}%
\bibitem [{\citenamefont {Pashitski\u\i}\ \emph {et~al.}(2001)\citenamefont
  {Pashitski\u\i}, \citenamefont {Vakaryuk}, \citenamefont {Ryabchenko},\ and\
  \citenamefont {Fedotov}}]{Pashitskii-LTP-2001}%
  \BibitemOpen
  \bibfield  {author} {\bibinfo {author} {\bibfnamefont {E.~A.}\ \bibnamefont
  {Pashitski\u\i}}, \bibinfo {author} {\bibfnamefont {V.~I.}\ \bibnamefont
  {Vakaryuk}}, \bibinfo {author} {\bibfnamefont {S.~M.}\ \bibnamefont
  {Ryabchenko}}, \ and\ \bibinfo {author} {\bibfnamefont {Y.~V.}\ \bibnamefont
  {Fedotov}},\ }\href {\doibase 10.1063/1.1353699} {\bibfield  {journal}
  {\bibinfo  {journal} {Low Temp. Phys.}\ }\textbf {\bibinfo {volume} {27}},\
  \bibinfo {pages} {96} (\bibinfo {year} {2001})}\BibitemShut {NoStop}%
\bibitem [{\citenamefont {Fedotov}\ \emph {et~al.}(2002)\citenamefont
  {Fedotov}, \citenamefont {Ryabchenko}, \citenamefont {Pashitski\u\i},
  \citenamefont {Semenov}, \citenamefont {Vakaryuk}, \citenamefont {Pan},\ and\
  \citenamefont {Flis}}]{Fedotov-LTP-2002}%
  \BibitemOpen
  \bibfield  {author} {\bibinfo {author} {\bibfnamefont {Y.~V.}\ \bibnamefont
  {Fedotov}}, \bibinfo {author} {\bibfnamefont {S.~M.}\ \bibnamefont
  {Ryabchenko}}, \bibinfo {author} {\bibfnamefont {E.~A.}\ \bibnamefont
  {Pashitski\u\i}}, \bibinfo {author} {\bibfnamefont {A.~V.}\ \bibnamefont
  {Semenov}}, \bibinfo {author} {\bibfnamefont {V.~I.}\ \bibnamefont
  {Vakaryuk}}, \bibinfo {author} {\bibfnamefont {V.~M.}\ \bibnamefont {Pan}}, \
  and\ \bibinfo {author} {\bibfnamefont {V.~S.}\ \bibnamefont {Flis}},\ }\href
  {\doibase 10.1063/1.1468520} {\bibfield  {journal} {\bibinfo  {journal} {Low
  Temp. Phys.}\ }\textbf {\bibinfo {volume} {28}},\ \bibinfo {pages} {172}
  (\bibinfo {year} {2002})}\BibitemShut {NoStop}%
\bibitem [{\citenamefont {Djupmyr}\ \emph {et~al.}(2005)\citenamefont
  {Djupmyr}, \citenamefont {Cristiani}, \citenamefont {Habermeier},\ and\
  \citenamefont {Albrecht}}]{Djupmyr-PRB-2005}%
  \BibitemOpen
  \bibfield  {author} {\bibinfo {author} {\bibfnamefont {M.}~\bibnamefont
  {Djupmyr}}, \bibinfo {author} {\bibfnamefont {G.}~\bibnamefont {Cristiani}},
  \bibinfo {author} {\bibfnamefont {H.-U.}\ \bibnamefont {Habermeier}}, \ and\
  \bibinfo {author} {\bibfnamefont {J.}~\bibnamefont {Albrecht}},\ }\href
  {\doibase 10.1103/PhysRevB.72.220507} {\bibfield  {journal} {\bibinfo
  {journal} {Phys. Rev.~B}\ }\textbf {\bibinfo {volume} {72}},\ \bibinfo
  {pages} {220507(R)} (\bibinfo {year} {2005})}\BibitemShut {NoStop}%
\bibitem [{\citenamefont {Albrecht}\ \emph {et~al.}(2007)\citenamefont
  {Albrecht}, \citenamefont {Djupmyr},\ and\ \citenamefont
  {Br\"uck}}]{Albrecht-JP:CM-2007}%
  \BibitemOpen
  \bibfield  {author} {\bibinfo {author} {\bibfnamefont {J.}~\bibnamefont
  {Albrecht}}, \bibinfo {author} {\bibfnamefont {M.}~\bibnamefont {Djupmyr}}, \
  and\ \bibinfo {author} {\bibfnamefont {S.}~\bibnamefont {Br\"uck}},\ }\href
  {\doibase 10.1088/0953-8984/19/21/216211} {\bibfield  {journal} {\bibinfo
  {journal} {J.~Phys.: Condens. Matter}\ }\textbf {\bibinfo {volume} {19}},\
  \bibinfo {pages} {216211(7)} (\bibinfo {year} {2007})}\BibitemShut {NoStop}%
\bibitem [{\citenamefont {Jung}\ \emph {et~al.}(1999)\citenamefont {Jung},
  \citenamefont {Yan}, \citenamefont {Darhmaoui}, \citenamefont {Abdelhadi},
  \citenamefont {Boyce},\ and\ \citenamefont {Lemberger}}]{Jung-SST-1999}%
  \BibitemOpen
  \bibfield  {author} {\bibinfo {author} {\bibfnamefont {J.}~\bibnamefont
  {Jung}}, \bibinfo {author} {\bibfnamefont {H.}~\bibnamefont {Yan}}, \bibinfo
  {author} {\bibfnamefont {H.}~\bibnamefont {Darhmaoui}}, \bibinfo {author}
  {\bibfnamefont {M.}~\bibnamefont {Abdelhadi}}, \bibinfo {author}
  {\bibfnamefont {B.}~\bibnamefont {Boyce}}, \ and\ \bibinfo {author}
  {\bibfnamefont {T.}~\bibnamefont {Lemberger}},\ }\href {\doibase
  10.1088/0953-2048/12/12/312} {\bibfield  {journal} {\bibinfo  {journal}
  {Supercond. Sci. Technol.}\ }\textbf {\bibinfo {volume} {12}},\ \bibinfo
  {pages} {1086} (\bibinfo {year} {1999})}\BibitemShut {NoStop}%
\bibitem [{\citenamefont {Yan}\ \emph {et~al.}(2000)\citenamefont {Yan}, ,
  \citenamefont {Jung}, \citenamefont {Darhmaoui}, \citenamefont {Ren},
  \citenamefont {Wang},\ and\ \citenamefont {Kwok}}]{Yan-PRB-2000}%
  \BibitemOpen
  \bibfield  {author} {\bibinfo {author} {\bibfnamefont {H.}~\bibnamefont
  {Yan}}, , \bibinfo {author} {\bibfnamefont {J.}~\bibnamefont {Jung}},
  \bibinfo {author} {\bibfnamefont {H.}~\bibnamefont {Darhmaoui}}, \bibinfo
  {author} {\bibfnamefont {Z.~F.}\ \bibnamefont {Ren}}, \bibinfo {author}
  {\bibfnamefont {J.~H.}\ \bibnamefont {Wang}}, \ and\ \bibinfo {author}
  {\bibfnamefont {W.-K.}\ \bibnamefont {Kwok}},\ }\href {\doibase
  10.1103/PhysRevB.61.11711} {\bibfield  {journal} {\bibinfo  {journal} {Phys.
  Rev.~B}\ }\textbf {\bibinfo {volume} {61}},\ \bibinfo {pages} {11 711}
  (\bibinfo {year} {2000})}\BibitemShut {NoStop}%
\bibitem [{\citenamefont {Fruchter}\ \emph {et~al.}(1991)\citenamefont
  {Fruchter}, \citenamefont {Malozemoff}, \citenamefont {Campbell},
  \citenamefont {Sanchez}, \citenamefont {Konczykowski}, \citenamefont
  {Griessen},\ and\ \citenamefont {Holtzberg}}]{Fruchter-PRB-1991}%
  \BibitemOpen
  \bibfield  {author} {\bibinfo {author} {\bibfnamefont {L.}~\bibnamefont
  {Fruchter}}, \bibinfo {author} {\bibfnamefont {A.~P.}\ \bibnamefont
  {Malozemoff}}, \bibinfo {author} {\bibfnamefont {I.~A.}\ \bibnamefont
  {Campbell}}, \bibinfo {author} {\bibfnamefont {J.}~\bibnamefont {Sanchez}},
  \bibinfo {author} {\bibfnamefont {M.}~\bibnamefont {Konczykowski}}, \bibinfo
  {author} {\bibfnamefont {R.}~\bibnamefont {Griessen}}, \ and\ \bibinfo
  {author} {\bibfnamefont {F.}~\bibnamefont {Holtzberg}},\ }\href {\doibase
  10.1103/PhysRevB.43.8709} {\bibfield  {journal} {\bibinfo  {journal} {Phys.
  Rev.~B}\ }\textbf {\bibinfo {volume} {43}},\ \bibinfo {pages} {8709}
  (\bibinfo {year} {1991})}\BibitemShut {NoStop}%
\bibitem [{\citenamefont {van Dalen}\ \emph {et~al.}(1996)\citenamefont {van
  Dalen}, \citenamefont {Griessen}, \citenamefont {Libbrecht}, \citenamefont
  {Bruynseraede},\ and\ \citenamefont {Osquiguil}}]{vanDalen-PRB-1996}%
  \BibitemOpen
  \bibfield  {author} {\bibinfo {author} {\bibfnamefont {A.~J.~J.}\
  \bibnamefont {van Dalen}}, \bibinfo {author} {\bibfnamefont {R.}~\bibnamefont
  {Griessen}}, \bibinfo {author} {\bibfnamefont {S.}~\bibnamefont {Libbrecht}},
  \bibinfo {author} {\bibfnamefont {Y.}~\bibnamefont {Bruynseraede}}, \ and\
  \bibinfo {author} {\bibfnamefont {E.}~\bibnamefont {Osquiguil}},\ }\href
  {\doibase 10.1103/PhysRevB.54.1366} {\bibfield  {journal} {\bibinfo
  {journal} {Phys. Rev.~B}\ }\textbf {\bibinfo {volume} {54}},\ \bibinfo
  {pages} {1366} (\bibinfo {year} {1996})}\BibitemShut {NoStop}%
\bibitem [{\citenamefont {Hoekstra}\ \emph {et~al.}(1999)\citenamefont
  {Hoekstra}, \citenamefont {Testa}, \citenamefont {Doornbos}, \citenamefont
  {Martinez}, \citenamefont {Dam}, \citenamefont {Griessen}, \citenamefont
  {Ivlev}, \citenamefont {Brinkmann}, \citenamefont {Westerholt}, \citenamefont
  {Kwok},\ and\ \citenamefont {Crabtree}}]{Hoekstra-PRB-1999}%
  \BibitemOpen
  \bibfield  {author} {\bibinfo {author} {\bibfnamefont {A.~F.~T.}\
  \bibnamefont {Hoekstra}}, \bibinfo {author} {\bibfnamefont {A.~M.}\
  \bibnamefont {Testa}}, \bibinfo {author} {\bibfnamefont {G.}~\bibnamefont
  {Doornbos}}, \bibinfo {author} {\bibfnamefont {J.~C.}\ \bibnamefont
  {Martinez}}, \bibinfo {author} {\bibfnamefont {B.}~\bibnamefont {Dam}},
  \bibinfo {author} {\bibfnamefont {R.}~\bibnamefont {Griessen}}, \bibinfo
  {author} {\bibfnamefont {B.~I.}\ \bibnamefont {Ivlev}}, \bibinfo {author}
  {\bibfnamefont {M.}~\bibnamefont {Brinkmann}}, \bibinfo {author}
  {\bibfnamefont {K.}~\bibnamefont {Westerholt}}, \bibinfo {author}
  {\bibfnamefont {W.~K.}\ \bibnamefont {Kwok}}, \ and\ \bibinfo {author}
  {\bibfnamefont {G.~W.}\ \bibnamefont {Crabtree}},\ }\href {\doibase
  10.1103/PhysRevB.59.7222} {\bibfield  {journal} {\bibinfo  {journal} {Phys.
  Rev.~B}\ }\textbf {\bibinfo {volume} {59}},\ \bibinfo {pages} {7222}
  (\bibinfo {year} {1999})}\BibitemShut {NoStop}%
\bibitem [{\citenamefont {Landau}\ and\ \citenamefont
  {Ott}(2000)}]{Landau-PC-2000:2}%
  \BibitemOpen
  \bibfield  {author} {\bibinfo {author} {\bibfnamefont {I.}~\bibnamefont
  {Landau}}\ and\ \bibinfo {author} {\bibfnamefont {H.}~\bibnamefont {Ott}},\
  }\href {\doibase http://dx.doi.org/10.1016/S0921-4534(00)01510-0} {\bibfield
  {journal} {\bibinfo  {journal} {Physica~C}\ }\textbf {\bibinfo {volume}
  {340}},\ \bibinfo {pages} {251} (\bibinfo {year} {2000})}\BibitemShut
  {NoStop}%
\bibitem [{\citenamefont {Farrell}\ \emph {et~al.}(1990)\citenamefont
  {Farrell}, \citenamefont {Rice}, \citenamefont {Ginsberg},\ and\
  \citenamefont {Liu}}]{Farrell-PRL-1990}%
  \BibitemOpen
  \bibfield  {author} {\bibinfo {author} {\bibfnamefont {D.~E.}\ \bibnamefont
  {Farrell}}, \bibinfo {author} {\bibfnamefont {J.~P.}\ \bibnamefont {Rice}},
  \bibinfo {author} {\bibfnamefont {D.~M.}\ \bibnamefont {Ginsberg}}, \ and\
  \bibinfo {author} {\bibfnamefont {J.~Z.}\ \bibnamefont {Liu}},\ }\href
  {\doibase 10.1103/PhysRevLett.64.1573} {\bibfield  {journal} {\bibinfo
  {journal} {Phys. Rev. Lett.}\ }\textbf {\bibinfo {volume} {64}},\ \bibinfo
  {pages} {1573} (\bibinfo {year} {1990})}\BibitemShut {NoStop}%
\bibitem [{\citenamefont {Blatter}\ \emph {et~al.}(2004)\citenamefont
  {Blatter}, \citenamefont {Geshkenbein},\ and\ \citenamefont
  {Koopmann}}]{Blatter-PRL-2004}%
  \BibitemOpen
  \bibfield  {author} {\bibinfo {author} {\bibfnamefont {G.}~\bibnamefont
  {Blatter}}, \bibinfo {author} {\bibfnamefont {V.~B.}\ \bibnamefont
  {Geshkenbein}}, \ and\ \bibinfo {author} {\bibfnamefont {J.~A.~G.}\
  \bibnamefont {Koopmann}},\ }\href {\doibase 10.1103/PhysRevLett.92.067009}
  {\bibfield  {journal} {\bibinfo  {journal} {Phys. Rev. Lett.}\ }\textbf
  {\bibinfo {volume} {62}},\ \bibinfo {pages} {067009} (\bibinfo {year}
  {2004})}\BibitemShut {NoStop}%
\bibitem [{\citenamefont {Ovchinnikov}\ and\ \citenamefont
  {Ivlev}(1991)}]{Ovchinnikov-PRB-1991}%
  \BibitemOpen
  \bibfield  {author} {\bibinfo {author} {\bibfnamefont {Y.~N.}\ \bibnamefont
  {Ovchinnikov}}\ and\ \bibinfo {author} {\bibfnamefont {B.~I.}\ \bibnamefont
  {Ivlev}},\ }\href {\doibase 10.1103/PhysRevB.43.8024} {\bibfield  {journal}
  {\bibinfo  {journal} {Phys. Rev.~B}\ }\textbf {\bibinfo {volume} {43}},\
  \bibinfo {pages} {8024} (\bibinfo {year} {1991})}\BibitemShut {NoStop}%
\bibitem [{\citenamefont {Gurevich}\ and\ \citenamefont
  {Pashitskii}(1998)}]{Gurevich-PRB-1998}%
  \BibitemOpen
  \bibfield  {author} {\bibinfo {author} {\bibfnamefont {A.}~\bibnamefont
  {Gurevich}}\ and\ \bibinfo {author} {\bibfnamefont {E.~A.}\ \bibnamefont
  {Pashitskii}},\ }\href {\doibase 10.1103/PhysRevB.57.13878} {\bibfield
  {journal} {\bibinfo  {journal} {Phys. Rev.~B}\ }\textbf {\bibinfo {volume}
  {57}},\ \bibinfo {pages} {13 878 } (\bibinfo {year} {1998})}\BibitemShut
  {NoStop}%
\bibitem [{\citenamefont {Pan}\ \emph {et~al.}(2003)\citenamefont {Pan},
  \citenamefont {Pashitskii}, \citenamefont {Ryabchenko}, \citenamefont
  {Komashko}, \citenamefont {Pan}, \citenamefont {Dou}, \citenamefont
  {Semenov}, \citenamefont {Tretiatchenko},\ and\ \citenamefont
  {Fedotov}}]{Pan-IEEETAS-2003}%
  \BibitemOpen
  \bibfield  {author} {\bibinfo {author} {\bibfnamefont {V.~M.}\ \bibnamefont
  {Pan}}, \bibinfo {author} {\bibfnamefont {E.~A.}\ \bibnamefont {Pashitskii}},
  \bibinfo {author} {\bibfnamefont {S.~M.}\ \bibnamefont {Ryabchenko}},
  \bibinfo {author} {\bibfnamefont {V.~A.}\ \bibnamefont {Komashko}}, \bibinfo
  {author} {\bibfnamefont {A.~V.}\ \bibnamefont {Pan}}, \bibinfo {author}
  {\bibfnamefont {S.~X.}\ \bibnamefont {Dou}}, \bibinfo {author} {\bibfnamefont
  {A.~V.}\ \bibnamefont {Semenov}}, \bibinfo {author} {\bibfnamefont {C.~G.}\
  \bibnamefont {Tretiatchenko}}, \ and\ \bibinfo {author} {\bibfnamefont
  {Y.~V.}\ \bibnamefont {Fedotov}},\ }\href {\doibase 10.1109/TASC.2003.812523}
  {\bibfield  {journal} {\bibinfo  {journal} {IEEE Trans. Appl. Supercond.}\
  }\textbf {\bibinfo {volume} {13}},\ \bibinfo {pages} {054508} (\bibinfo
  {year} {2003})}\BibitemShut {NoStop}%
\bibitem [{\citenamefont {Kosse}\ \emph {et~al.}(2008)\citenamefont {Kosse},
  \citenamefont {Prokhorov}, \citenamefont {Khokhlov}, \citenamefont
  {Levchenko}, \citenamefont {Semenov}, \citenamefont {Kovalchuk},
  \citenamefont {Chernomorets},\ and\ \citenamefont
  {Mikheenko}}]{Kosse-SST-2008}%
  \BibitemOpen
  \bibfield  {author} {\bibinfo {author} {\bibfnamefont {A.~I.}\ \bibnamefont
  {Kosse}}, \bibinfo {author} {\bibfnamefont {A.~Y.}\ \bibnamefont
  {Prokhorov}}, \bibinfo {author} {\bibfnamefont {V.~A.}\ \bibnamefont
  {Khokhlov}}, \bibinfo {author} {\bibfnamefont {G.~G.}\ \bibnamefont
  {Levchenko}}, \bibinfo {author} {\bibfnamefont {A.~V.}\ \bibnamefont
  {Semenov}}, \bibinfo {author} {\bibfnamefont {D.~G.}\ \bibnamefont
  {Kovalchuk}}, \bibinfo {author} {\bibfnamefont {M.~P.}\ \bibnamefont
  {Chernomorets}}, \ and\ \bibinfo {author} {\bibfnamefont {P.~N.}\
  \bibnamefont {Mikheenko}},\ }\href {\doibase 10.1088/0953-2048/21/7/075015}
  {\bibfield  {journal} {\bibinfo  {journal} {Supercond. Sci. Technol.}\
  }\textbf {\bibinfo {volume} {21}},\ \bibinfo {pages} {075015} (\bibinfo
  {year} {2008})}\BibitemShut {NoStop}%
\bibitem [{\citenamefont {Ivanov}\ \emph {et~al.}(1991)\citenamefont {Ivanov},
  \citenamefont {Galkin}, \citenamefont {Kuznetsov},\ and\ \citenamefont
  {Menushenkov}}]{Ivanov-PC-1991}%
  \BibitemOpen
  \bibfield  {author} {\bibinfo {author} {\bibfnamefont {A.~A.}\ \bibnamefont
  {Ivanov}}, \bibinfo {author} {\bibfnamefont {S.~G.}\ \bibnamefont {Galkin}},
  \bibinfo {author} {\bibfnamefont {A.~V.}\ \bibnamefont {Kuznetsov}}, \ and\
  \bibinfo {author} {\bibfnamefont {A.~P.}\ \bibnamefont {Menushenkov}},\
  }\href {\doibase 10.1016/0921-4534(91)90638-F} {\bibfield  {journal}
  {\bibinfo  {journal} {Physica~C}\ }\textbf {\bibinfo {volume} {180}},\
  \bibinfo {pages} {69} (\bibinfo {year} {1991})}\BibitemShut {NoStop}%
\bibitem [{\citenamefont {Pechen}\ \emph {et~al.}(1995)\citenamefont {Pechen},
  \citenamefont {Varlashkin}, \citenamefont {Krasnosvobodtsev}, \citenamefont
  {Brunner},\ and\ \citenamefont {Renk}}]{Pechen-APL-1995}%
  \BibitemOpen
  \bibfield  {author} {\bibinfo {author} {\bibfnamefont {E.~V.}\ \bibnamefont
  {Pechen}}, \bibinfo {author} {\bibfnamefont {A.~V.}\ \bibnamefont
  {Varlashkin}}, \bibinfo {author} {\bibfnamefont {S.~I.}\ \bibnamefont
  {Krasnosvobodtsev}}, \bibinfo {author} {\bibfnamefont {B.}~\bibnamefont
  {Brunner}}, \ and\ \bibinfo {author} {\bibfnamefont {K.~F.}\ \bibnamefont
  {Renk}},\ }\href {\doibase 10.1063/1.113264} {\bibfield  {journal} {\bibinfo
  {journal} {Appl. Phys. Lett.}\ }\textbf {\bibinfo {volume} {66}},\ \bibinfo
  {pages} {2292} (\bibinfo {year} {1995})}\BibitemShut {NoStop}%
\bibitem [{\citenamefont {Trofimov}(1992)}]{Trofimov92:1}%
  \BibitemOpen
  \bibfield  {author} {\bibinfo {author} {\bibfnamefont {V.~N.}\ \bibnamefont
  {Trofimov}},\ }\href {\doibase 10.1016/0011-2275(92)90218-Y} {\bibfield
  {journal} {\bibinfo  {journal} {Cyogenics}\ }\textbf {\bibinfo {volume}
  {32}},\ \bibinfo {pages} {513} (\bibinfo {year} {1992})}\BibitemShut
  {NoStop}%
\bibitem [{\citenamefont {Kuznetsov}\ \emph {et~al.}(1995)\citenamefont
  {Kuznetsov}, \citenamefont {Ivanov}, \citenamefont {Eremenko},\ and\
  \citenamefont {Trofimov}}]{Kuznetsov-PRB-1995}%
  \BibitemOpen
  \bibfield  {author} {\bibinfo {author} {\bibfnamefont {A.~V.}\ \bibnamefont
  {Kuznetsov}}, \bibinfo {author} {\bibfnamefont {A.~A.}\ \bibnamefont
  {Ivanov}}, \bibinfo {author} {\bibfnamefont {D.~V.}\ \bibnamefont
  {Eremenko}}, \ and\ \bibinfo {author} {\bibfnamefont {V.~N.}\ \bibnamefont
  {Trofimov}},\ }\href {http://link.aps.org/doi/10.1103/PhysRevB.52.9637}
  {\bibfield  {journal} {\bibinfo  {journal} {Phys. Rev. B}\ }\textbf {\bibinfo
  {volume} {52}},\ \bibinfo {pages} {9637} (\bibinfo {year}
  {1995})}\BibitemShut {NoStop}%
\bibitem [{\citenamefont {Landau}\ and\ \citenamefont
  {Ott}(2001)}]{Landau-PRB-2001}%
  \BibitemOpen
  \bibfield  {author} {\bibinfo {author} {\bibfnamefont {I.~L.}\ \bibnamefont
  {Landau}}\ and\ \bibinfo {author} {\bibfnamefont {H.~R.}\ \bibnamefont
  {Ott}},\ }\href {\doibase 10.1103/PhysRevB.63.184516} {\bibfield  {journal}
  {\bibinfo  {journal} {Phys. Rev.~B}\ }\textbf {\bibinfo {volume} {63}},\
  \bibinfo {pages} {184516} (\bibinfo {year} {2001})}\BibitemShut {NoStop}%
\bibitem [{\citenamefont {Mikheenko}\ and\ \citenamefont
  {Kuzovlev}(1993)}]{Mikheenko-PC-1993}%
  \BibitemOpen
  \bibfield  {author} {\bibinfo {author} {\bibfnamefont {P.~N.}\ \bibnamefont
  {Mikheenko}}\ and\ \bibinfo {author} {\bibfnamefont {Y.~E.}\ \bibnamefont
  {Kuzovlev}},\ }\href
  {http://www.sciencedirect.com/science/article/pii/092145349391004F}
  {\bibfield  {journal} {\bibinfo  {journal} {Physica~C}\ }\textbf {\bibinfo
  {volume} {204}},\ \bibinfo {pages} {229} (\bibinfo {year}
  {1993})}\BibitemShut {NoStop}%
\bibitem [{\citenamefont {Clem}\ and\ \citenamefont
  {Sanchez}(1994)}]{Clem-PRB-1994}%
  \BibitemOpen
  \bibfield  {author} {\bibinfo {author} {\bibfnamefont {J.~R.}\ \bibnamefont
  {Clem}}\ and\ \bibinfo {author} {\bibfnamefont {A.}~\bibnamefont {Sanchez}},\
  }\href {http://link.aps.org/doi/10.1103/PhysRevB.50.9355} {\bibfield
  {journal} {\bibinfo  {journal} {Phys. Rev. B}\ }\textbf {\bibinfo {volume}
  {50}},\ \bibinfo {pages} {9355} (\bibinfo {year} {1994})}\BibitemShut
  {NoStop}%
\bibitem [{\citenamefont {Babaei~Brojeny}\ and\ \citenamefont
  {Clem}(2005)}]{Babaei_Brojeny-SST-2005}%
  \BibitemOpen
  \bibfield  {author} {\bibinfo {author} {\bibfnamefont {A.~A.}\ \bibnamefont
  {Babaei~Brojeny}}\ and\ \bibinfo {author} {\bibfnamefont {J.~R.}\
  \bibnamefont {Clem}},\ }\href {\doibase 10.1088/0953-2048/18/6/016}
  {\bibfield  {journal} {\bibinfo  {journal} {Supercond. Sci. Technol.}\
  }\textbf {\bibinfo {volume} {18}},\ \bibinfo {pages} {888} (\bibinfo {year}
  {2005})}\BibitemShut {NoStop}%
\bibitem [{\citenamefont {Bernstein}(2012)}]{Bernstein-JAP-2012}%
  \BibitemOpen
  \bibfield  {author} {\bibinfo {author} {\bibfnamefont {P.}~\bibnamefont
  {Bernstein}},\ }\href {\doibase 10.1063/1.4721363} {\bibfield  {journal}
  {\bibinfo  {journal} {J.~Appl. Phys.}\ }\textbf {\bibinfo {volume} {111}},\
  \bibinfo {pages} {103913} (\bibinfo {year} {2012})}\BibitemShut {NoStop}%
\bibitem [{\citenamefont {Kuznetsov}\ \emph {et~al.}(2016)\citenamefont
  {Kuznetsov}, \citenamefont {Sannikov}, \citenamefont {Ivanov}, \citenamefont
  {Menushenkov},\ and\ \citenamefont {Churkin}}]{Kuznetsov-IEEETAS-2016}%
  \BibitemOpen
  \bibfield  {author} {\bibinfo {author} {\bibfnamefont {A.~V.}\ \bibnamefont
  {Kuznetsov}}, \bibinfo {author} {\bibfnamefont {I.~I.}\ \bibnamefont
  {Sannikov}}, \bibinfo {author} {\bibfnamefont {A.~A.}\ \bibnamefont
  {Ivanov}}, \bibinfo {author} {\bibfnamefont {A.~P.}\ \bibnamefont
  {Menushenkov}}, \ and\ \bibinfo {author} {\bibfnamefont {O.~A.}\ \bibnamefont
  {Churkin}},\ }\href {\doibase 10.1109/TASC.2016.2525939} {\bibfield
  {journal} {\bibinfo  {journal} {IEEE Trans. Appl. Sup.}\ }\textbf {\bibinfo
  {volume} {26}},\ \bibinfo {pages} {8000505} (\bibinfo {year}
  {2016})}\BibitemShut {NoStop}%
\bibitem [{\citenamefont {Beasley}\ \emph {et~al.}(1969)\citenamefont
  {Beasley}, \citenamefont {Labusch},\ and\ \citenamefont
  {Webb}}]{Beasley-PR-1969}%
  \BibitemOpen
  \bibfield  {author} {\bibinfo {author} {\bibfnamefont {M.~R.}\ \bibnamefont
  {Beasley}}, \bibinfo {author} {\bibfnamefont {R.}~\bibnamefont {Labusch}}, \
  and\ \bibinfo {author} {\bibfnamefont {W.~W.}\ \bibnamefont {Webb}},\ }\href
  {http://link.aps.org/doi/10.1103/PhysRev.181.682} {\bibfield  {journal}
  {\bibinfo  {journal} {Phys. Rev.}\ }\textbf {\bibinfo {volume} {181}},\
  \bibinfo {pages} {682} (\bibinfo {year} {1969})}\BibitemShut {NoStop}%
\bibitem [{\citenamefont {Nider\"ost}\ \emph {et~al.}(1996)\citenamefont
  {Nider\"ost}, \citenamefont {Suter}, \citenamefont {Visani}, \citenamefont
  {Mota},\ and\ \citenamefont {Blatter}}]{Niderost-PRB-1996}%
  \BibitemOpen
  \bibfield  {author} {\bibinfo {author} {\bibfnamefont {M.}~\bibnamefont
  {Nider\"ost}}, \bibinfo {author} {\bibfnamefont {A.}~\bibnamefont {Suter}},
  \bibinfo {author} {\bibfnamefont {P.}~\bibnamefont {Visani}}, \bibinfo
  {author} {\bibfnamefont {A.~C.}\ \bibnamefont {Mota}}, \ and\ \bibinfo
  {author} {\bibfnamefont {G.}~\bibnamefont {Blatter}},\ }\href {\doibase
  10.1103/PhysRevB.53.9286} {\bibfield  {journal} {\bibinfo  {journal} {Phys.
  Rev.~B}\ }\textbf {\bibinfo {volume} {53}},\ \bibinfo {pages} {9286}
  (\bibinfo {year} {1996})}\BibitemShut {NoStop}%
\bibitem [{\citenamefont {Krylov}\ \emph {et~al.}(1998)\citenamefont {Krylov},
  \citenamefont {Maritz},\ and\ \citenamefont {Nyeanchi}}]{Krylov-PRB-1998}%
  \BibitemOpen
  \bibfield  {author} {\bibinfo {author} {\bibfnamefont {I.~P.}\ \bibnamefont
  {Krylov}}, \bibinfo {author} {\bibfnamefont {E.~J.}\ \bibnamefont {Maritz}},
  \ and\ \bibinfo {author} {\bibfnamefont {E.~B.}\ \bibnamefont {Nyeanchi}},\
  }\href {\doibase 10.1103/PhysRevB.58.14609} {\bibfield  {journal} {\bibinfo
  {journal} {Phys. Rev.~B}\ }\textbf {\bibinfo {volume} {58}},\ \bibinfo
  {pages} {14609} (\bibinfo {year} {1998})}\BibitemShut {NoStop}%
\bibitem [{\citenamefont {Thompson}\ \emph {et~al.}(2005)\citenamefont
  {Thompson}, \citenamefont {Sorge}, \citenamefont {Cantoni}, \citenamefont
  {Kerchner}, \citenamefont {Christen},\ and\ \citenamefont
  {Paranthaman}}]{Thompson-SST-2005}%
  \BibitemOpen
  \bibfield  {author} {\bibinfo {author} {\bibfnamefont {J.~R.}\ \bibnamefont
  {Thompson}}, \bibinfo {author} {\bibfnamefont {K.~D.}\ \bibnamefont {Sorge}},
  \bibinfo {author} {\bibfnamefont {C.}~\bibnamefont {Cantoni}}, \bibinfo
  {author} {\bibfnamefont {H.~R.}\ \bibnamefont {Kerchner}}, \bibinfo {author}
  {\bibfnamefont {D.~K.}\ \bibnamefont {Christen}}, \ and\ \bibinfo {author}
  {\bibfnamefont {M.}~\bibnamefont {Paranthaman}},\ }\href
  {http://iopscience.iop.org/0953-2048/18/7/008} {\bibfield  {journal}
  {\bibinfo  {journal} {Supercond. Sci. Technol.}\ }\textbf {\bibinfo {volume}
  {18}},\ \bibinfo {pages} {970} (\bibinfo {year} {2005})}\BibitemShut
  {NoStop}%
\bibitem [{\citenamefont {Sannikov}\ \emph {et~al.}(2014)\citenamefont
  {Sannikov}, \citenamefont {Ivanov}, \citenamefont {Kuznetsov}, \citenamefont
  {Menushenkov},\ and\ \citenamefont {Churkin}}]{Sannikov-BLPI-2014}%
  \BibitemOpen
  \bibfield  {author} {\bibinfo {author} {\bibfnamefont {I.~I.}\ \bibnamefont
  {Sannikov}}, \bibinfo {author} {\bibfnamefont {A.~A.}\ \bibnamefont
  {Ivanov}}, \bibinfo {author} {\bibfnamefont {A.~V.}\ \bibnamefont
  {Kuznetsov}}, \bibinfo {author} {\bibfnamefont {A.~P.}\ \bibnamefont
  {Menushenkov}}, \ and\ \bibinfo {author} {\bibfnamefont {O.~A.}\ \bibnamefont
  {Churkin}},\ }\href
  {http://link.springer.com/article/10.3103/S1068335614080016} {\bibfield
  {journal} {\bibinfo  {journal} {Bull. of the Lebedev Phys. Inst.}\ }\textbf
  {\bibinfo {volume} {41}},\ \bibinfo {pages} {215} (\bibinfo {year}
  {2014})}\BibitemShut {NoStop}%
\bibitem [{\citenamefont {Sannikov}\ \emph {et~al.}(2015)\citenamefont
  {Sannikov}, \citenamefont {Ivanov}, \citenamefont {Kuznetsov}, \citenamefont
  {Menushenkov},\ and\ \citenamefont {Churkin}}]{Sannikov-PP-2015-1}%
  \BibitemOpen
  \bibfield  {author} {\bibinfo {author} {\bibfnamefont {I.~I.}\ \bibnamefont
  {Sannikov}}, \bibinfo {author} {\bibfnamefont {A.~A.}\ \bibnamefont
  {Ivanov}}, \bibinfo {author} {\bibfnamefont {A.~V.}\ \bibnamefont
  {Kuznetsov}}, \bibinfo {author} {\bibfnamefont {A.~P.}\ \bibnamefont
  {Menushenkov}}, \ and\ \bibinfo {author} {\bibfnamefont {O.~A.}\ \bibnamefont
  {Churkin}},\ }\href
  {http://www.sciencedirect.com/science/article/pii/S1875389215003016}
  {\bibfield  {journal} {\bibinfo  {journal} {Physics Procedia}\ }\textbf
  {\bibinfo {volume} {65}},\ \bibinfo {pages} {113} (\bibinfo {year}
  {2015})}\BibitemShut {NoStop}%
\bibitem [{hys()}]{hysteresis}%
  \BibitemOpen
  \href@noop {} {}\bibinfo {note} {Only the copper tube temperature is
  controlled in our magnetometer. Overlapping tube inner cross section, large
  samples make convection of the heat-exchange gas more difficult. As a
  sequence sample's and tube's temperatures can differ if $T$ changes
  rapidly.}\BibitemShut {Stop}%
\bibitem [{EcJ()}]{EcJc}%
  \BibitemOpen
  \href@noop {} {}\bibinfo {note} {At the critical current the Joule heat
  density is calculated as $W_c =E_c J_c$. The electric field is estimated as
  $E_c=\mathcal{A}\rho_\mathrm{FF} J_c/$ where $\rho_\mathrm{FF}(H,T)$ is the
  resistivity of the vortex motion and $\mathcal{A}$ is a constant of the order
  of unity.\cite{Blatter-RMP-1994, Brandt-RPP-1995} Taking for magnetic field
  of 1~kOe $\rho_\mathrm{FF} \gtrsim 10-60\text{ n}\Omega\cdot$cm
  (Refs.~\onlinecite{Golosovsky-PRB-1994, Maeda-JPCM-2005}) and
  $J_c>10^7-10^6$~A/cm$^2$ (Refs.~\onlinecite{Dam-N-1999, Peurla-SST-2005,
  Ijaduola-PRB-2006, Foltyn-NM-2007, Polat-PRB-2011}) in the range from the
  liquid helium to the liquid nitrogen temperatures we estimated $E_c\gtrsim
  (100-60)$~mV/cm and $W_c\gtrsim 1-0.06$~MW/cm$^3$.}\BibitemShut {Stop}%
\bibitem [{E-J()}]{E-J}%
  \BibitemOpen
  \href@noop {} {}\bibinfo {note} {The transport current is determined as a
  rule using the electric field criterion 0.1--1~$\mu$V/cm. For induced
  currents the field depends on type of measurements. $E \simeq
  0.01$--0.1~$\mu$V/cm for vibrating sample magnetometry if $H$ is swept with
  the rate of 10--100~Oe/s.\cite{Polat-PRB-2011, Golovchanskiy-JAP-2013} $E
  \simeq 0.01$--0.1~nV/cm for standard SQUID magnetometry. During relaxation
  measurements in the time window about 1 minute to 10 hours $E$ changes in the
  range 100--0.01~pV/cm.}\BibitemShut {Stop}%
\bibitem [{\citenamefont {Lairson}\ \emph {et~al.}(1990)\citenamefont
  {Lairson}, \citenamefont {Sun}, \citenamefont {Bravman},\ and\ \citenamefont
  {Geballe}}]{Lairson-PRB-1990}%
  \BibitemOpen
  \bibfield  {author} {\bibinfo {author} {\bibfnamefont {B.~M.}\ \bibnamefont
  {Lairson}}, \bibinfo {author} {\bibfnamefont {J.~Z.}\ \bibnamefont {Sun}},
  \bibinfo {author} {\bibfnamefont {J.~C.}\ \bibnamefont {Bravman}}, \ and\
  \bibinfo {author} {\bibfnamefont {T.~H.}\ \bibnamefont {Geballe}},\ }\href
  {\doibase 10.1103/PhysRevB.42.1008} {\bibfield  {journal} {\bibinfo
  {journal} {Phys. Rev.~B}\ }\textbf {\bibinfo {volume} {42}},\ \bibinfo
  {pages} {1008} (\bibinfo {year} {1990})}\BibitemShut {NoStop}%
\bibitem [{Jos()}]{Josephson}%
  \BibitemOpen
  \href@noop {} {}\bibinfo {note} {Explanation of $J(T)$ behavior based on
  Josephson coupling of nano-sized domains and presence of nano-sized
  inclusions of underdoped phase in YBa$_2$Cu$_3$O$_{7-\delta}$
  films\cite{Darhmaoui-PRB-1996, Darhmaoui-PRB-1998, Jung-SST-1999,
  Yan-PRB-2000} is not discussed in this paper since it completely ignores
  pinning.}\BibitemShut {Stop}%
\bibitem [{\citenamefont {Klaassen}\ \emph {et~al.}(2001)\citenamefont
  {Klaassen}, \citenamefont {Doornbos}, \citenamefont {Huijbregtse},
  \citenamefont {van~der Geest}, \citenamefont {Dam},\ and\ \citenamefont
  {Griessen}}]{Klaassen-PRB-2001}%
  \BibitemOpen
  \bibfield  {author} {\bibinfo {author} {\bibfnamefont {F.~C.}\ \bibnamefont
  {Klaassen}}, \bibinfo {author} {\bibfnamefont {G.}~\bibnamefont {Doornbos}},
  \bibinfo {author} {\bibfnamefont {J.~M.}\ \bibnamefont {Huijbregtse}},
  \bibinfo {author} {\bibfnamefont {R.~C.~F.}\ \bibnamefont {van~der Geest}},
  \bibinfo {author} {\bibfnamefont {B.}~\bibnamefont {Dam}}, \ and\ \bibinfo
  {author} {\bibfnamefont {R.}~\bibnamefont {Griessen}},\ }\href {\doibase
  10.1103/PhysRevB.64.184523} {\bibfield  {journal} {\bibinfo  {journal} {Phys.
  Rev.~B}\ }\textbf {\bibinfo {volume} {64}},\ \bibinfo {pages} {184523}
  (\bibinfo {year} {2001})}\BibitemShut {NoStop}%
\bibitem [{\citenamefont {Griessen}\ \emph {et~al.}(1994)\citenamefont
  {Griessen}, \citenamefont {Hai-hu}, \citenamefont {van Dalen}, \citenamefont
  {Dam}, \citenamefont {Rector}, \citenamefont {Schnack}, \citenamefont
  {Libbrecht}, \citenamefont {Osquiguil},\ and\ \citenamefont
  {Bruynseraede}}]{Griessen-PRL-1994}%
  \BibitemOpen
  \bibfield  {author} {\bibinfo {author} {\bibfnamefont {R.}~\bibnamefont
  {Griessen}}, \bibinfo {author} {\bibfnamefont {W.}~\bibnamefont {Hai-hu}},
  \bibinfo {author} {\bibfnamefont {A.~J.~J.}\ \bibnamefont {van Dalen}},
  \bibinfo {author} {\bibfnamefont {B.}~\bibnamefont {Dam}}, \bibinfo {author}
  {\bibfnamefont {J.}~\bibnamefont {Rector}}, \bibinfo {author} {\bibfnamefont
  {H.~G.}\ \bibnamefont {Schnack}}, \bibinfo {author} {\bibfnamefont
  {S.}~\bibnamefont {Libbrecht}}, \bibinfo {author} {\bibfnamefont
  {E.}~\bibnamefont {Osquiguil}}, \ and\ \bibinfo {author} {\bibfnamefont
  {Y.}~\bibnamefont {Bruynseraede}},\ }\href
  {http://link.aps.org/doi/10.1103/PhysRevLett.72.1910} {\bibfield  {journal}
  {\bibinfo  {journal} {Phys. Rev. Lett.}\ }\textbf {\bibinfo {volume} {72}},\
  \bibinfo {pages} {1910} (\bibinfo {year} {1994})}\BibitemShut {NoStop}%
\bibitem [{dTc()}]{dTc_dl}%
  \BibitemOpen
  \href@noop {} {}\bibinfo {note} {Spatial variations in the critical
  temperature $T_c$ and in the charge carrier mean free path $\ell$ near
  lattice defects lead respectively to $\delta T_c$ and $\delta\ell$
  pinning\cite{Blatter-RMP-1994, Griessen-PRL-1994}.}\BibitemShut {Stop}%
\bibitem [{\citenamefont {Schneemeyer}\ \emph {et~al.}(1987)\citenamefont
  {Schneemeyer}, \citenamefont {Gyorgy},\ and\ \citenamefont
  {Waszczak}}]{Schneemeyer-PRB-1987}%
  \BibitemOpen
  \bibfield  {author} {\bibinfo {author} {\bibfnamefont {L.~F.}\ \bibnamefont
  {Schneemeyer}}, \bibinfo {author} {\bibfnamefont {E.~M.}\ \bibnamefont
  {Gyorgy}}, \ and\ \bibinfo {author} {\bibfnamefont {J.~V.}\ \bibnamefont
  {Waszczak}},\ }\href {\doibase 10.1103/PhysRevB.36.8804} {\bibfield
  {journal} {\bibinfo  {journal} {Phys. Rev.~B}\ }\textbf {\bibinfo {volume}
  {36}},\ \bibinfo {pages} {8804} (\bibinfo {year} {1987})}\BibitemShut
  {NoStop}%
\bibitem [{\citenamefont {Senoussi}\ \emph {et~al.}(1988)\citenamefont
  {Senoussi}, \citenamefont {Oussena}, \citenamefont {Collin},\ and\
  \citenamefont {Campbell}}]{Senoussi-PRB-1988}%
  \BibitemOpen
  \bibfield  {author} {\bibinfo {author} {\bibfnamefont {S.}~\bibnamefont
  {Senoussi}}, \bibinfo {author} {\bibfnamefont {M.}~\bibnamefont {Oussena}},
  \bibinfo {author} {\bibfnamefont {G.}~\bibnamefont {Collin}}, \ and\ \bibinfo
  {author} {\bibfnamefont {I.~A.}\ \bibnamefont {Campbell}},\ }\href {\doibase
  10.1103/PhysRevB.37.9792} {\bibfield  {journal} {\bibinfo  {journal} {Phys.
  Rev.~B}\ }\textbf {\bibinfo {volume} {37}},\ \bibinfo {pages} {9792}
  (\bibinfo {year} {1988})}\BibitemShut {NoStop}%
\bibitem [{\citenamefont {Christen}\ and\ \citenamefont
  {Thomson}(1993)}]{Christen-N-1993}%
  \BibitemOpen
  \bibfield  {author} {\bibinfo {author} {\bibfnamefont {D.~K.}\ \bibnamefont
  {Christen}}\ and\ \bibinfo {author} {\bibfnamefont {J.~R.}\ \bibnamefont
  {Thomson}},\ }\href {\doibase 10.1038/364098a0} {\bibfield  {journal}
  {\bibinfo  {journal} {Nature}\ }\textbf {\bibinfo {volume} {366}},\ \bibinfo
  {pages} {98} (\bibinfo {year} {1993})}\BibitemShut {NoStop}%
\bibitem [{\citenamefont {Mart\'{\i}nez}\ \emph {et~al.}(1996)\citenamefont
  {Mart\'{\i}nez}, \citenamefont {Obradors}, \citenamefont {Gou}, \citenamefont
  {Gomis}, \citenamefont {Pi\~nol}, \citenamefont {Fontcuberta},\ and\
  \citenamefont {Van~Tol}}]{Martinez-PRB-1996}%
  \BibitemOpen
  \bibfield  {author} {\bibinfo {author} {\bibfnamefont {B.}~\bibnamefont
  {Mart\'{\i}nez}}, \bibinfo {author} {\bibfnamefont {X.}~\bibnamefont
  {Obradors}}, \bibinfo {author} {\bibfnamefont {A.}~\bibnamefont {Gou}},
  \bibinfo {author} {\bibfnamefont {V.}~\bibnamefont {Gomis}}, \bibinfo
  {author} {\bibfnamefont {S.}~\bibnamefont {Pi\~nol}}, \bibinfo {author}
  {\bibfnamefont {J.}~\bibnamefont {Fontcuberta}}, \ and\ \bibinfo {author}
  {\bibfnamefont {H.}~\bibnamefont {Van~Tol}},\ }\href {\doibase
  10.1103/PhysRevB.53.2797} {\bibfield  {journal} {\bibinfo  {journal} {Phys.
  Rev.~B}\ }\textbf {\bibinfo {volume} {53}},\ \bibinfo {pages} {2797}
  (\bibinfo {year} {1996})}\BibitemShut {NoStop}%
\bibitem [{\citenamefont {Plain}\ \emph {et~al.}(2002)\citenamefont {Plain},
  \citenamefont {Puig}, \citenamefont {Sandiumenge}, \citenamefont {Obradors},\
  and\ \citenamefont {Rabier}}]{Plain-PRB-2002}%
  \BibitemOpen
  \bibfield  {author} {\bibinfo {author} {\bibfnamefont {J.}~\bibnamefont
  {Plain}}, \bibinfo {author} {\bibfnamefont {T.}~\bibnamefont {Puig}},
  \bibinfo {author} {\bibfnamefont {F.}~\bibnamefont {Sandiumenge}}, \bibinfo
  {author} {\bibfnamefont {X.}~\bibnamefont {Obradors}}, \ and\ \bibinfo
  {author} {\bibfnamefont {J.}~\bibnamefont {Rabier}},\ }\href {\doibase
  10.1103/PhysRevB.65.104526} {\bibfield  {journal} {\bibinfo  {journal} {Phys.
  Rev.~B}\ }\textbf {\bibinfo {volume} {65}},\ \bibinfo {pages} {104526}
  (\bibinfo {year} {2002})}\BibitemShut {NoStop}%
\bibitem [{\citenamefont {Deak}\ \emph {et~al.}(1994)\citenamefont {Deak},
  \citenamefont {McElfresh}, \citenamefont {Clem}, \citenamefont {Hao},
  \citenamefont {M.~Konczykowski}, \citenamefont {Foltyn},\ and\ \citenamefont
  {Dye}}]{Deak-PRB-1994}%
  \BibitemOpen
  \bibfield  {author} {\bibinfo {author} {\bibfnamefont {J.}~\bibnamefont
  {Deak}}, \bibinfo {author} {\bibfnamefont {M.}~\bibnamefont {McElfresh}},
  \bibinfo {author} {\bibfnamefont {J.~R.}\ \bibnamefont {Clem}}, \bibinfo
  {author} {\bibfnamefont {Z.}~\bibnamefont {Hao}}, \bibinfo {author}
  {\bibfnamefont {R.~M.}\ \bibnamefont {M.~Konczykowski}}, \bibinfo {author}
  {\bibfnamefont {S.}~\bibnamefont {Foltyn}}, \ and\ \bibinfo {author}
  {\bibfnamefont {R.}~\bibnamefont {Dye}},\ }\href {\doibase
  10.1103/PhysRevB.49.6270} {\bibfield  {journal} {\bibinfo  {journal} {Phys.
  Rev.~B}\ }\textbf {\bibinfo {volume} {49}},\ \bibinfo {pages} {6270}
  (\bibinfo {year} {1994})}\BibitemShut {NoStop}%
\bibitem [{\citenamefont {Schilling}\ \emph {et~al.}(1990)\citenamefont
  {Schilling}, \citenamefont {Hulliger},\ and\ \citenamefont
  {Ott}}]{Schilling-PC-1990}%
  \BibitemOpen
  \bibfield  {author} {\bibinfo {author} {\bibfnamefont {A.}~\bibnamefont
  {Schilling}}, \bibinfo {author} {\bibfnamefont {F.}~\bibnamefont {Hulliger}},
  \ and\ \bibinfo {author} {\bibfnamefont {H.}~\bibnamefont {Ott}},\ }\href
  {\doibase 10.1016/0921-4534(90)90516-H} {\bibfield  {journal} {\bibinfo
  {journal} {Physica~C}\ }\textbf {\bibinfo {volume} {168}},\ \bibinfo {pages}
  {272} (\bibinfo {year} {1990})}\BibitemShut {NoStop}%
\bibitem [{\citenamefont {Zimmermann}\ \emph {et~al.}(1995)\citenamefont
  {Zimmermann}, \citenamefont {Keller}, \citenamefont {Lee}, \citenamefont
  {Savi\'c}, \citenamefont {Warden}, \citenamefont {Zech}, \citenamefont
  {Cubitt}, \citenamefont {Forgan}, \citenamefont {Kaldis}, \citenamefont
  {Karpinski},\ and\ \citenamefont {Kr\"uger}}]{Zimmermann-PRB-1995}%
  \BibitemOpen
  \bibfield  {author} {\bibinfo {author} {\bibfnamefont {P.}~\bibnamefont
  {Zimmermann}}, \bibinfo {author} {\bibfnamefont {H.}~\bibnamefont {Keller}},
  \bibinfo {author} {\bibfnamefont {S.~L.}\ \bibnamefont {Lee}}, \bibinfo
  {author} {\bibfnamefont {I.~M.}\ \bibnamefont {Savi\'c}}, \bibinfo {author}
  {\bibfnamefont {M.}~\bibnamefont {Warden}}, \bibinfo {author} {\bibfnamefont
  {D.}~\bibnamefont {Zech}}, \bibinfo {author} {\bibfnamefont {R.}~\bibnamefont
  {Cubitt}}, \bibinfo {author} {\bibfnamefont {E.~M.}\ \bibnamefont {Forgan}},
  \bibinfo {author} {\bibfnamefont {E.}~\bibnamefont {Kaldis}}, \bibinfo
  {author} {\bibfnamefont {J.}~\bibnamefont {Karpinski}}, \ and\ \bibinfo
  {author} {\bibfnamefont {C.}~\bibnamefont {Kr\"uger}},\ }\href {\doibase
  10.1103/PhysRevB.52.541} {\bibfield  {journal} {\bibinfo  {journal} {Phys.
  Rev.~B}\ }\textbf {\bibinfo {volume} {52}},\ \bibinfo {pages} {541} (\bibinfo
  {year} {1995})}\BibitemShut {NoStop}%
\bibitem [{\citenamefont {Hao}\ \emph {et~al.}(1991)\citenamefont {Hao},
  \citenamefont {Clem}, \citenamefont {McElfresh}, \citenamefont {Civale},
  \citenamefont {Malozemoff},\ and\ \citenamefont {Holtzberg}}]{Hao91:1}%
  \BibitemOpen
  \bibfield  {author} {\bibinfo {author} {\bibfnamefont {Z.}~\bibnamefont
  {Hao}}, \bibinfo {author} {\bibfnamefont {J.~R.}\ \bibnamefont {Clem}},
  \bibinfo {author} {\bibfnamefont {M.~W.}\ \bibnamefont {McElfresh}}, \bibinfo
  {author} {\bibfnamefont {L.}~\bibnamefont {Civale}}, \bibinfo {author}
  {\bibfnamefont {A.~P.}\ \bibnamefont {Malozemoff}}, \ and\ \bibinfo {author}
  {\bibfnamefont {F.}~\bibnamefont {Holtzberg}},\ }\href {\doibase
  10.1103/PhysRevB.43.2844} {\bibfield  {journal} {\bibinfo  {journal} {Phys.
  Rev. B}\ }\textbf {\bibinfo {volume} {43}},\ \bibinfo {pages} {2844}
  (\bibinfo {year} {1991})}\BibitemShut {NoStop}%
\bibitem [{\citenamefont {Jorgensen}\ \emph {et~al.}(1990)\citenamefont
  {Jorgensen}, \citenamefont {Veal}, \citenamefont {Paulikas}, \citenamefont
  {Nowicki}, \citenamefont {Crabtree}, \citenamefont {Claus},\ and\
  \citenamefont {Kwok}}]{Jorgensen-PRB-1990}%
  \BibitemOpen
  \bibfield  {author} {\bibinfo {author} {\bibfnamefont {J.~D.}\ \bibnamefont
  {Jorgensen}}, \bibinfo {author} {\bibfnamefont {B.~W.}\ \bibnamefont {Veal}},
  \bibinfo {author} {\bibfnamefont {A.~P.}\ \bibnamefont {Paulikas}}, \bibinfo
  {author} {\bibfnamefont {L.~J.}\ \bibnamefont {Nowicki}}, \bibinfo {author}
  {\bibfnamefont {G.~W.}\ \bibnamefont {Crabtree}}, \bibinfo {author}
  {\bibfnamefont {H.}~\bibnamefont {Claus}}, \ and\ \bibinfo {author}
  {\bibfnamefont {W.~K.}\ \bibnamefont {Kwok}},\ }\href {\doibase
  10.1103/PhysRevB.41.1863} {\bibfield  {journal} {\bibinfo  {journal} {Phys.
  Rev.~B}\ }\textbf {\bibinfo {volume} {41}},\ \bibinfo {pages} {1863}
  (\bibinfo {year} {1990})}\BibitemShut {NoStop}%
\bibitem [{\citenamefont {Bosma}\ \emph {et~al.}(2011)\citenamefont {Bosma},
  \citenamefont {Weyeneth}, \citenamefont {Puzniak}, \citenamefont {Erb},
  \citenamefont {Schilling},\ and\ \citenamefont {Keller}}]{Bosma-PRB-2011}%
  \BibitemOpen
  \bibfield  {author} {\bibinfo {author} {\bibfnamefont {S.}~\bibnamefont
  {Bosma}}, \bibinfo {author} {\bibfnamefont {S.}~\bibnamefont {Weyeneth}},
  \bibinfo {author} {\bibfnamefont {R.}~\bibnamefont {Puzniak}}, \bibinfo
  {author} {\bibfnamefont {A.}~\bibnamefont {Erb}}, \bibinfo {author}
  {\bibfnamefont {A.}~\bibnamefont {Schilling}}, \ and\ \bibinfo {author}
  {\bibfnamefont {H.}~\bibnamefont {Keller}},\ }\href {\doibase
  10.1103/PhysRevB.84.024514} {\bibfield  {journal} {\bibinfo  {journal} {Phys.
  Rev.~B}\ }\textbf {\bibinfo {volume} {84}},\ \bibinfo {pages} {024514}
  (\bibinfo {year} {2011})}\BibitemShut {NoStop}%
\bibitem [{\citenamefont {Barone}\ \emph {et~al.}(1990)\citenamefont {Barone},
  \citenamefont {Larkin},\ and\ \citenamefont {Ovchinnikov}}]{Barone-JS-1990}%
  \BibitemOpen
  \bibfield  {author} {\bibinfo {author} {\bibfnamefont {A.}~\bibnamefont
  {Barone}}, \bibinfo {author} {\bibfnamefont {A.~I.}\ \bibnamefont {Larkin}},
  \ and\ \bibinfo {author} {\bibfnamefont {Y.~N.}\ \bibnamefont
  {Ovchinnikov}},\ }\href {\doibase 10.1007/BF00624502} {\bibfield  {journal}
  {\bibinfo  {journal} {J.~Supersond.}\ }\textbf {\bibinfo {volume} {3}},\
  \bibinfo {pages} {155} (\bibinfo {year} {1990})}\BibitemShut {NoStop}%
\bibitem [{\citenamefont {Ivlev}\ and\ \citenamefont
  {Kopnin}(1990)}]{Ivlev-JLTP-1990}%
  \BibitemOpen
  \bibfield  {author} {\bibinfo {author} {\bibfnamefont {B.~I.}\ \bibnamefont
  {Ivlev}}\ and\ \bibinfo {author} {\bibfnamefont {N.~B.}\ \bibnamefont
  {Kopnin}},\ }\href {\doibase 10.1007/BF00683483} {\bibfield  {journal}
  {\bibinfo  {journal} {J.~Low Temp. Phys.}\ }\textbf {\bibinfo {volume}
  {80}},\ \bibinfo {pages} {161} (\bibinfo {year} {1990})}\BibitemShut
  {NoStop}%
\bibitem [{\citenamefont {Feinberg}\ and\ \citenamefont
  {Villard}(1990{\natexlab{a}})}]{Feinberg-MPLB-1990}%
  \BibitemOpen
  \bibfield  {author} {\bibinfo {author} {\bibfnamefont {D.}~\bibnamefont
  {Feinberg}}\ and\ \bibinfo {author} {\bibfnamefont {C.}~\bibnamefont
  {Villard}},\ }\href {\doibase 10.1142/S0217984990000039} {\bibfield
  {journal} {\bibinfo  {journal} {Mod. Phys. Lett.~B}\ }\textbf {\bibinfo
  {volume} {4}},\ \bibinfo {pages} {9} (\bibinfo {year}
  {1990}{\natexlab{a}})}\BibitemShut {NoStop}%
\bibitem [{\citenamefont {Feinberg}\ and\ \citenamefont
  {Villard}(1990{\natexlab{b}})}]{Feinberg-PRl-1990}%
  \BibitemOpen
  \bibfield  {author} {\bibinfo {author} {\bibfnamefont {D.}~\bibnamefont
  {Feinberg}}\ and\ \bibinfo {author} {\bibfnamefont {C.}~\bibnamefont
  {Villard}},\ }\href {\doibase 10.1103/PhysRevLett.65.919} {\bibfield
  {journal} {\bibinfo  {journal} {Phys. Rev. Lett.}\ }\textbf {\bibinfo
  {volume} {65}},\ \bibinfo {pages} {919} (\bibinfo {year}
  {1990}{\natexlab{b}})}\BibitemShut {NoStop}%
\bibitem [{\citenamefont {Brandt}(1998)}]{Brandt-PRB-1998-1}%
  \BibitemOpen
  \bibfield  {author} {\bibinfo {author} {\bibfnamefont {E.~H.}\ \bibnamefont
  {Brandt}},\ }\href {http://link.aps.org/doi/10.1103/PhysRevB.58.6506}
  {\bibfield  {journal} {\bibinfo  {journal} {Phys. Rev. B}\ }\textbf {\bibinfo
  {volume} {58}},\ \bibinfo {pages} {6506} (\bibinfo {year}
  {1998})}\BibitemShut {NoStop}%
\bibitem [{EDP()}]{EDP}%
  \BibitemOpen
  \href@noop {} {}\bibinfo {note} {Eq.~\protect\eqref{eq:Jd(TB)} is valid for
  applied fields $H<10$~kOe, at high fields the dependence $J_{cd}(T,B)$ is
  more complicated since a dislocation density becomes greater than the vortex
  density and some function is needed to accommodate vortices and
  dislocations\cite{Pan-PRB-2006, Kosse-SST-2008}}\BibitemShut {NoStop}%
\bibitem [{vdB()}]{vdBeek}%
  \BibitemOpen
  \href@noop {} {}\bibinfo {note} {Expression similar
  to~\protect\eqref{eq:Ji(TB)} was first obtained from OI theory in
  Ref.~\onlinecite{vanderBeek-PRB-2002} but authors incorrectly approximated
  field dependence of the current as $J_{ci}\propto b^{-5/8}$.}\BibitemShut
  {Stop}%
\bibitem [{\citenamefont {Golosovsky}\ \emph {et~al.}(1994)\citenamefont
  {Golosovsky}, \citenamefont {Tsindlekht}, \citenamefont {Chayet},\ and\
  \citenamefont {Davidov}}]{Golosovsky-PRB-1994}%
  \BibitemOpen
  \bibfield  {author} {\bibinfo {author} {\bibfnamefont {M.}~\bibnamefont
  {Golosovsky}}, \bibinfo {author} {\bibfnamefont {M.}~\bibnamefont
  {Tsindlekht}}, \bibinfo {author} {\bibfnamefont {H.}~\bibnamefont {Chayet}},
  \ and\ \bibinfo {author} {\bibfnamefont {D.}~\bibnamefont {Davidov}},\ }\href
  {\doibase 10.1103/PhysRevB.50.470} {\bibfield  {journal} {\bibinfo  {journal}
  {Phys. Rev. B}\ }\textbf {\bibinfo {volume} {50}},\ \bibinfo {pages} {470}
  (\bibinfo {year} {1994})}\BibitemShut {NoStop}%
\bibitem [{\citenamefont {Maeda}\ \emph {et~al.}(2005)\citenamefont {Maeda},
  \citenamefont {Kitano},\ and\ \citenamefont {Inoue}}]{Maeda-JPCM-2005}%
  \BibitemOpen
  \bibfield  {author} {\bibinfo {author} {\bibfnamefont {A.}~\bibnamefont
  {Maeda}}, \bibinfo {author} {\bibfnamefont {H.}~\bibnamefont {Kitano}}, \
  and\ \bibinfo {author} {\bibfnamefont {R.}~\bibnamefont {Inoue}},\ }\href
  {http://iopscience.iop.org/0953-8984/17/4/R01} {\bibfield  {journal}
  {\bibinfo  {journal} {J.~Phys.: Condens. Matter}\ }\textbf {\bibinfo {volume}
  {17}},\ \bibinfo {pages} {R143} (\bibinfo {year} {2005})}\BibitemShut
  {NoStop}%
\bibitem [{\citenamefont {Golovchanskiy}\ \emph {et~al.}(2013)\citenamefont
  {Golovchanskiy}, \citenamefont {Pan}, \citenamefont {Shcherbakova},\ and\
  \citenamefont {Fedoseev}}]{Golovchanskiy-JAP-2013}%
  \BibitemOpen
  \bibfield  {author} {\bibinfo {author} {\bibfnamefont {I.~A.}\ \bibnamefont
  {Golovchanskiy}}, \bibinfo {author} {\bibfnamefont {A.~V.}\ \bibnamefont
  {Pan}}, \bibinfo {author} {\bibfnamefont {O.~V.}\ \bibnamefont
  {Shcherbakova}}, \ and\ \bibinfo {author} {\bibfnamefont {S.~A.}\
  \bibnamefont {Fedoseev}},\ }\href {\doibase 10.1063/1.4826531} {\bibfield
  {journal} {\bibinfo  {journal} {J.~Appl. Phys.}\ }\textbf {\bibinfo {volume}
  {114}},\ \bibinfo {pages} {164910} (\bibinfo {year} {2013})}\BibitemShut
  {NoStop}%
\bibitem [{\citenamefont {Darhmaoui}\ and\ \citenamefont
  {Jung}(1996)}]{Darhmaoui-PRB-1996}%
  \BibitemOpen
  \bibfield  {author} {\bibinfo {author} {\bibfnamefont {H.}~\bibnamefont
  {Darhmaoui}}\ and\ \bibinfo {author} {\bibfnamefont {J.}~\bibnamefont
  {Jung}},\ }\href {\doibase 10.1103/PhysRevB.53.14621} {\bibfield  {journal}
  {\bibinfo  {journal} {Phys. Rev.~B}\ }\textbf {\bibinfo {volume} {53}},\
  \bibinfo {pages} {14621} (\bibinfo {year} {1996})}\BibitemShut {NoStop}%
\bibitem [{\citenamefont {Darhmaoui}\ and\ \citenamefont
  {Jung}(1998)}]{Darhmaoui-PRB-1998}%
  \BibitemOpen
  \bibfield  {author} {\bibinfo {author} {\bibfnamefont {H.}~\bibnamefont
  {Darhmaoui}}\ and\ \bibinfo {author} {\bibfnamefont {J.}~\bibnamefont
  {Jung}},\ }\href {\doibase 10.1103/PhysRevB.57.8009} {\bibfield  {journal}
  {\bibinfo  {journal} {Phys. Rev.~B}\ }\textbf {\bibinfo {volume} {57}},\
  \bibinfo {pages} {8009} (\bibinfo {year} {1998})}\BibitemShut {NoStop}%
\end{thebibliography}
\end{document}